\documentclass[aps,pre,longbibliography,showpacs,amsmath,amssymb]{revtex4-1}
\usepackage{amsmath,amssymb,latexsym,epsfig,epsf,bm}
\usepackage{graphicx}
\usepackage{float}

\newcommand{\bR}{\bold R}

\newcommand{\bRa}{{\bf R}_{\rm a}}

\newcommand{\bRb}{{\bf R}_{\rm b}}

\newcommand{\tbR}{{\tilde{\bold R}}}

\newcommand{\be}{\begin{equation}}
\newcommand{\ee}{\end{equation}}
\newcommand{\fig}[1]{Fig.~\ref{#1}}
\newcommand{\Fig}[1]{Figure~\ref{#1}}
\newcommand{\eq}[1]{Eq.~(\ref{#1})}
\newcommand{\Eq}[1]{Equation~(\ref{#1})}

\newcommand{\tcVex}{{\tilde{c}_V^{\rm ex}}}
\newcommand{\Sex}{{S}_{\rm ex}}

\usepackage{nicefrac}

\begin{document}
	\title{The EXP pair-potential system. II. Fluid phase isomorphs}
	\date{\today}
	\author{Andreas Kvist Bacher}\email{ankvba@gmail.com}
	\author{Thomas B. Schr{\o}der}
	\author{Jeppe C. Dyre}\email{dyre@ruc.dk}
	\affiliation{Glass and Time, IMFUFA, Department of Science and Environment, Roskilde University, P.O. Box 260, DK-4000 Roskilde, Denmark}
	
\begin{abstract}
This paper continues the investigation of the exponentially repulsive EXP pair-potential system  of Paper I with a focus on isomorphs in the low-temperature gas and liquid phases. As expected from the EXP system's strong virial potential-energy correlations, the system's reduced-unit structure and dynamics are isomorph invariant to a good approximation. Three methods for generating isomorphs are compared: the small-step method that is exact in the limit of small density changes and two versions of the direct-isomorph-check method that allows for much larger density changes. Results from approximate methods are compared to those of the small-step method for each of three isomorphs generated by 230 one percent density increases, covering in all one decade of density variation. Both approximate methods work well.
\end{abstract}
\maketitle

\section{Introduction}\label{sec:intro}

This paper and its companion (Paper I, \cite{EXP_I_arXiv}) present an investigation of the EXP pair-potential system consisting of identical particles interacting via the purely repulsive pair potential,

\be\label{EXPppX}
v_{\rm EXP}(r)\,=\,\varepsilon\, e^{-r/\sigma}\,.
\ee
Both papers investigate the region of the thermodynamic phase diagram where temperature is so low that the finite value $v_{\rm EXP}(0)$ plays little role for the physics, i.e., where $k_BT\ll\varepsilon$. The focus is, moreover, on the low-density gas and liquid phases, i.e., where  $\rho\sigma^3\ll 1$. While the densities considered are low in relation to the $\sigma$ parameter of \eq{EXPppX}, it should be emphasized that at low temperature the densities are not, in fact, low relative to the effective hard-sphere radius, but actually typical for usual studies of simple fluid gas-solid systems.

The EXP pair potential has not been studied much on its own right, in fact even less than other purely repulsive pair potentials like the family of inverse-power law pair potentials \cite{hoo72,hiw74,hey07,hey08,lan09,bra11,pie14,din15}. In most cases, the exponential function appears as a term in mathematically more involved potentials, for instance: 1) giving the repulsive part of the Born-Meyer pair potential from 1932 \cite{bor32} or in embedded-atom models of metals \cite{foi86,lad17}, 2) multiplied by a Coulomb term to give the Yukawa (screened Coulomb) potential \cite{yuk35,row89}, or 3) giving the attractive long-ranged part in a model that rigorously obeys the van der Waals equation of state in one dimension \cite{kac63}.

To a good approximation the EXP system conforms to the following ``hidden-scale invariance'' condition for uniform scaling of same-density configurations $\bRa$ and $\bRb$ \cite{sch14} ($\bR$ is the vector of all particle coordinates and $U(\bR)$ the system's potential-energy function):

\be\label{hs}
U(\bRa)<U(\bRb)\,\Rightarrow\, U(\lambda\bRa)<U(\lambda\bRb)\,.
\ee
Here $\lambda>0$ is a scaling parameter. \Eq{hs} expresses that if the potential energy of configuration $\bRa$ is lower than that of configuration $\bRb$, this applies also after a uniform scaling of both configurations. For most systems, including the EXP system, \eq{hs} is approximate and not obvious from the mathematical expression for the potential energy, hence the term ``hidden scale invariance'' \cite{sch09,dyr14}. 

Paper I demonstrated one of the consequences of \eq{hs} for the EXP pair potential, namely strong virial potential-energy correlations in the constant-volume thermal equilibrium fluctuations. The EXP pair-potential system has stronger such correlations than, e.g., the Lennard-Jones system \cite{ped08,I}. In fact, the EXP system's virial potential-energy Pearson correlation coefficient $R$ is larger than 99\% in a large part of its phase diagram (see Paper I and Refs. \onlinecite{I,bac14a}). A system is termed ``R-simple'' if it has better than 90\% correlation.

The fact that the EXP pair-potential system obeys \eq{hs} implies that it has isomorphs, which are curves in the phase diagram along which structure and dynamics are invariant to a good approximation when given in reduced units \cite{IV,sch14}. Isomorphs are defined as curves of constant excess entropy, i.e., they are the system's configurational adiabats. While all systems have configurational adiabats, only R-simple systems, i.e., those obeying the hidden-scale-invariance condition \eq{hs} to a good approximation, have invariant physics along their configurational adiabats \cite{ros77,IV,ing12,sch14,dyr16}. 

A derivation of simple liquids' quasiuniversality based on the EXP pair potential was given in Refs. \onlinecite{bac14a,dyr16}; an alternative proof utilizing constant-potential-energy ($NVU$) dynamics \cite{NVU_I} was presented in Paper I. Both proofs are based on the fact that under certain conditions a sum of two EXP pair potentials describes a system, which to a good approximation has the same physics as that of a single EXP pair-potential system. From this one easily shows that quasiuniversality applies for any system with a pair potential that to a good approximation may be written as a sum of EXP terms with numerically large prefactors in reduced units (see below) \cite{bac14a}. The EXP pair-potential system is thus central for explaining the physics of simple liquids, which justifies a closer investigation of the properties of the EXP system itself.

The isomorph theory is based on the use of reduced units \cite{and31,hoo72,ros77,IV}, which are different from those usually applied for presenting simulation data involving the potential-energy function's characteristic energy and length. Instead reduced units utilize \textit{macroscopic} parameters that vary with the thermodynamic state point. Consider a state point of temperature $T$ and density $\rho=N/V$ ($N$ is the number of particles and $V$ the system volume). If the average particle mass is $m$, reduced units make quantities dimensionless by scaling with the length $l_0=\rho^{-1/3}$, the energy $e_0=k_BT$, and the time $t_0=\rho^{-1/3}\sqrt{m/k_BT}$ \cite{IV,ing12,dyr16}. Reduced quantities are denoted by a tilde, for instance $\tbR\equiv\bR/l_0=\rho^{1/3}\bR$ is the reduced configuration vector. We can now make precise the statement that the isomorphs of an R-simple system are lines of virtually identical physic; this refers to the system's reduced-unit structure and dynamics.

Besides demonstrating approximate isomorph invariance of the EXP system's structure and dynamics, the present paper discusses methods for generating the EXP system's isomorphs in computer simulations. Before doing this we demonstrate numerically in Sec. \ref{sec:hsinv} examples of the system's hidden scale invariance, and Sec. \ref{sec:gammaisochores} presents results for the density-scaling exponent's variation throughout the thermodynamic phase diagram (the isomorph slope in the log-log density-temperature phase diagram). The next sections report results from computer simulations along isomorphs traced out in different ways. Sec. \ref{sec:isom} presents the  ``small-step'' method which, in the limit of infinitely small density changes, rigorously identifies configurational adiabats; Sec. \ref{sec:DIC} discusses two versions of the so-called direct isomorph check method, both allowing for much larger density changes. Finally, Sec. \ref{sec:sum} gives a brief discussion.

\section{The EXP system's hidden scale invariance}\label{sec:hsinv}

The excess entropy $\Sex$ of a thermodynamic state point is defined as the entropy minus that of an ideal gas at the same density and temperature \cite{han13} (note that $\Sex<0$ since no system is more disordered than an ideal gas). $\Sex$ is the non-trivial part of a system's entropy, the contribution deriving from interactions. In general, excess thermodynamic quantities are calculated by leaving out the momentum degrees of freedom in the partition function \cite{han13}; they obey all standard thermodynamic relations like $T=(\partial U/\partial\Sex)_\rho$, etc. 

If one defines the microscopic excess entropy function $\Sex(\bR)$ as the thermodynamic equilibrium excess entropy of the state point with the density $\rho$ of the configuration $\bR$ and with average potential energy equal to $U(\bR)$, it is straightforward to show that the hidden-scale-invariance condition \eq{hs} implies a scale-invariant entropy function, i.e., $\Sex(\bR) =\Sex(\lambda\bR)$ in which $\lambda$ is a scaling parameter \cite{sch14}. This means that $\Sex(\bR)$ depends only on the configuration's reduced coordinates, $\tbR=\rho^{1/3}\bR$. Inserting this into the identity $U(\bR)=U(\rho,\Sex(\bR))$ that defines $\Sex(\bR)$ (the function $U(\rho,\Sex)$ is the thermodynamic average potential energy as a function of density and excess entropy), one arrives \cite{sch14} at 

\be\label{Rfundeq}
U(\bR)
\,=\, U(\rho,\Sex(\tbR)) \,.
\ee
This is the basic identity characterizing an R-simple system from which the isomorph invariance of structure and dynamics follow \cite{sch14}. This does not imply, however, that all of the excess thermodynamics is isomorph invariant since there is also a density dependence in \eq{Rfundeq}. For instance, the Helmholtz and Gibbs free energies are not isomorph invariant. A recent paper detailing how to use the isomorph theory for predicting how a number of quantities vary along the melting line gives an example of how the Gibbs free energy variation along a Lennard-Jones system isomorph is determined \cite{ped16}.  

Physically, the hidden-scale-invariance condition \eq{hs} states that if configurations at some density are ordered according to their potential energy, this ordering is maintained if the configurations are scaled uniformly to a different density. Note that \eq{hs} -- and thus \eq{Rfundeq} and the entire isomorph theory -- are approximate except for systems with an Euler-homogeneous potential-energy function (plus a constant). 

A consequence of \eq{Rfundeq} is that an R-simple system has strong correlations between its constant-density thermal-equilibrium fluctuations of virial and potential energy. This property was documented for the EXP system in Ref. \cite{bac14a} and in Paper I, and previously also for many other systems, including molecular and polymeric systems \cite{ped08,I,ing12b,vel14}. Recall that the microscopic virial $W(\bR)$ is defined from the change of potential energy upon a uniform scaling of all particle coordinates, i.e., $W(\bR)\equiv(\partial U(\bR)/\partial\ln\rho)_\tbR$ since a uniform scaling leaves $\tbR$ unchanged. Substituting \eq{Rfundeq} into this expression one finds

\be\label{virial_ident}
W(\bR)
\,=\,\left.\frac{\partial U(\rho,\Sex)}{\partial\ln\rho}\right|_{\Sex=\Sex(\tbR)}\,.
\ee
In other words, $W(\bR)=W(\rho,\Sex(\tbR))$ in which the function $W(\rho,\Sex)$ is the thermodynamic virial, i.e., the average of the microscopic virial at the state point with density $\rho$ and excess entropy $\Sex$. In conjunction with \eq{Rfundeq}, the identity $W(\bR)=W(\rho,\Sex(\tbR))$ implies perfect correlation between the virial and the potential energy at fixed density in the sense that one of these two quantities uniquely determines the other. Note that this one-to-one relation between $W$ and $U$ is predicted to apply whether or not configurations are selected from an equilibrium simulation, it applies for instance also during aging \cite{III}.

\begin{figure}[!htbp]
	\centering
	\includegraphics[width=0.4\textwidth]{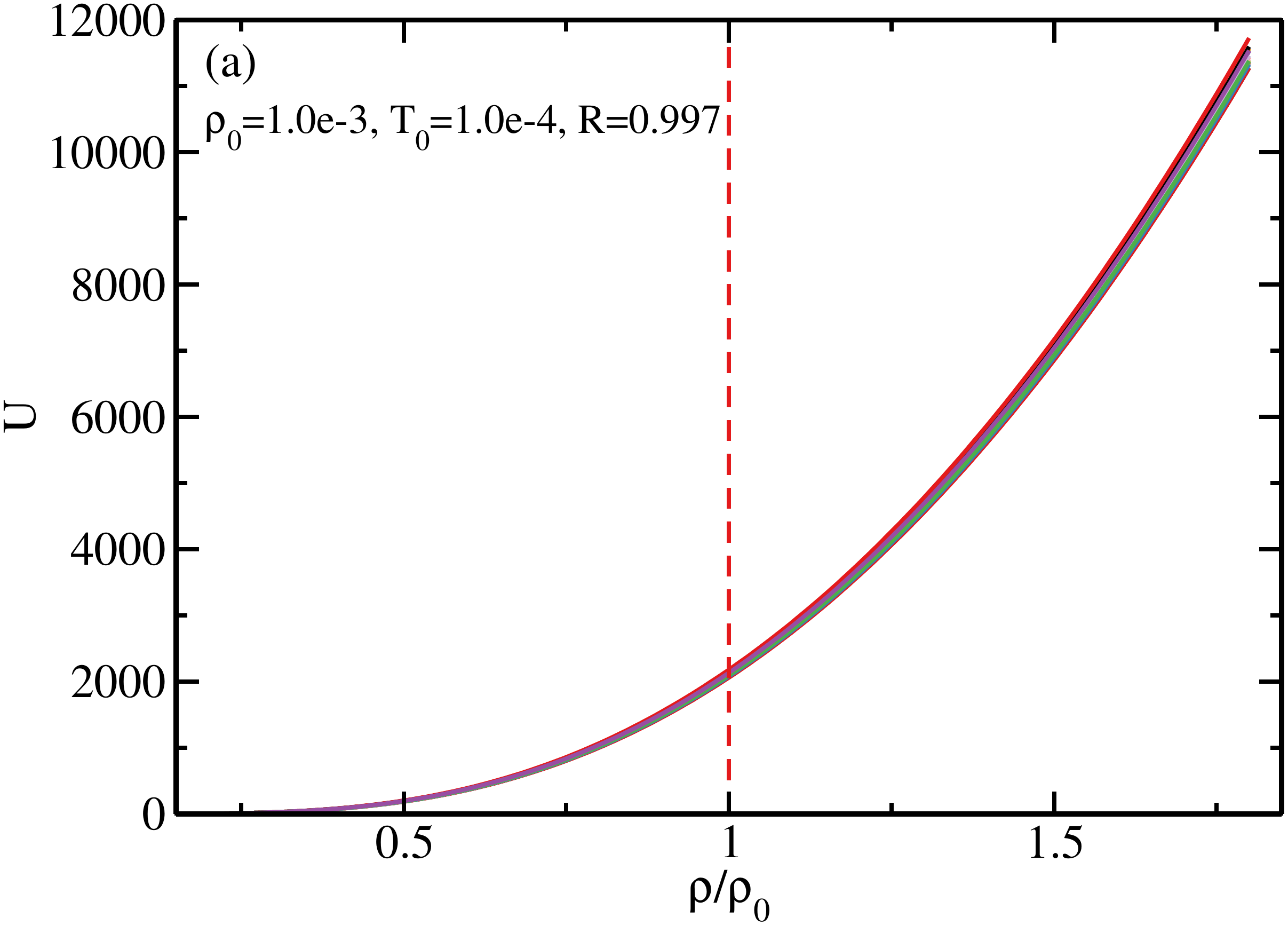}
	\includegraphics[width=0.4\textwidth]{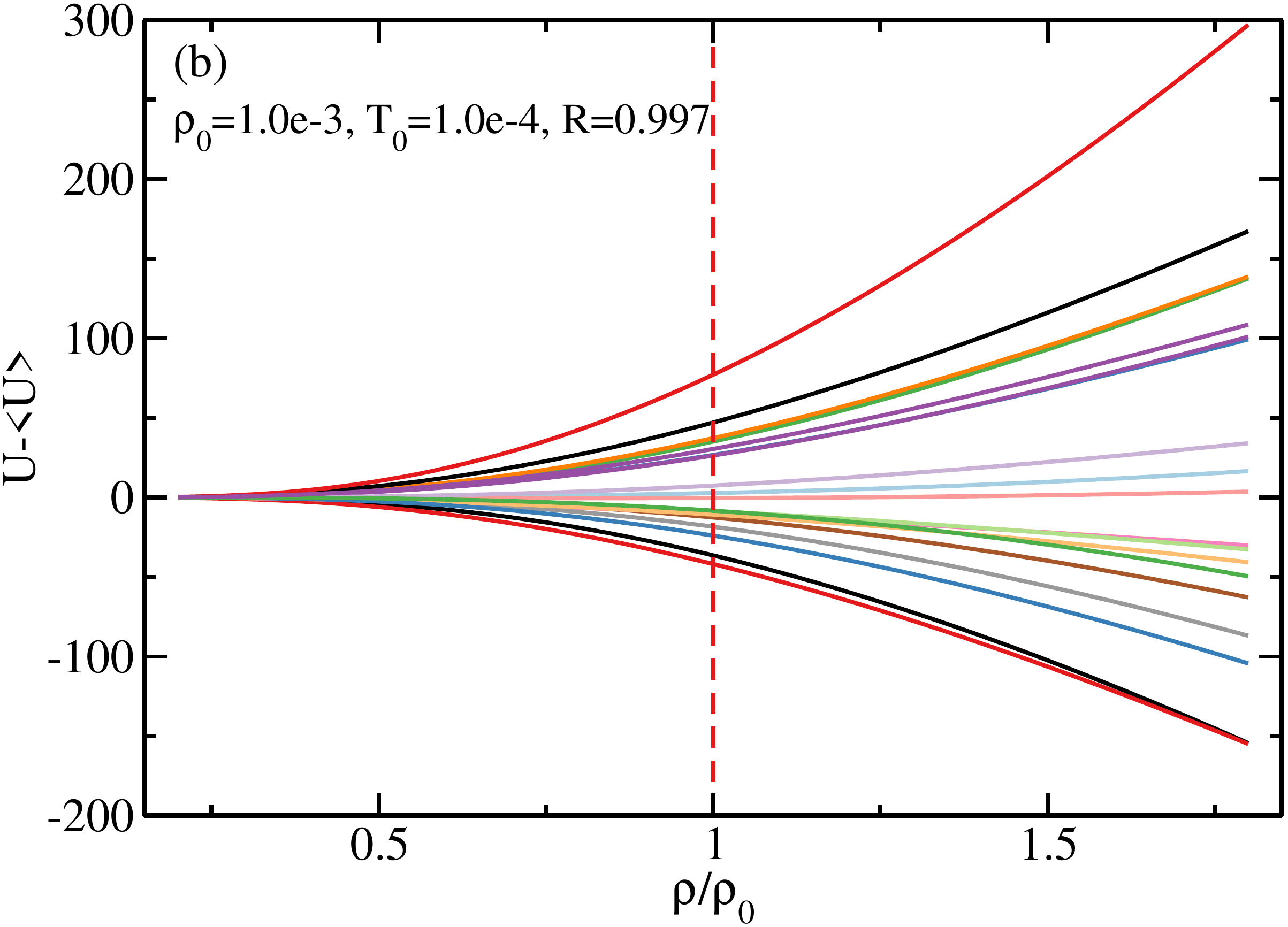}
	\includegraphics[width=0.4\textwidth]{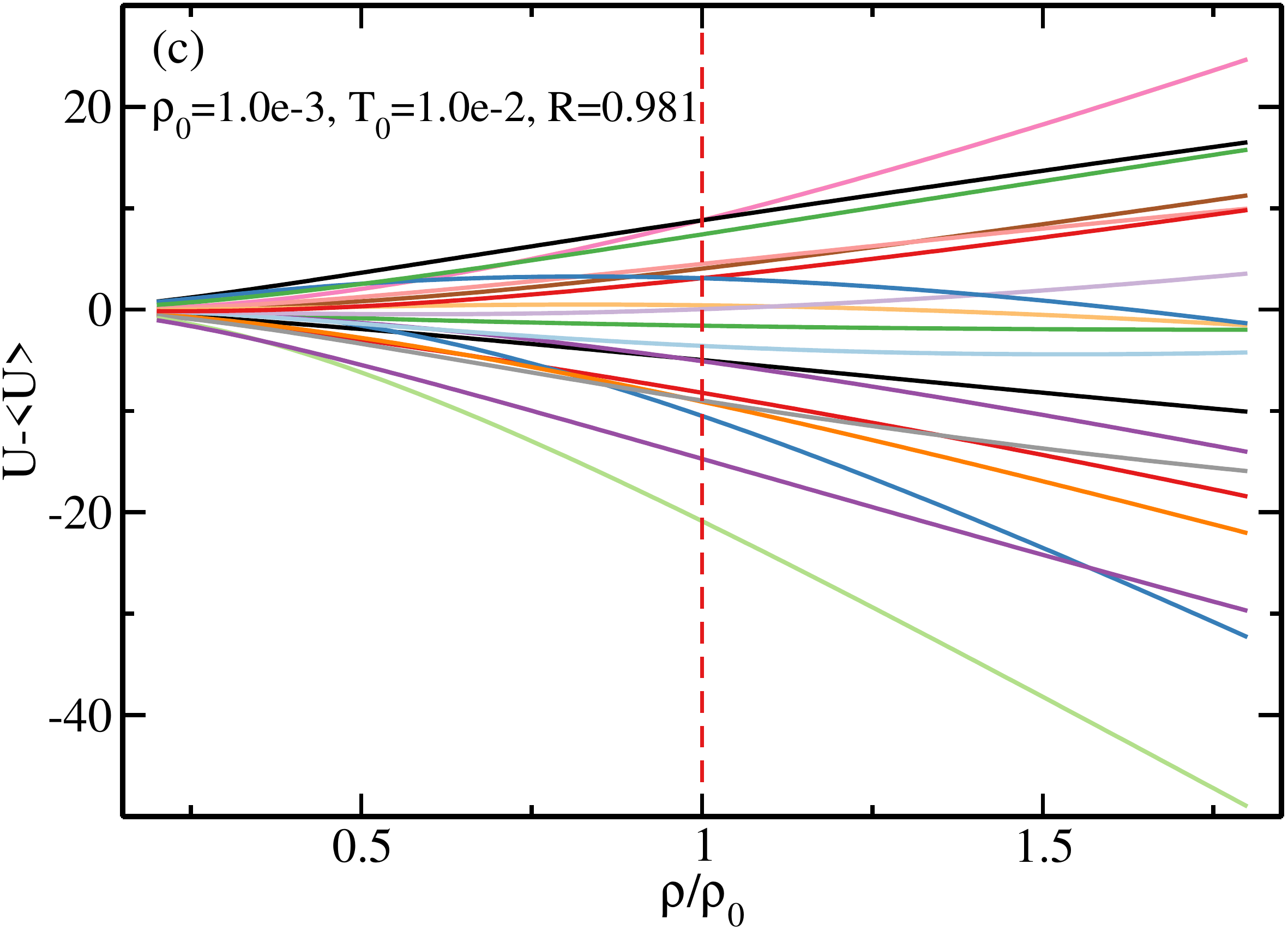}
	\includegraphics[width=0.4\textwidth]{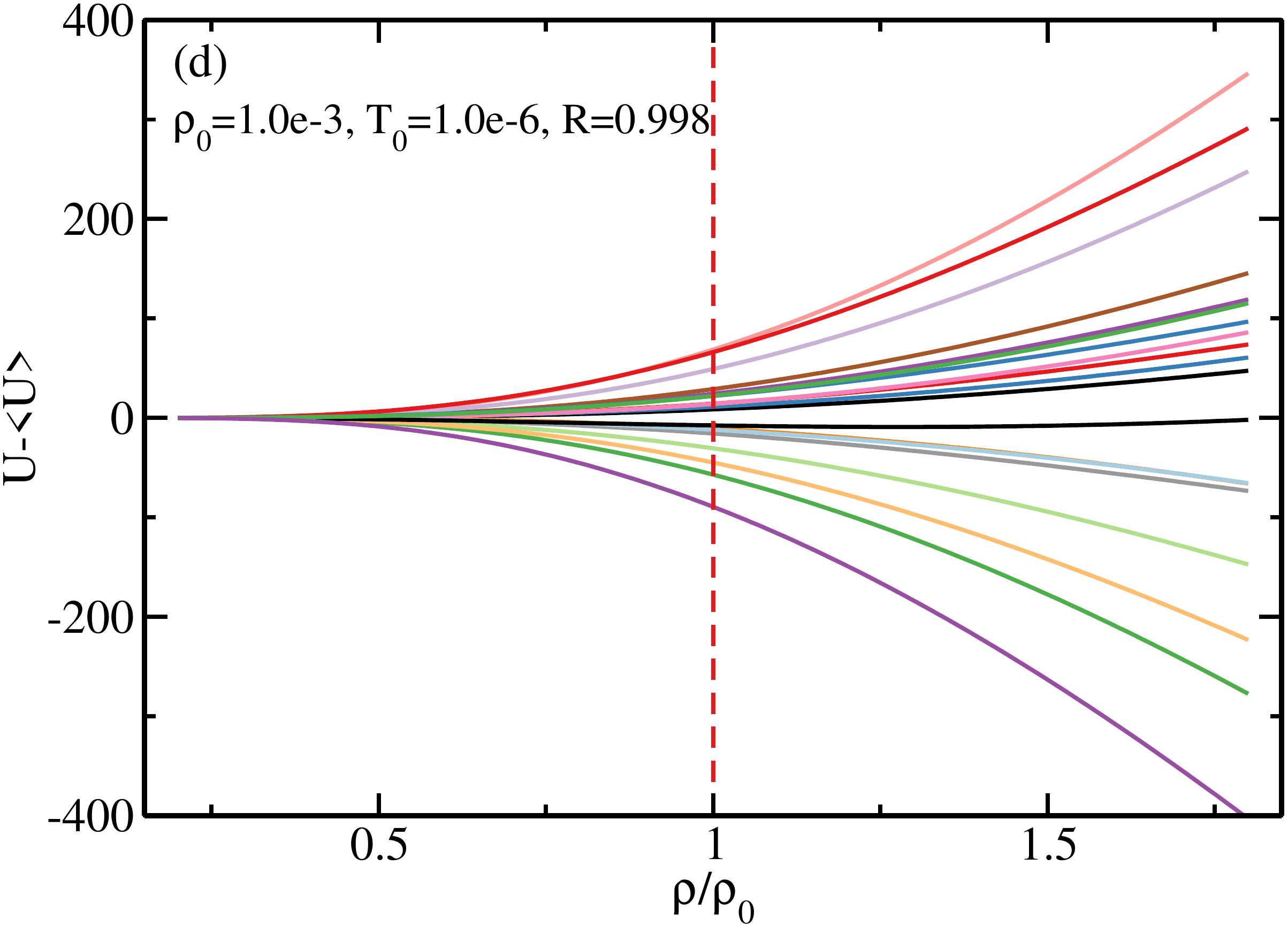}
	\caption{Investigating \eq{hs} for the EXP pair-potential system. $R$ is the virial potential-energy Pearson correlation coefficient. 
		(a) Potential energies of 20 statistically independent configurations from an equilibrium simulation of a system with 1000 particles at density $\rho=10^{-3}$ and temperature $T=10^{-4}$ (red dashed line), scaled uniformly to different densities and plotted as a function of density. 
		(b) Same data with the average potential energy subtracted at each density, making it easier to investigate the implication of \eq{hs} that no crossings take place. This is seen to apply to a good approximation.
		(c) Same as in (b) for configurations selected from a simulation at $\rho=10^{-3}$ and $T=10^{-2}$. There are here more crossings, meaning that \eq{hs} is less accurately obeyed, consistent with the lower $R$. 
		(d) Same as in (b) for configurations from a simulations at $\rho=10^{-3}$ and $T=10^{-6}$; here $R=99.8$\% and \eq{hs} is very well obeyed.}
	\label{fi:Dyre}
\end{figure}

We proceed to demonstrate numerically that the EXP pair-potential system obeys the hidden-scale-invariance condition \eq{hs} to a good approximation. While most quantities below are reported in reduced units, density and temperature are by definition constant in reduced units. This makes it impossible to specify a state point using reduced units for density and temperature so numerical values of the density are reported below in units of $1/\sigma^3$ and numerical temperatures in units of $\varepsilon/k_B$. This is referred to as the ``EXP unit system'' (Paper I).

Figures \ref{fi:Dyre}(a) and (b) show results from an equilibrium simulation at density $\rho=10^{-3}$ and temperature $T=10^{-4}$, a liquid state point close to the melting line (simulation details are provided in Paper I). From the simulations we selected 20 statistically independent configurations (separated by $5\cdot 10^5$ time steps). Each configuration was scaled uniformly to a different density $\rho$ in the range $0.25\cdot 10^{-3}<\rho<1.75\cdot 10^{-3}$, i.e., a factor of seven density variation is involved. \Fig{fi:Dyre}(a) plots the potential energies of the scaled configurations as a function of their density. Note that no new simulations were performed to generate this figure, we merely scaled the 20 configurations uniformly and then evaluated their potential energies. Not surprisingly, for all configurations the potential energy increases strongly with increasing density. While they visually follow each other closely in \fig{fi:Dyre}(a), the figure does not allow for checking \eq{hs}. To do this, \fig{fi:Dyre}(b) plots the same data by subtracting at each density the average potential energy of the 20 scaled configurations, making it possible to use a much smaller unit on the potential-energy axis. There are few crossings of the curves. This confirms that the EXP system obeys \eq{hs} to a good approximation, i.e., that it is R-simple in this region of the thermodynamic phase diagram.

\Fig{fi:Dyre}(c) shows a plot like (b) at the same density but the higher temperature $T=10^{-2}$. Here crossings are more common, implying that \eq{hs} is less accurately obeyed. This is consistent with the finding of Paper I that the virial potential-energy correlation coefficient decreases as temperature increases. Moving in the opposite direction, \fig{fi:Dyre}(d) shows data for $T=10^{-6}$ where the pattern is similar to that of (b), but with even fewer crossings. In summary, \eq{hs} works well at low temperatures, but breaks down gradually as higher temperatures are approached. 

The absence of crossings upon a uniform scaling of all particle coordinates is not trivial. Hidden scale invariance is exact only for inverse power-law pair potentials, but it applies to a good approximation for many pair potential systems, e.g., Lennard-Jones type systems \cite{sch14}. In contrast, there are many crossings, e.g., for the Lennard-Jones Gaussian pair potential for a density change of merely $20$\% \cite{sch14}.

\section{The density-scaling exponent}\label{sec:gammaisochores}

At any given state point one defines the so-called density-scaling exponent $\gamma$ \cite{IV} by

\be\label{dsexp}
\gamma
\,\equiv\,\left(\frac{\partial\ln T }{\partial\ln\rho}\right)_{\Sex}\,.
\ee
For an R-simple system, since isomorphs are curves of constant $\Sex$ \cite{IV}, $\gamma$ gives the slope of the isomorph through a given state point in the log-log density-temperature phase diagram. If $\gamma$ were constant, according to the isomorph theory there would be invariance of the reduced-unit structure and dynamics along the phase diagram's lines of constant $\rho^\gamma/T$. This is the origin of the name ``density-scaling exponent''. Before isomorph theory was developed density scaling was demonstrated experimentally for many glass-forming liquids \cite{rol05}.

\begin{figure}[!htbp]
	\centering
	\includegraphics[width=0.4\textwidth]{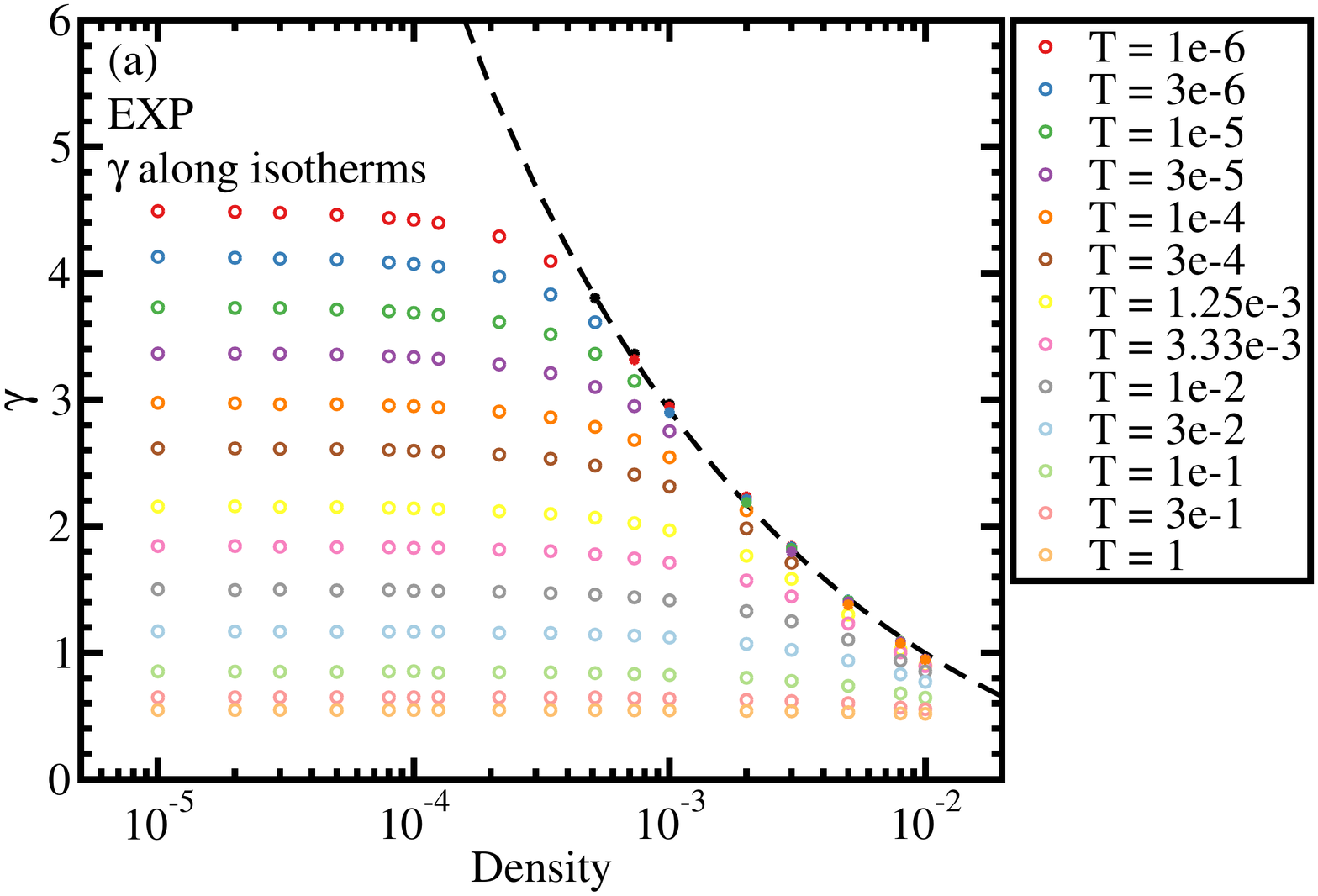}
	\includegraphics[width=0.4\textwidth]{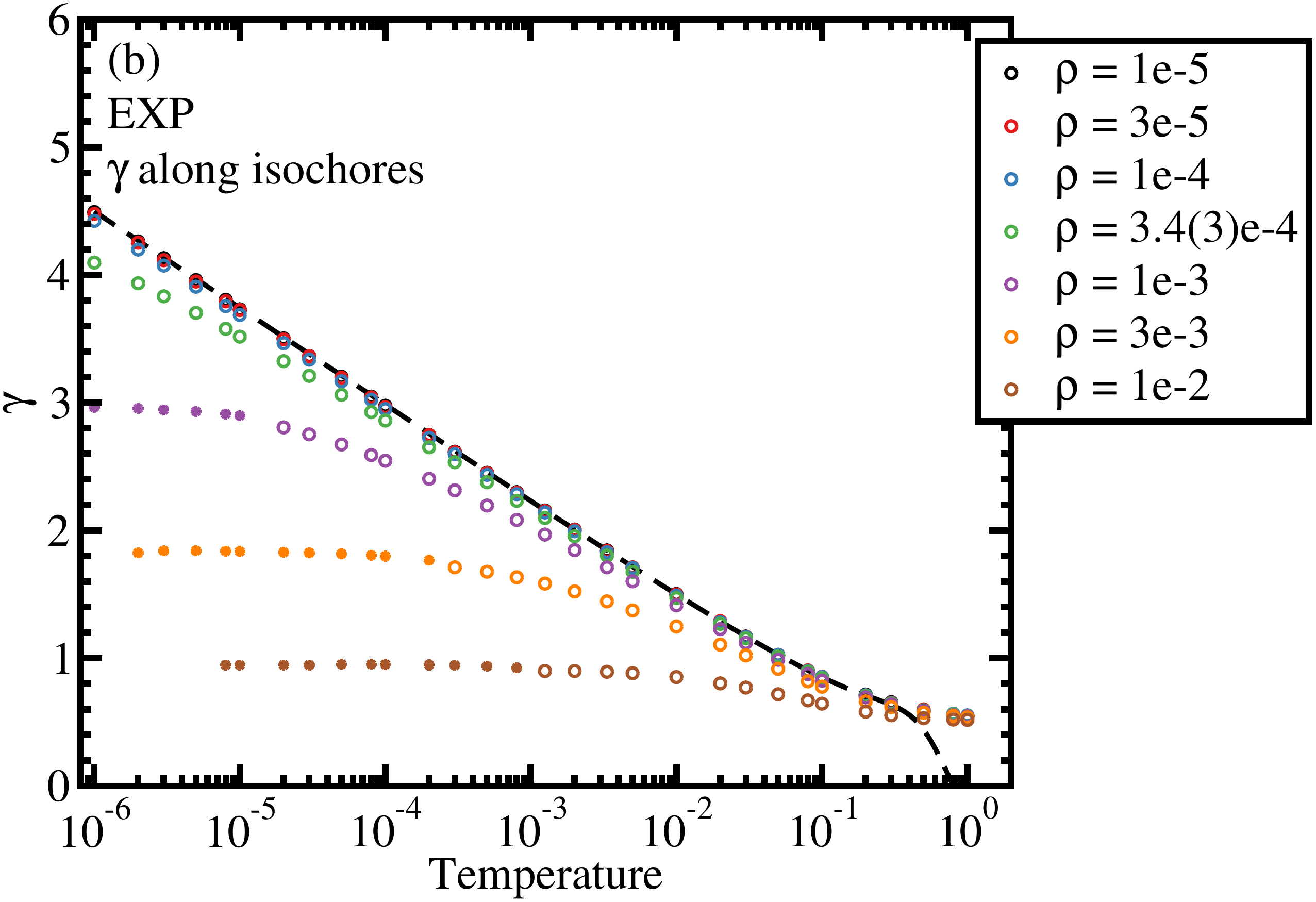}
	\includegraphics[width=0.4\textwidth]{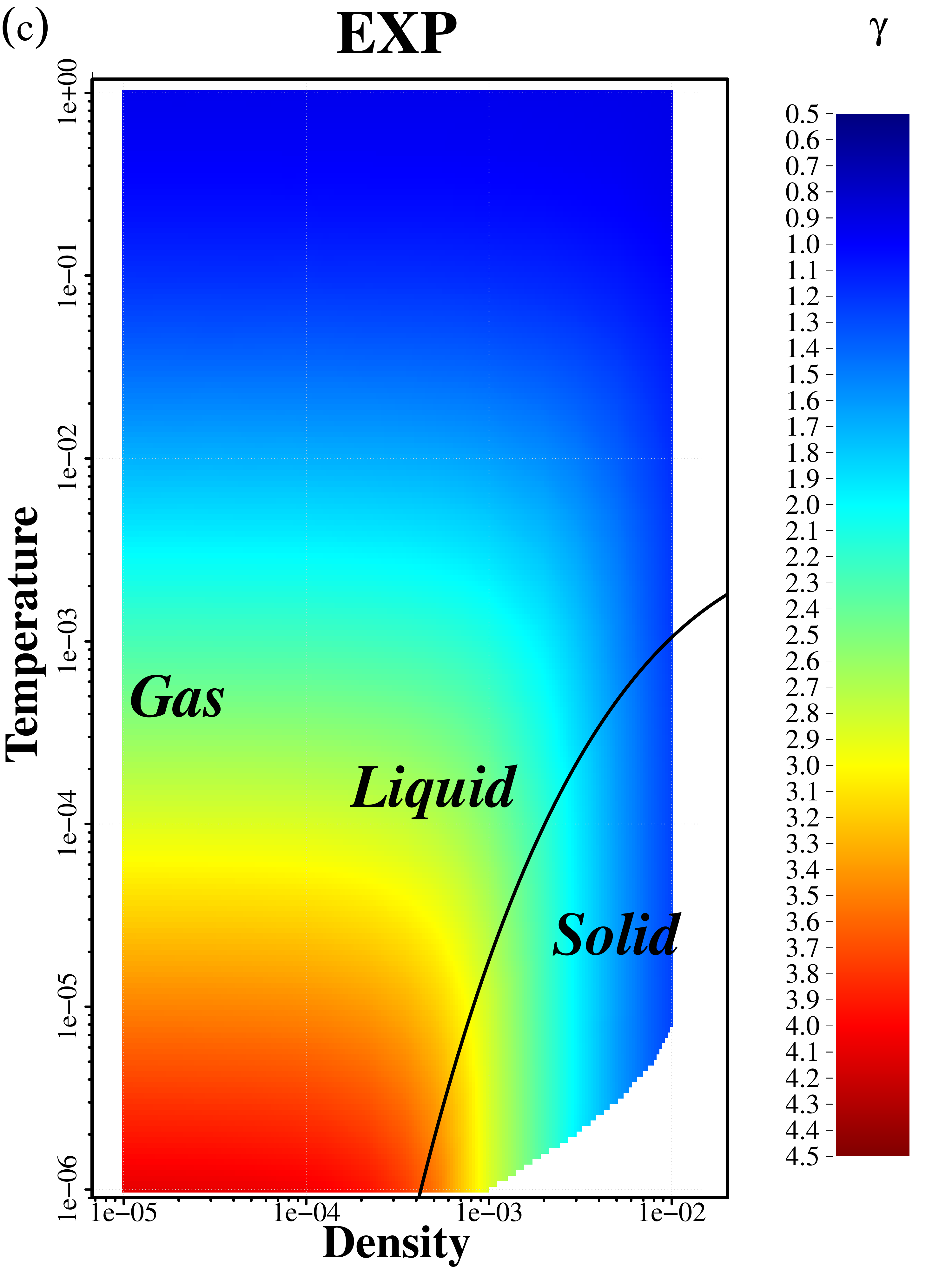}
	\includegraphics[width=0.4\textwidth]{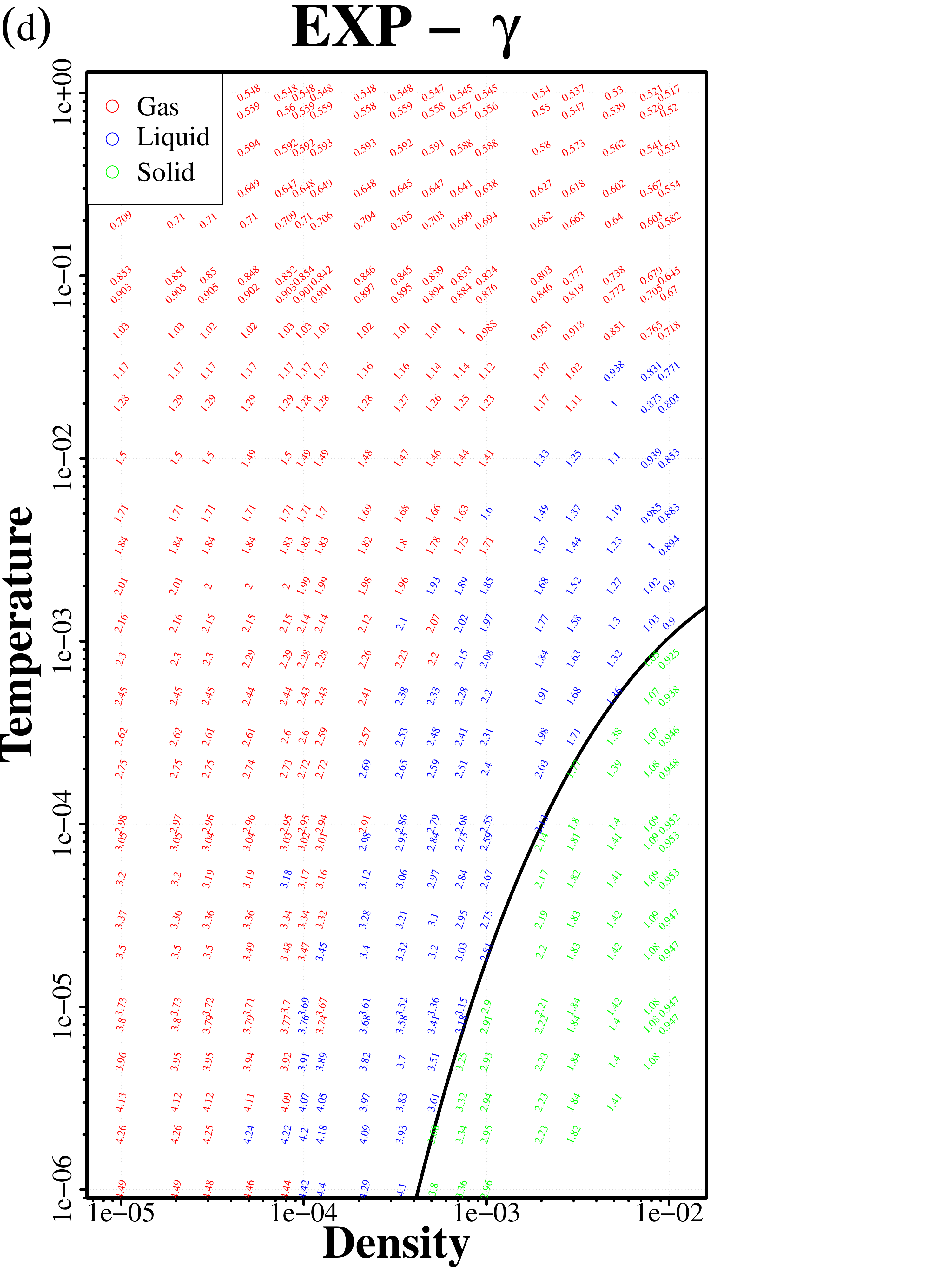}	
	\caption{Density and temperature variation of the density-scaling exponent $\gamma$ (\eq{dsexp}).
	(a) shows $\gamma$ along isotherms. At low densities $\gamma$ is mainly temperature dependent. The dashed line gives the prediction of the eIPL approximation \eq{gamma_est} \cite{II} with $\Lambda=1.075$, which describes the condensed liquid phase well.	
	(b) shows $\gamma$ along isochores. The dashed line is here the prediction of the analytical theory for the gas phase, \eq{gamma_gas3}, for which $\gamma$ depends only on the temperature.
	(c) shows $\gamma$ in a continuous color plot, visualizing the fact that $\gamma$ at low densities is mainly temperature dependent, whereas it at high densities and low temperatures, i.e., in the liquid and solid phases, is mainly density dependent.
	(d)	reports numerical values of $\gamma$ (visible upon magnification) written into the phase diagram at the densities and temperatures listed in Appendix II of Paper I. At each state point $\gamma$ is written with a slope marking the direction of the isomorph through the state point in question; red indicates gas, blue liquid, and green solid phase state points.
	\label{fi:isochore_gamma}}
\end{figure}

By means of standard thermodynamic fluctuation theory $\gamma$ may be calculated from the canonical-ensemble constant-density ($NVT$) equilibrium virial and potential-energy fluctuations \cite{IV},

\be\label{gamma_eq}
\gamma
\,=\,\frac{\langle\Delta U\Delta W\rangle}{\langle(\Delta U)^2\rangle}\,.
\ee
This identity makes it straightforward to determine $\gamma$ in simulations. \Fig{fi:isochore_gamma} shows results for the density-scaling exponent with (a) giving $\gamma$'s density variation along isotherms and (b) giving $\gamma$'s temperature variation along isochores. In the figure we mark three phases: gas, liquid, and solid, although the EXP system like any purely repulsive system has no liquid-gas phase transition, but merely a single fluid phase. The rationale for distinguishing between gas and liquid states -- depending on whether or not the radial distribution function has a well-defined minimum above its nearest-neighbor peak, see Paper I -- is the quasiuniversality of simple liquids' structure and dynamics that includes the EXP system (Sec. VII of Paper I). There is a large transition region between typical gas and typical liquid states; the least dense isomorph studied below is located in this region and referred to as the ``gas-liquid'' isomorph. The data in \fig{fi:isochore_gamma} reveal two regimes: At high densities and low temperatures (liquid and solid phases) $\gamma$ is mainly density dependent, at low densities (gas phase) $\gamma$ is mainly temperature dependent.

It is possible to construct approximate analytical theories for the two limiting behaviors. Consider first the high-density case corresponding to the liquid and solid phases, both of which are characterized by strong interactions between a given particle and its several nearest neighbors. To explain the strong virial potential-energy correlations of the Lennard-Jones (LJ) pair-potential liquid, Ref. \onlinecite{II} developed an approach based on the ``extended inverse power-law'' (eIPL) approximation that works as follows. At typical liquid or solid densities, within the first coordination shell the LJ pair potential is very well approximated \cite{II} by the eIPL pair potential

\be\label{eIPLv}
v_{\rm eIPL}(r)
\,=\,Ar^{-n}+B+Cr\,.
\ee
If one imagines a particle being displaced within its nearest-neighbor cage, some distances increase and some decrease, but the sum of all nearest-neighbor distances remains almost unchanged. This means that the $B+Cr$ term is almost constant, so the eIPL pair potential may effectively be replaced by $v_{\rm eIPL}(r)\cong Ar^{-n}+D$. When added over all particle pairs, this implies that the hidden-scale-invariance condition \eq{hs} applies to a good approximation.

The above explains the strong virial potential-energy correlations found for the LJ and similar pair-potential systems \cite{I,vel15,cos16a}. It also suggests a means for approximately calculating the density-scaling exponent $\gamma$. Note first that 

\be\label{gammaIPL}
\gamma=n/3
\ee
for the IPL system $v(r)= Ar^{-n}+D$; this follows from \eq{gamma_eq} and the definition of the pair virial $w(r)\equiv (-1/3)ru'(r)$ \cite{II}. How to identify an effective IPL exponent $n$ at a given state point? For the eIPL pair potential one has 
$v_{\rm eIPL}'(r)=-nAr^{-(n+1)}+C$, 
$v_{\rm eIPL}''(r)=n(n+1)Ar^{-(n+2)}$, and 
$v_{\rm eIPL}'''(r)=-n(n+1)(n+2)Ar^{-(n+3)}$. 
This implies $n=-2-rv_{\rm eIPL}'''(r)/v_{\rm eIPL}''(r)$. Thus $n$ is given by $n=n_2(r)$ if one for any pair potential $v(r)$ defines the $r$-dependent effective IPL exponent $n_p(r)$ \cite{II} by ($v^{(p)}(r)$ is the $p$th derivative of $v(r)$)

\be\label{np}
n_p(r)
\,\equiv\, -p -r\frac{v^{(p+1)}(r)}{v^{(p)}(r)}\,.
\ee

Realistic values of $n$ are arrived at by using for $r$ a typical nearest-neighbor distance, of course. For the EXP unit system this means putting 

\be\label{rLambda}
r/\sigma
\,=\,\Lambda\, \rho^{-1/3}
\ee
in which $\Lambda\cong 1$ is a numerical constant. Using this value of $r$ for the EXP pair potential leads for $p=2$, via \eq{gammaIPL}, \eq{np}, and \eq{rLambda}, to the following estimate of the density-scaling exponent in the EXP system's condensed liquid and solid phases:

\be\label{gamma_est}
\gamma(\rho)
\,\cong\, \frac{-2+\Lambda \rho^{-1/3}}{3}\,.
\ee 
This is the black dashed line in \fig{fi:isochore_gamma}(a) in which $\Lambda=1.075$ was determined to get the best fit to data. At high densities the different isotherms collapse onto the line, a collapse that at low temperatures takes place earlier than at high temperatures.

Next we study the dilute gas limit. Here the system is characterized by much longer typical distances between the particles than the interaction range $\sigma$, i.e., $\rho\ll 1$. In this limit particle interactions predominantly take place via two-particles collisions. As shown in Paper I, it is possible to construct an analytical theory for the correlation coefficient $R$ in the gas phase by assuming that particle collisions are random and uncorrelated. It is likewise possible to calculate $\gamma$ analytically in the gas phase via \eq{gamma_eq} (compare Appendix I of Paper I). The relevant equation is

\be\label{gamma_gas}
\gamma
\,=\,\frac{\langle wv\rangle}{\langle v^2\rangle}\,
\ee
in which $v$ is the EXP pair potential treated as an independent variable, $w=(-1/3)rv'(r)=\ln(1/v)v/3$ is the pair virial and averages are taken over the non-normalizable $v$-probability distribution $p(v)\propto\ln^2(1/v)\exp(-\beta v)/v$. \Eq{gamma_gas} leads to  

\be\label{gamma_gas2}
\gamma
\,=\,\frac{A_3}{3A_2}\,
\ee
in which

\be
A_n\label{an}
\,=\,\int_{0}^{\infty} v\,\ln^n(1/v)\,e^{-\beta v}dv\,.
\ee
The integrals may be worked out analytically in terms of $\pi$, Euler's constant 0.577..., and the Riemann zeta function evaluated at $3$ (Appendix I of Paper I). Numerically, the result is 

\be\label{gamma_gas3}
\gamma
\,=\,\frac{\ln^3\beta-1.268\ln^2\beta+2.471\ln\beta-0.4895}{3\ln^2\beta-2.5368\ln\beta+2.4711}\,.
\ee

At low temperatures ($\beta\gg 1$) the dominant term is $\gamma\cong\ln\beta/3$. This corresponds to the $p=0$ effective IPL exponent in \eq{np} evaluated at the distance at which the pair potential equals $k_BT$, the typical distance of nearest approach in a collision. Our numerical data indicate that \eq{gamma_gas3} becomes exact at low densities, compare \fig{fi:isochore_gamma}(b) that shows \eq{gamma_gas3} as the dashed black line.

\begin{figure}[!htbp]
	\centering
	\includegraphics[width=0.4\textwidth]{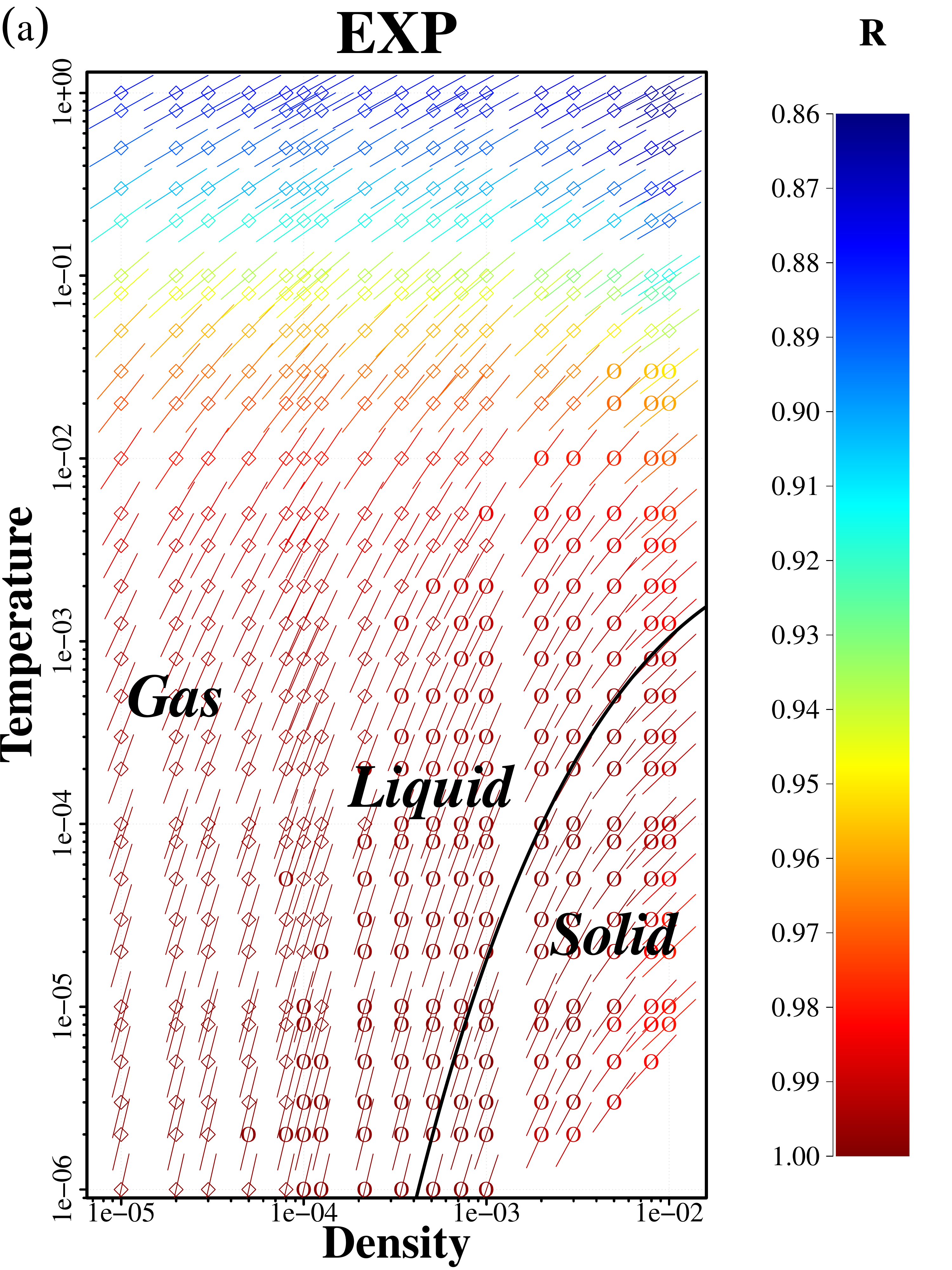}
	\includegraphics[width=0.4\textwidth]{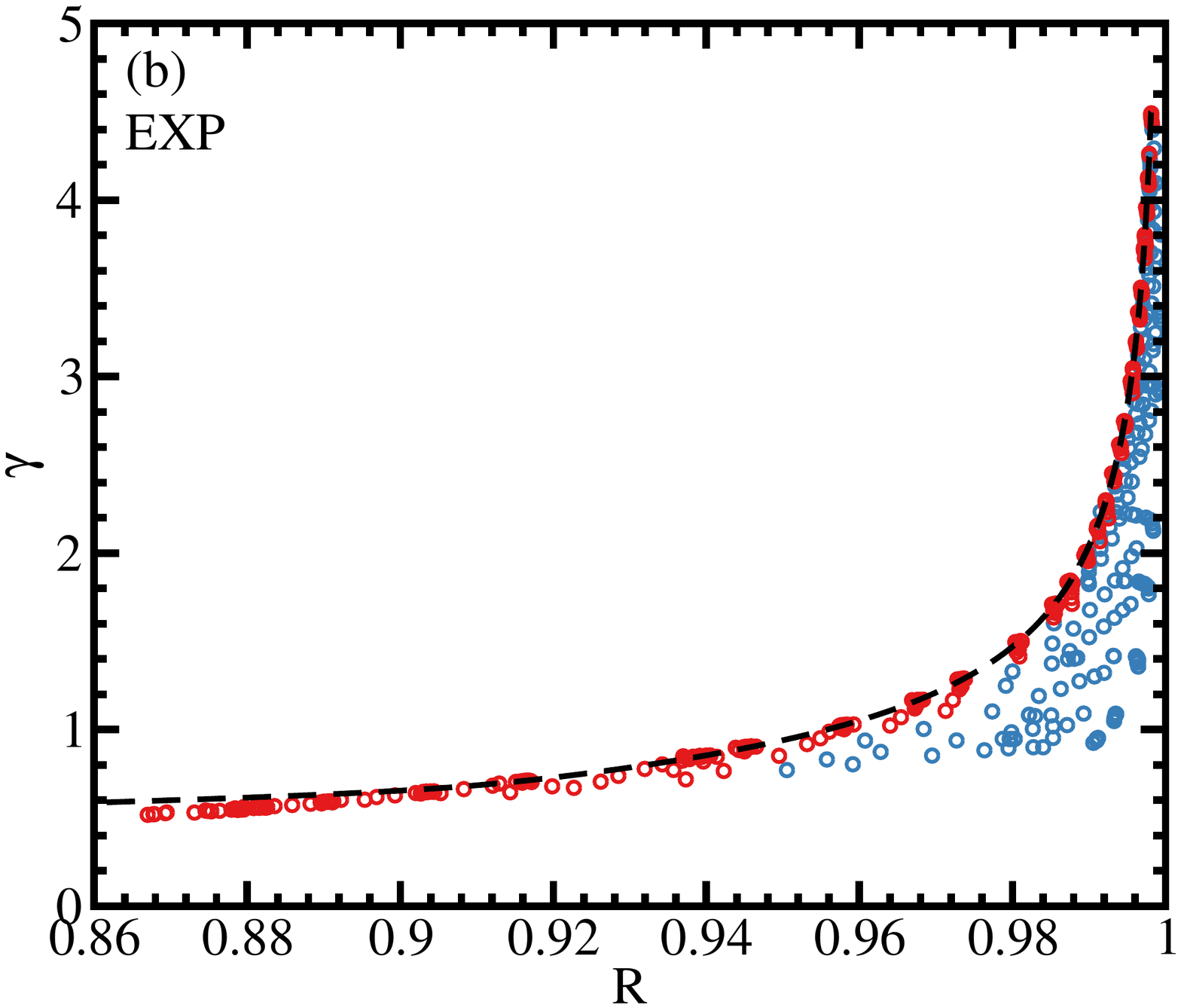}
	\caption{\label{fi:phase-diagram:cont}(a) shows the density-temperature phase diagram with line slopes given by the density-scaling exponent and color coding indicating the virial potential-energy correlation coefficient $R$. Diamonds mark gas-phase state points, circles mark liquid or solid state points. The line segments give the isomorph slopes, compare \eq{dsexp}. The black dashed line is the approximate melting-line isomorph (covering the entire coexistence region).
	(b) Density-scaling exponent $\gamma$ versus the virial potential-energy correlation coefficient $R$ for all state points simulated. Red symbols are gas state points, blue symbols are liquid  and solid state points. The dashed line is the prediction of the analytical gas-phase theory, which is obtained by combining \eq{gamma_gas2} with $R=A_3/\sqrt{A_2A_4}$ derived in Appendix I of Paper I. For a given value of $R$ the gas phase has the highest $\gamma$, for a given value of $\gamma$ the gas phase has the lowest $R$. 	
	}
\end{figure}

\Fig{fi:phase-diagram:cont}(a) summarizes our numerical findings for $R$ and $\gamma$ in a single phase diagram \cite{bac14a}. The color coding gives $R$, the line-segment slopes give $\gamma$. The line segments mark how the isomorphs run in the  phase diagram, basically parallel to the melting line that is itself an approximate isomorph \cite{IV,bac14a}. \Fig{fi:phase-diagram:cont}(b) plots all values of $(R,\gamma)$ with the dashed line marking the gas-phase analytical prediction. An important conclusion from this figure is that as $\gamma\rightarrow\infty$ one has $R\rightarrow 1$. Since large $\gamma$ corresponds to an effectively very strongly repulsive pair potential on the $k_BT$ energy scale, this means that one expects $R\rightarrow 1$ in the hard-sphere limit of the EXP system obtained by following an isomorph to zero temperature.

\section{Structure, dynamics, and specific heat along three isomorphs}\label{sec:isom}

The ``small-step'' method for tracing out an isomorph in the phase diagram is based directly on \eq{dsexp} and \eq{gamma_eq}. This section investigates predicted invariances along three isomorphs, which serve as reference ``true'' isomorphs in the next section dealing with faster, approximate methods for generating isomorphs. 

It is straightforward to calculate in an $NVT$ computer simulation the canonical averages on the right-hand side of \eq{gamma_eq}. Typical values of $\gamma$ for the EXP system are between 0.5 and 5 (\fig{fi:isochore_gamma}). If for instance $\gamma=3$, upon a 1\% density increase the temperature is to be increased by 3\% in order to keep $\Sex$ constant (\eq{gamma_eq}). After this change of density and temperature one recalculates the right-hand side of \eq{gamma_eq}, and so on. In this way an isomorph is traced out with a method that is, in principle, exact in the limit of infinitely small density changes. The method is tedious since many steps are needed if 1\% density changes are used to generate an isomorph covering a large density variation. The many steps required also mean that, in order to avoid accumulation of errors, each state-point simulation must be long enough to provide accurate data. Initial simulations for 5\%, 2\%, and 1\% density changes from various state points showed that the latter two give virtually indistinguishable results. We concluded that a 1\% density change is small enough to be reliable.

\begin{figure}[!htbp]
	\centering
	\includegraphics[width=0.4\textwidth]{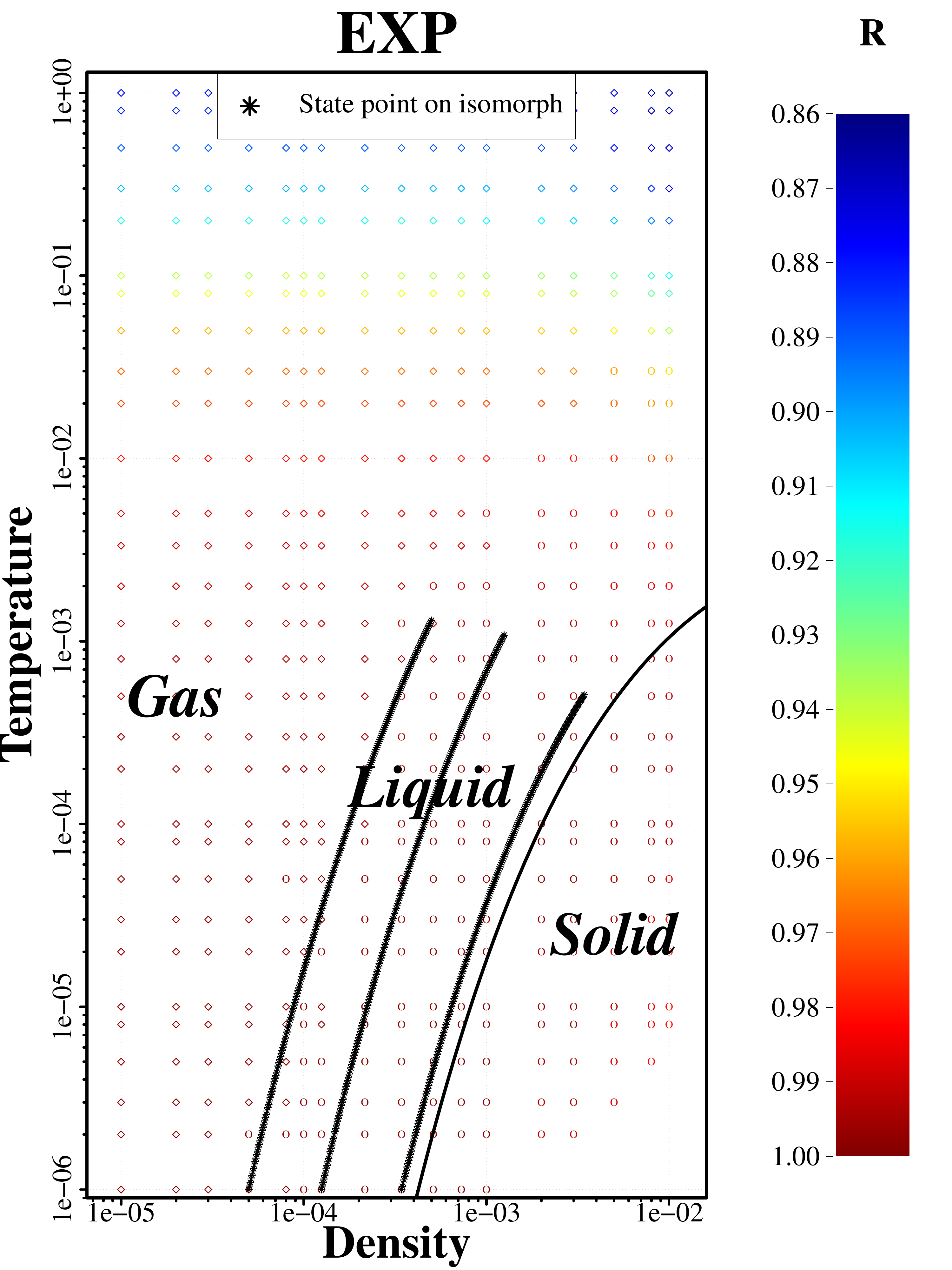}
	\caption{\label{Isofigur}The three isomorphs studied marked by black crosses merging into lines. The isomorphs were generated by the ``small-step method'' from 230 simulations, each increasing density by 1\% using \eq{dsexp} in conjunction with \eq{gamma_eq} to calculate the corresponding temperature change. The starting temperature for each isomorph was $T = 10^{-6}$, the starting densities were the following: gas-liquid isomorph (left): $5 \cdot 10^{-5}$, dilute-liquid isomorph (middle): $1.25 \cdot 10^{-4}$, dense-liquid isomorph (right): $3.43 \cdot 10^{-4}$.}
\end{figure}

Three isomorphs were traced out for systems of 1000 particles using $1$\% density changes to cover one decade of density based on 230 simulations, each involving 10 million time steps (\fig{Isofigur}). One isomorph is located in the gas-liquid transition region, one is in the dilute liquid phase, and a third one is the liquid phase near the melting line. Note that in the log-log phase diagram the isomorphs are virtually parallel to the melting line. This is because in a simplified version of the isomorph theory \cite{IV} the melting line is an isomorph and isomorphs are given by an expression of the form $h(\rho)/T=\,$Const \cite{ing12a}. Different isomorphs correspond to different constants, so the isomorphs are parallel to one another in the log-log density-temperature phase diagram. A more accurate melting theory is now available \cite{ped16}, but the corrections to the older understanding of Ref. \onlinecite{IV} are relatively small.

\begin{figure}[!htbp]
		\centering
		\includegraphics[width=0.375\textwidth]{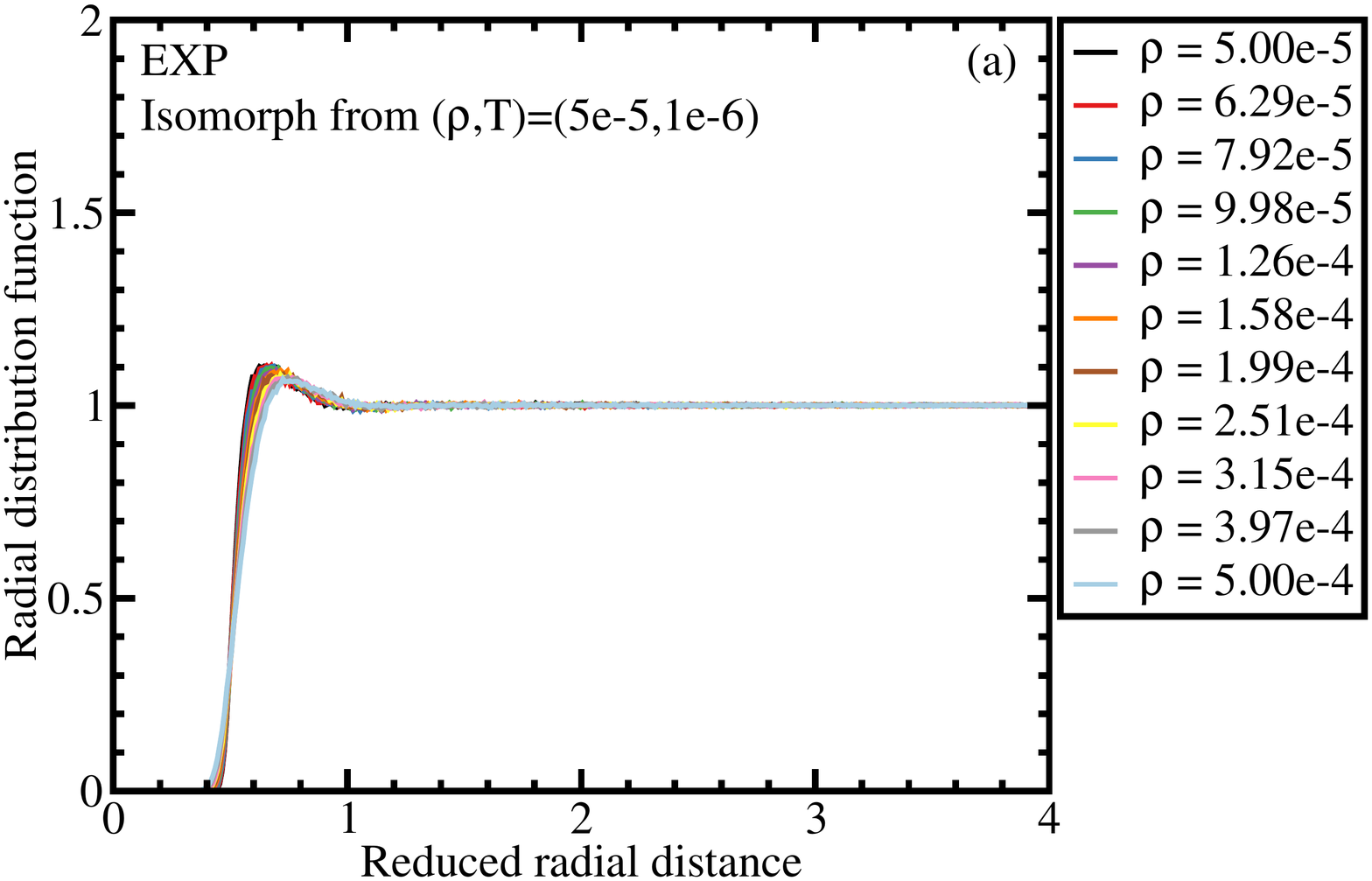}
		\includegraphics[width=0.375\textwidth]{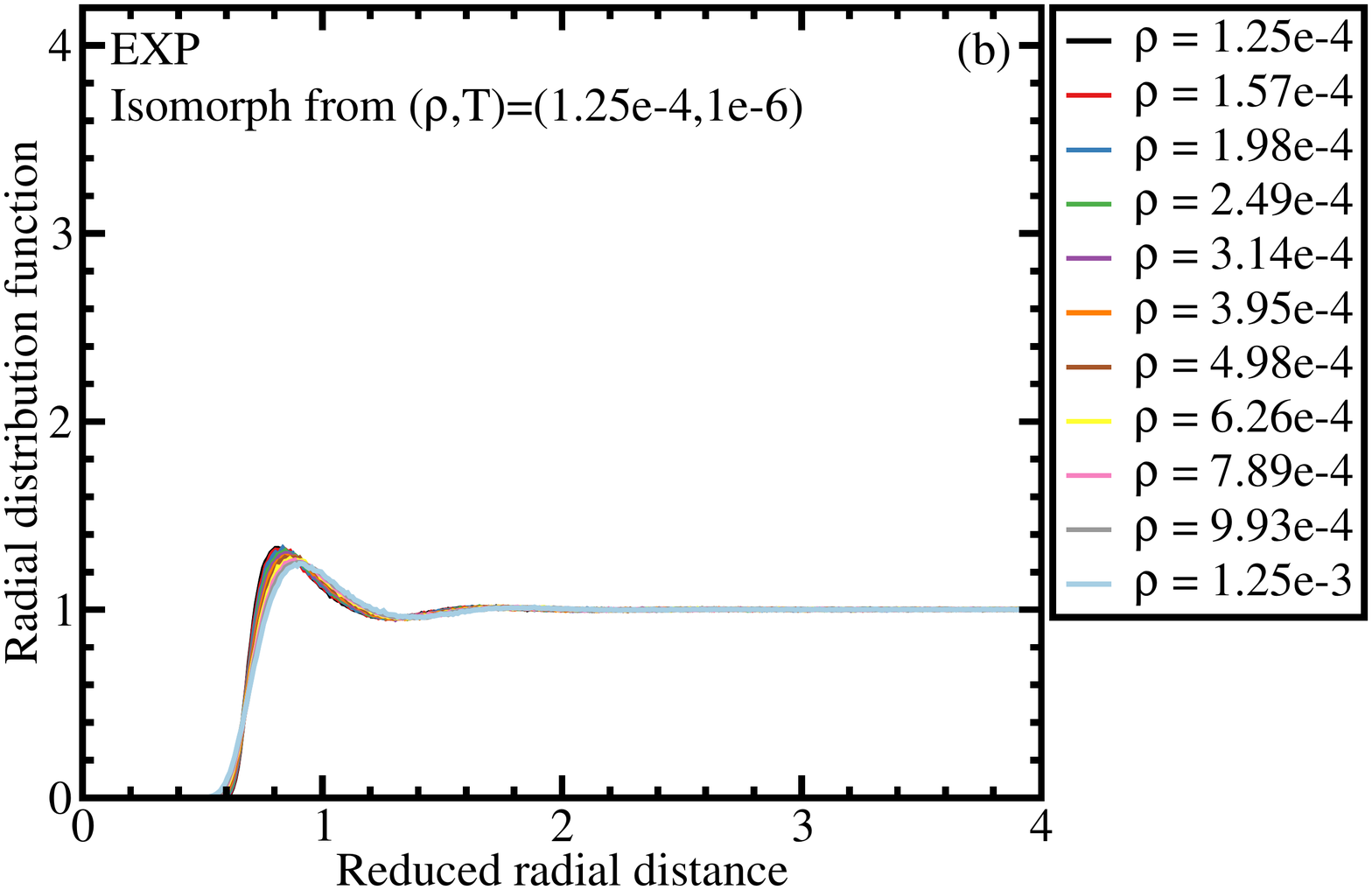}
		\includegraphics[width=0.375\textwidth]{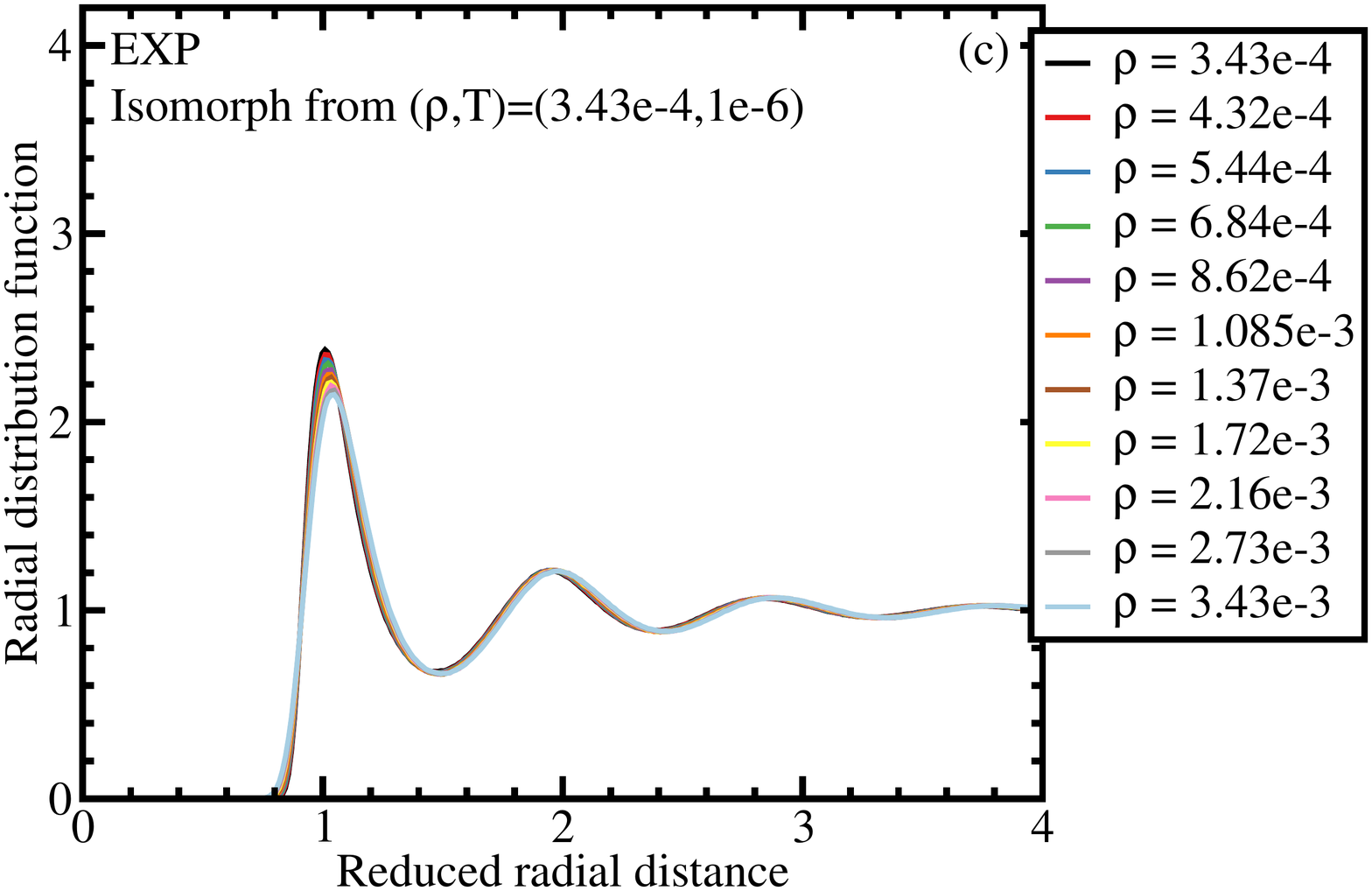}
		\caption{\label{fi:isom-rdf} Structure at selected state points along each of the three isomorphs shown in \fig{Isofigur} probed via the reduced radial distribution function (RDF).
		(a) Gas-liquid isomorph; this isomorph is located where the system has little structure but this is approximately invariant. 
		(b) Dilute-liquid isomorph; the structure here exhibits more structure and is still approximately invariant. 
		(c) Dense-liquid isomorph; the system has here a typical liquid-like structure that is well maintained along the isomorph.}
\end{figure}

In the following, data are presented for each isomorph for the starting state point and ten more state points evenly spaced on the logarithmic density axis (the coordinates of the selected isomorph state points are given in the Appendix). We first investigate how the structure changes along an isomorph in order to see whether structure is invariant. \Fig{fi:isom-rdf} shows the structure probed by the reduced radial distribution function (RDF) at the selected state points for each of the three isomorphs. The density change along each isomorph spans one decade, the temperatures span more than two decades. These variations are large compared to the first isomorph-theory simulations covering density variations of just a few percent \cite{IV}. Nevertheless, deviations from collapse are small; these are mainly observed around the first peak \cite{IV,ing12b}.

\begin{figure}[!htbp]
		\centering
		\includegraphics[width=0.4\textwidth]{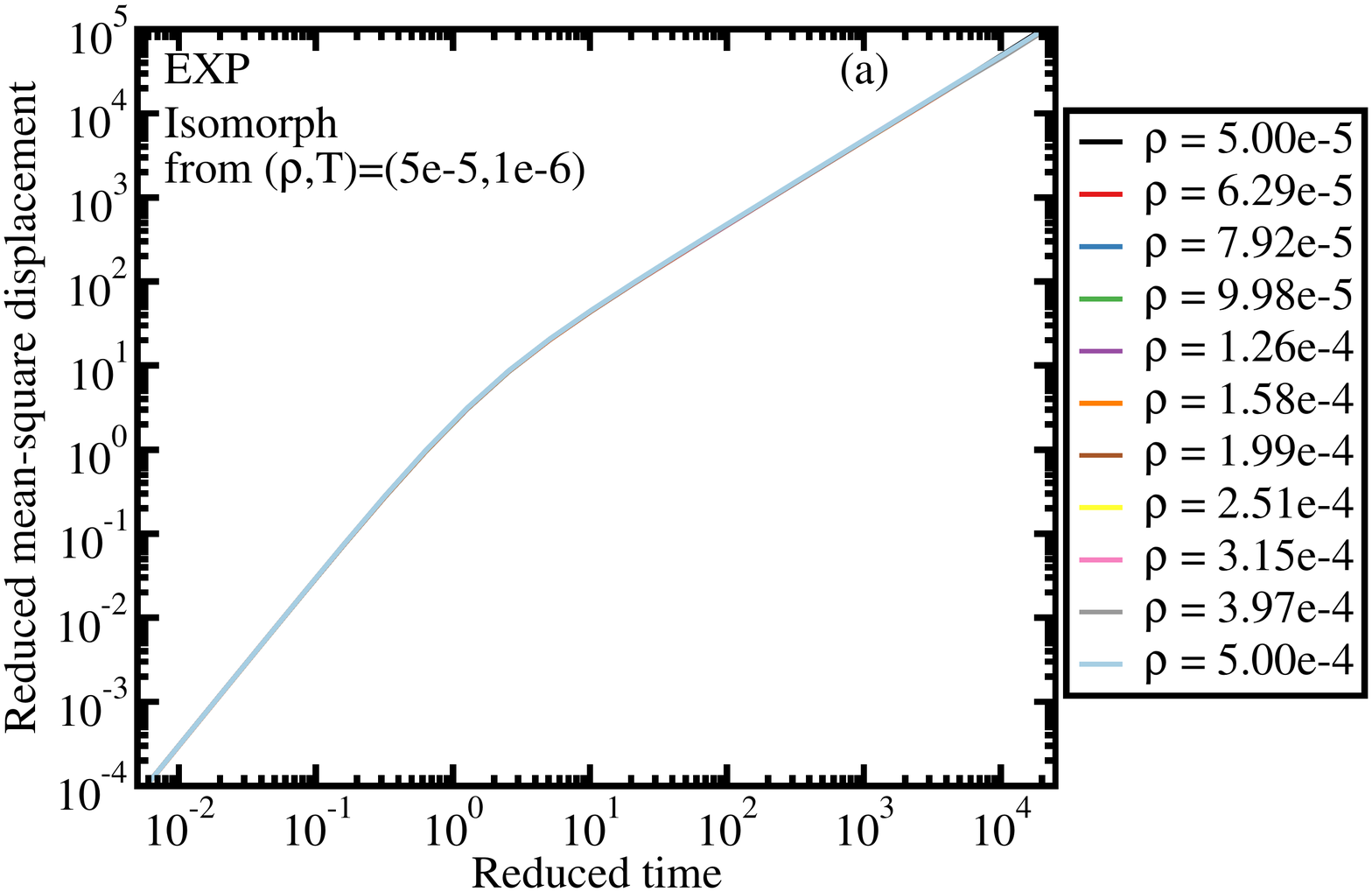}
		\includegraphics[width=0.4\textwidth]{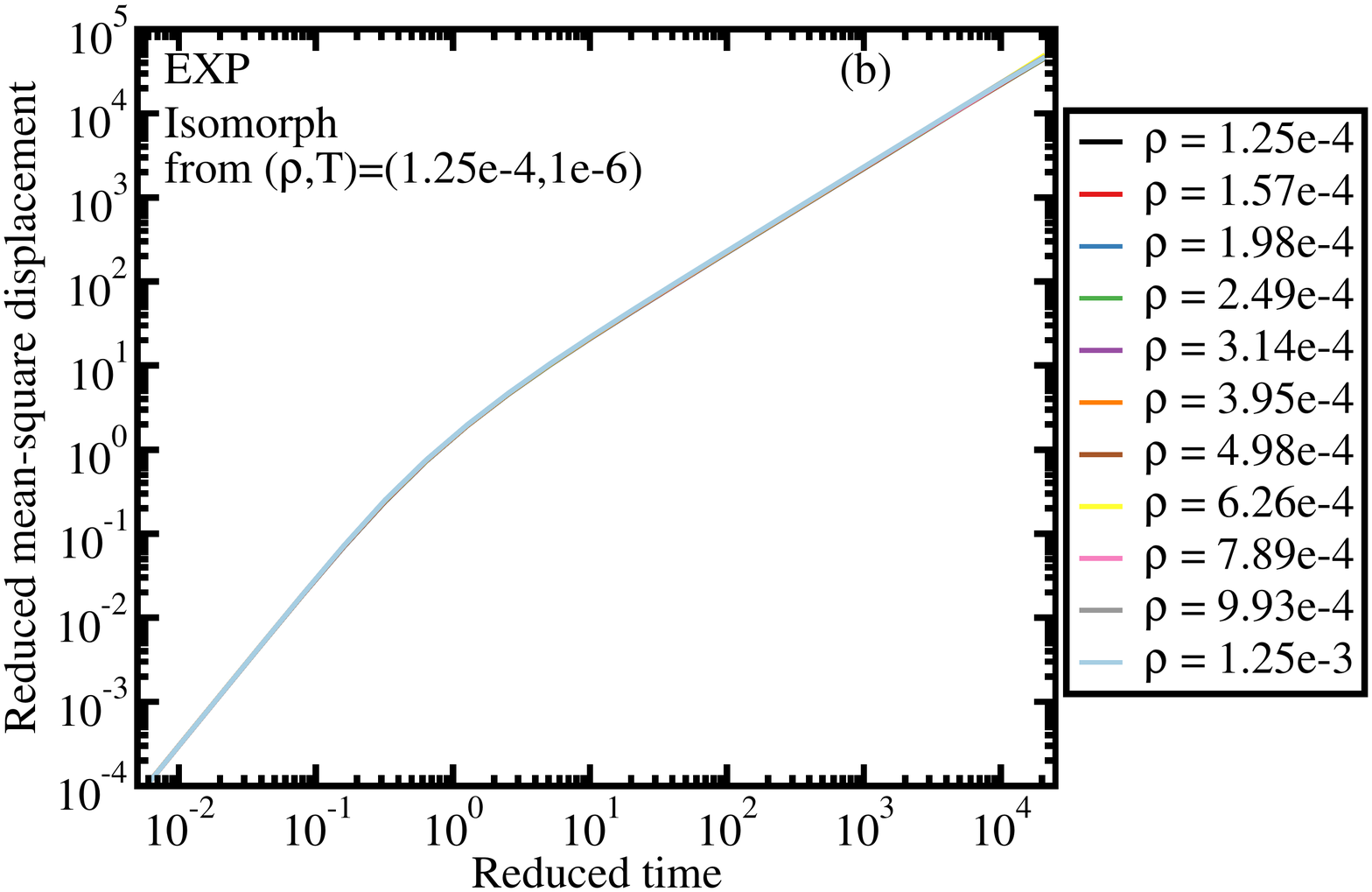}
		\includegraphics[width=0.4\textwidth]{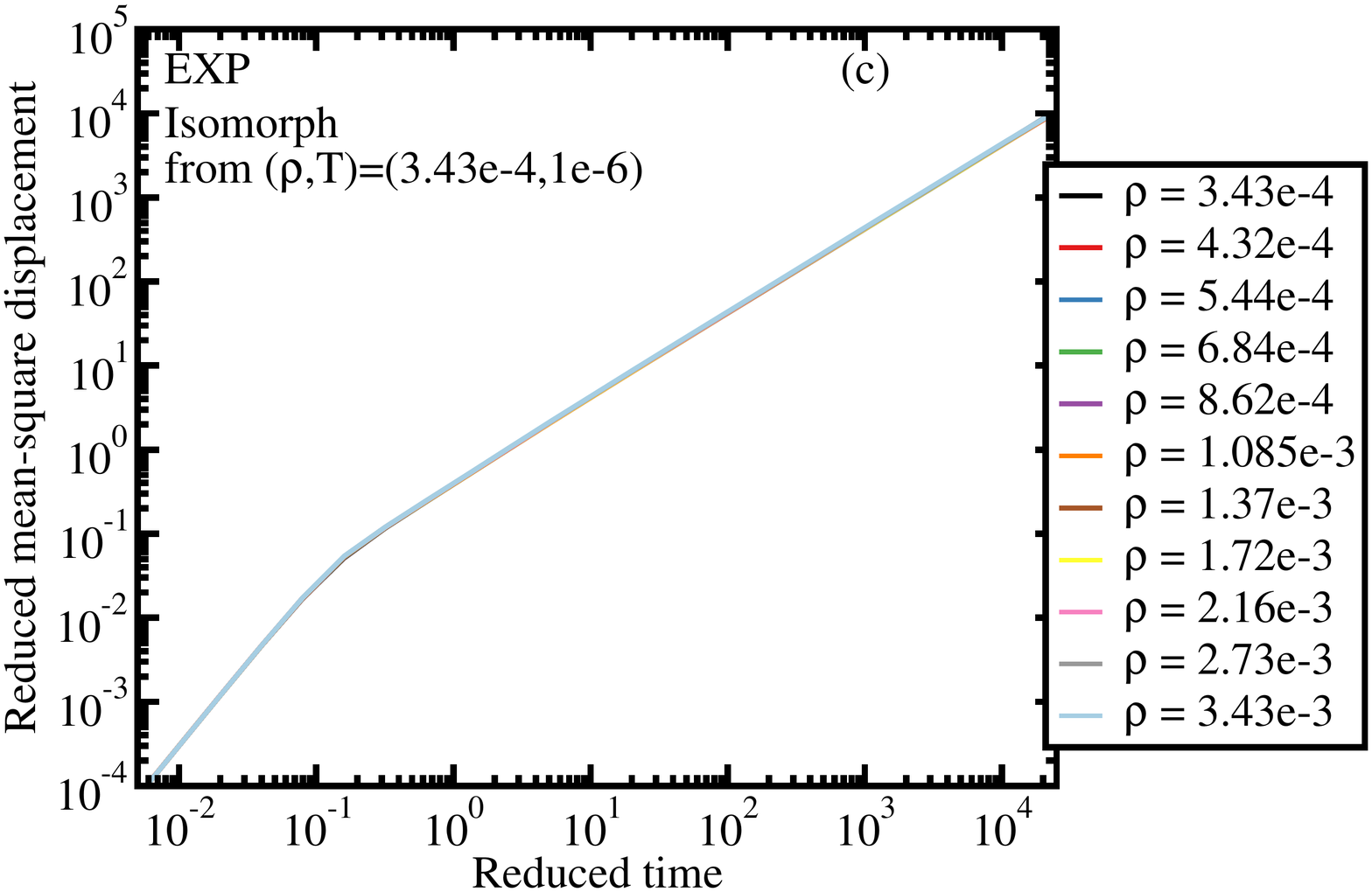}
		\caption{\label{fi:isom-msd}Reduced mean-square displacement (MSD) at selected state points along each of the three isomorphs plotted as a function of reduced time. At short times the MSD follows the ballistic prediction $3\,\tilde{ t}^2$ (Paper I), at long times it follows the diffusion prediction $\propto\tilde{ t}$. In all cases there is a good collapse.}
\end{figure}

\begin{figure}[!htbp]
		\centering
		\includegraphics[width=0.4\textwidth]{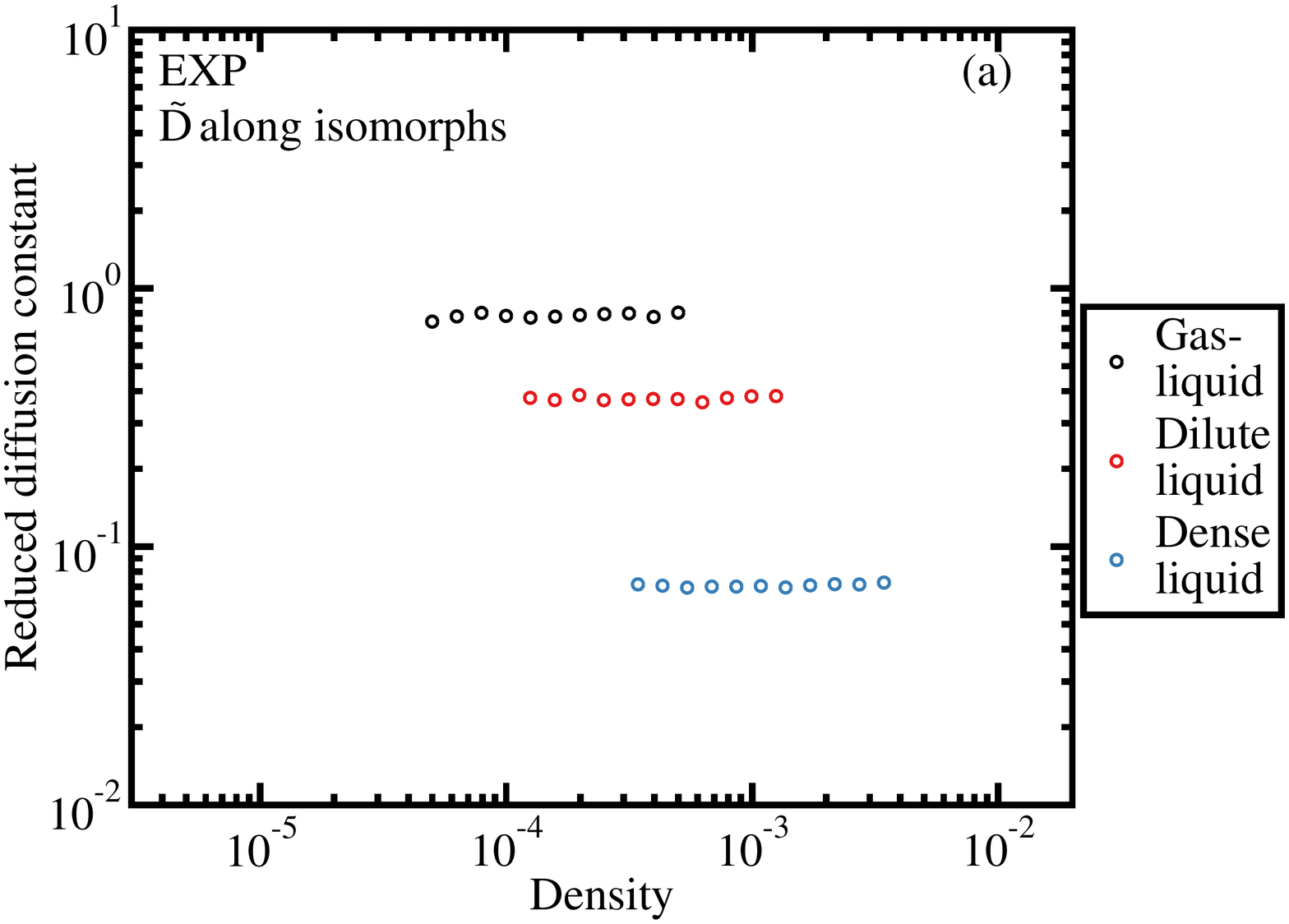}
		\includegraphics[width=0.4\textwidth]{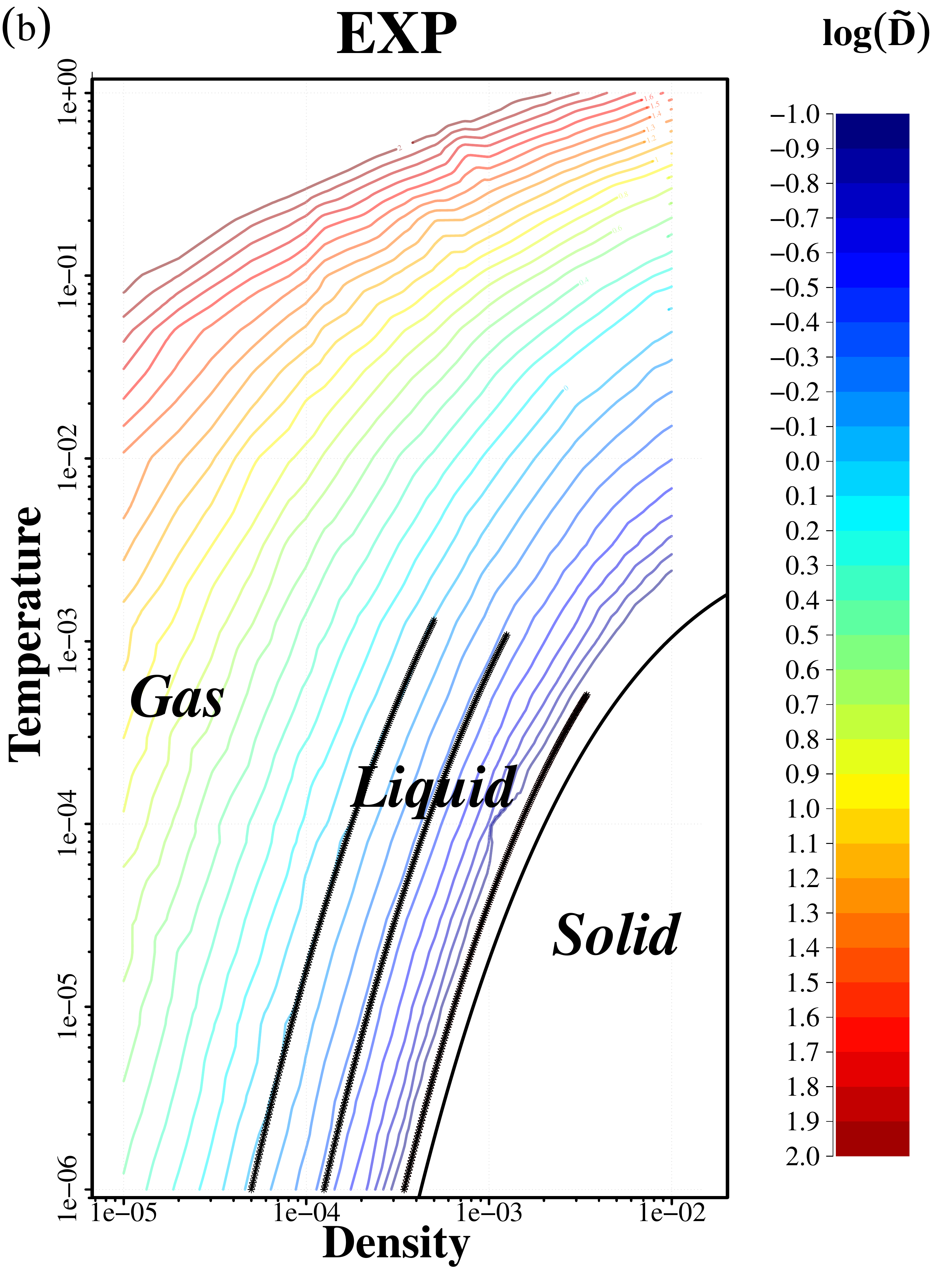}
		\caption{\label{fi:isom-D}
		(a) Reduced diffusion constant $\tilde{D}$ plotted as a function of density along each of the three isomorphs. The diffusion constant varies significantly between the isomorphs, but is virtually constant along each of them. 
		(b) Contour color plot giving the reduced diffusion constant's variation throughout the phase diagram in which lines of constant $\tilde{D}$ are drawn for logarithmically equally distributed values.}
\end{figure}

For each of the three isomorphs, \fig{fi:isom-msd} shows the reduced mean-square displacement (MSD) as a function of reduced time at the same eleven state points. Good collapse is observed. This may be compared to the lack of collapse of the reduced MSD along the EXP system's isotherms and isochores (Paper I). \Fig{fi:isom-D}(a) demonstrates that the reduced diffusion constant (derived from the long-time MSD) is invariant along each of the three isomorphs. The diffusion (in reduced units) is faster the more gas-like the isomorph is. \Fig{fi:isom-D}(b) shows a contour plot of the reduced diffusion constant that again illustrates its isomorph invariance.

\begin{figure}[!htbp]
		\centering
		\includegraphics[width=0.4\textwidth]{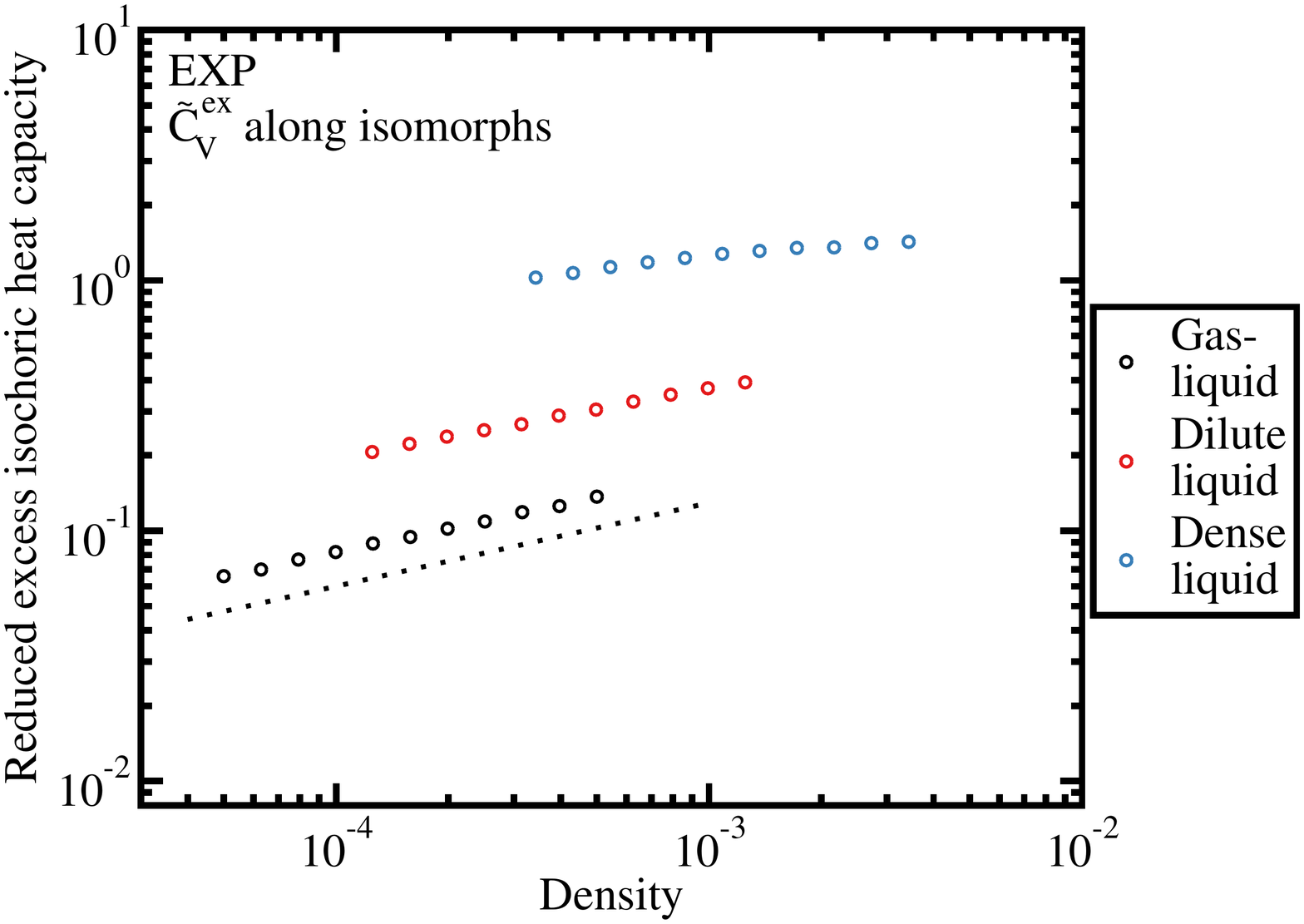}		
		\caption{\label{fi:isom-Cv} Reduced excess isochoric specific heat plotted as a function of density along each of the three isomorphs. Not surprisingly, the isomorph with least interparticle interactions -- the gas-liquid isomorph -- has the lowest excess specific heat. A systematic increase with density is observed for all three isomorphs. This is at variance with the original (2009) version of isomorph theory in which $C_V$ is an isomorph invariant \cite{IV}, but it is consistent with the 2014 version \cite{sch14}. The dashed line has slope 1/3, which is the prediction for the density variation along gas-phase isomorphs (see the text).}
\end{figure}

\Fig{fi:isom-Cv} gives the reduced excess isochoric specific heat per particle $\tcVex$ calculated from the fluctuations in potential energy in the $NVT$ (canonical) ensemble via the Einstein expression $\tcVex=\langle(\Delta U)^2\rangle/Nk_B^2T^2$. The more gas-like the structure is, the lower is the excess specific heat because interactions become infrequent. In the original (2009) version of isomorph theory \cite{IV} the excess specific heat was predicted to be an isomorph invariant. \Fig{fi:isom-Cv} shows that this is not the case; in fact $\tcVex\propto\rho^{1/3}$ in the gas phase. This confirms the need for the 2014  revision of the isomorph theory \cite{sch14} in which it was shown that the originally predicted isomorph invariance of $\tcVex$ results from a first-order approximation to a simpler and more correct theory, which starts from the hidden-scale-invariance condition \eq{hs}.

Along the gas-phase isomorph the density variation of $C_V$ is determined as follows. For individual particle collisions -- in particular at low temperature -- the EXP pair potential can be approximated by an effective inverse power-law pair potential $\propto r^{-n}$ with exponent given by 
$n=-d\ln v/d\ln r=-r\,v'_{\rm EXP}(r)/v_{\rm EXP}(r)$ evaluated at the distance $r=r_0$ of closest approach (this corresponds to the $p=0$ case of \eq{np}). The distance $r_0$ is estimated from $k_BT=v_{\rm EXP}(r_0)$. In the EXP unit system this leads to $\gamma=-\ln T/3$ since the density-scaling exponent is given by $\gamma=n/3$ (\eq{gammaIPL}). Substituting $\gamma=-\ln T/3$ into \eq{dsexp} leads to a simple first-order differential equation for $\ln T$ as a function of $\ln\rho$. Its solution is $\ln T=A(\Sex)\rho^{-1/3}$ in which $A(\Sex)$ is an integration constant. This implies that $(\partial\ln T/\partial\Sex)_\rho=A'(\Sex)\rho^{-1/3}$, which via the identity $C_V=(\partial\Sex/\partial\ln T)_\rho$ implies that $C_V\propto\rho^{1/3}$ along the gas-phase isomorphs. The dashed line in \fig{fi:isom-Cv} has slope 1/3; it fits well to the density variation of $C_V$ along the gas-liquid isomorph.

\begin{figure}[!htbp]
		\centering
		\includegraphics[width=0.4\textwidth]{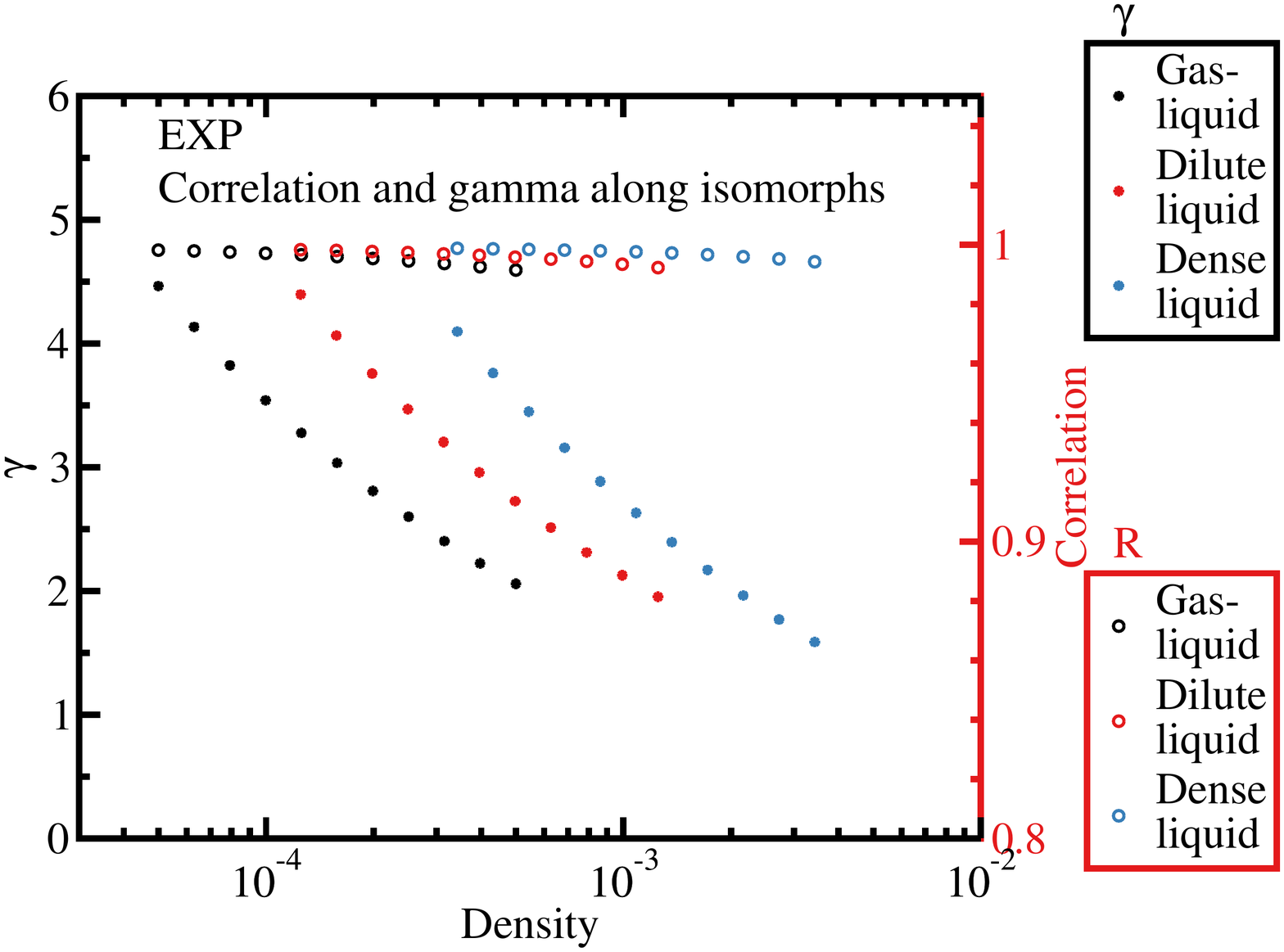}
		\caption{The density-scaling exponent $\gamma$ (full circles, left) and the virial potential-energy correlation coefficient $R$ (open circles, right) plotted as a function of density along each of the three isomorphs. Neither $\gamma$ nor $R$ are predicted to be isomorph invariant. Note that strong correlations are maintained even as $\gamma$ decreases significantly with increasing density.}
		\label{fi:isom-rg}
\end{figure}

Figure \ref{fi:isom-rg} shows that the density-scaling exponent $\gamma$ decreases significantly with increasing density along each isomorph. Notably, the system has $R>0.9$, i.e., is strongly correlating even with $\gamma$ values as low as $1.5$.

\section{Direct-isomorph-check approximate isomorph}\label{sec:DIC}

In this section we use the same three starting state points as above to trace out approximate isomorphs using two versions of the so-called direct isomorph check. This method, which allows for much larger density jumps than 1\%, is justified as follows.

Suppose simulations at the state point $(\rho_1,T_1)$ generate a series of configurations. For a different density $\rho_2$ we wish to determine the temperature $T_2$ for which the state point $(\rho_2,T_2)$ is on the same isomorph as $(\rho_1,T_1)$. Each of the generated configurations $\bR_1$ is scaled uniformly to density $\rho_2$ via $\bR_2=(\rho_1/\rho_2)^{1/3}\bR_1$. We denote the potential energies of $\bR_1$ and $\bR_2$ by $U_1$ and $U_2$, respectively. Because $\bR_1$ and $\bR_2$ have the same reduced coordinates and $\Sex(\bR)$ for any R-simple system depends only on the configuration's reduced coordinate \cite{sch14}, one has $\Sex(\rho_1,U_1)=\Sex(\rho_2,U_2)$. Consequently, \eq{Rfundeq} implies

\be
U_2
\,=\,U(\rho_2,\Sex(\rho_1,U_1))\,.
\ee
Since $(\partial U/\partial\Sex)_\rho=T$ and $\rho_1$ and $\rho_2$ are both fixed, this implies for ratio of the potential-energy variations among different $\bR_1$ configurations, $\Delta U_1$, to the variation among the scaled configurations' potential energy, $\Delta U_2$, that

\be
\frac{\Delta U_2}{\Delta U_1}
\,=\,\left(\frac{\partial U(\rho_2,\Sex(\rho_1,U_1))}{\partial \Sex}\right)_{\rho_2}
\left(\frac{\partial\Sex(\rho_1,U_1)}{\partial U_1}\right)_{\rho_1}
\,=\,\frac{T_2}{T_1}\,.
\ee
In other words, the slope of a $U(\bR_2)$ versus $U(\bR_1)$ scatter plot is $T_2/T_1$, which allows for an easy way to determine $T_2$. This is the direct isomorph check (DIC) \cite{IV,sch14}. For it to work properly it is important that the potential energies of the original and the scaled configurations are well correlated \cite{IV}; for the EXP system we find correlation coefficients above 99.8\% when the density is doubled. 

\begin{figure}[!htbp]
	\centering
	\includegraphics[width=0.4\textwidth]{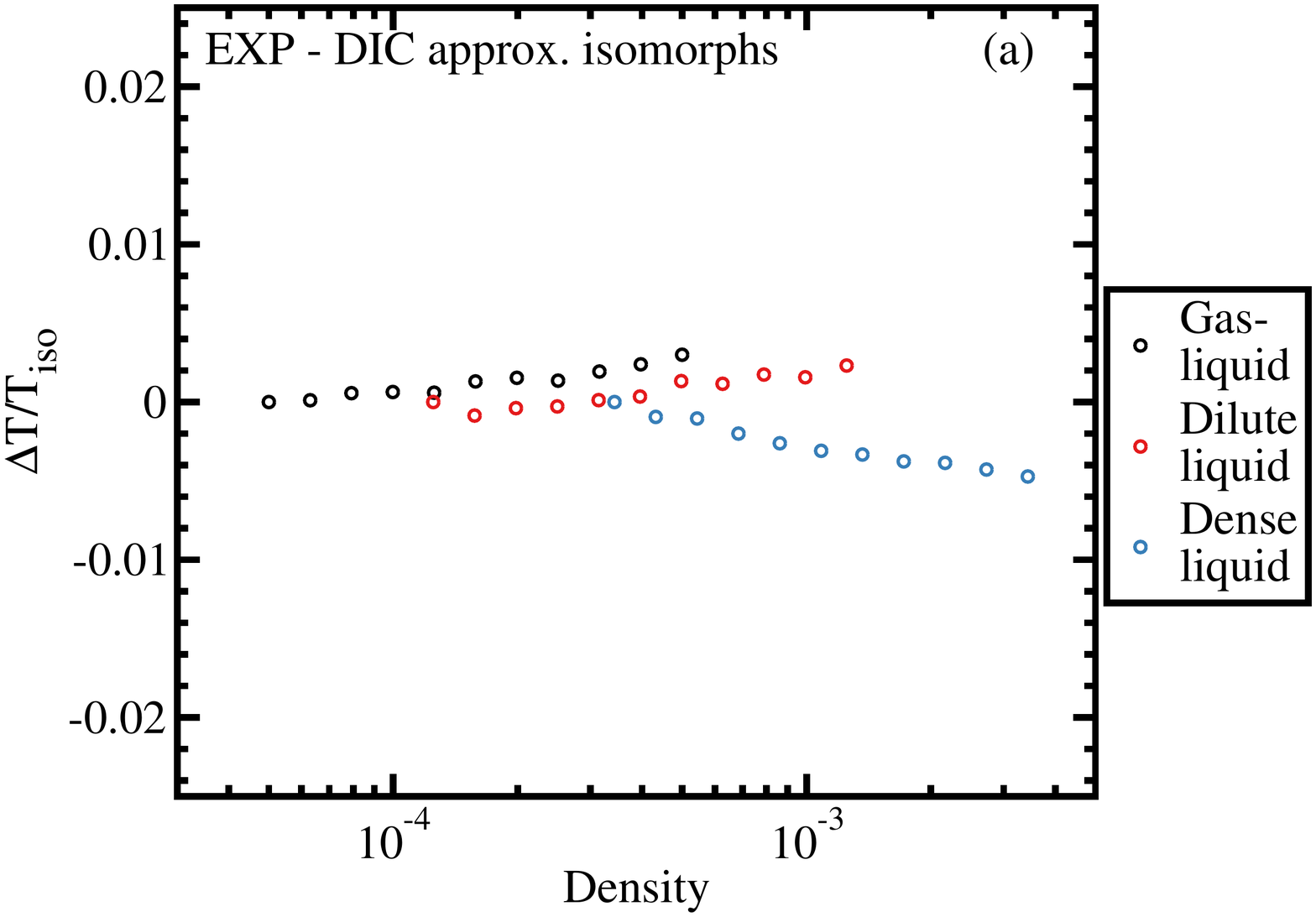}
	\includegraphics[width=0.4\textwidth]{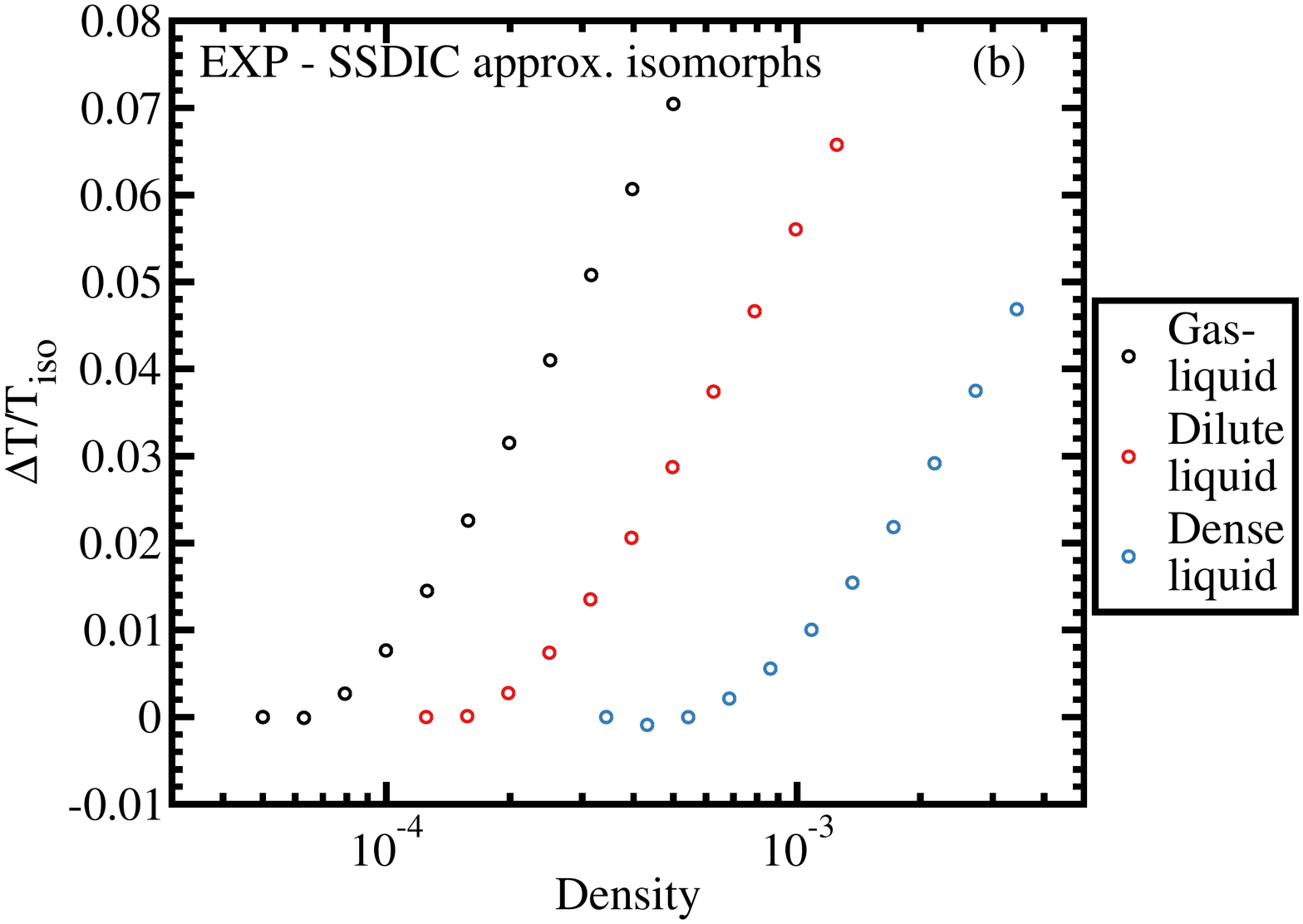}
	\caption{\label{fi:Delta_T}Relative differences in temperature along three approximate isomorphs generated by the direct isomorph check (DIC) and the single-state-point direct isomorph check (SSDIC) methods, compared to the temperatures of the ``exact'' isomorphs of Sec. \ref{sec:isom}. 
	(a) shows results for the DIC method based on ten consecutive density increases, each of 25.9\%, covering in all one density decade. 
	(b) shows results where each of the ten jumps were generated by the SSDIC method, i.e., jumping from the same low-density starting point. Not surprisingly, the latter method is less accurate than the DIC method, in particular for the largest density jumps, but in both cases deviations from the exact isomorph temperatures are small.
	}
\end{figure}

We proceed to compare isomorphs generated from ten successive DIC jumps to what is termed ``single state point DIC'' (SSDIC)-generated isomorphs that start from the same density $\rho_1$. The latter method ultimately increases density by one order of magnitude, whereas the DIC-generated isomorphs cover one decade of densities via ten jumps that are equally spaced on the logarithmic density axis, each increasing density by $25.9\%$.

\begin{figure}[!htbp]
	\centering
	\includegraphics[width=0.4\textwidth]{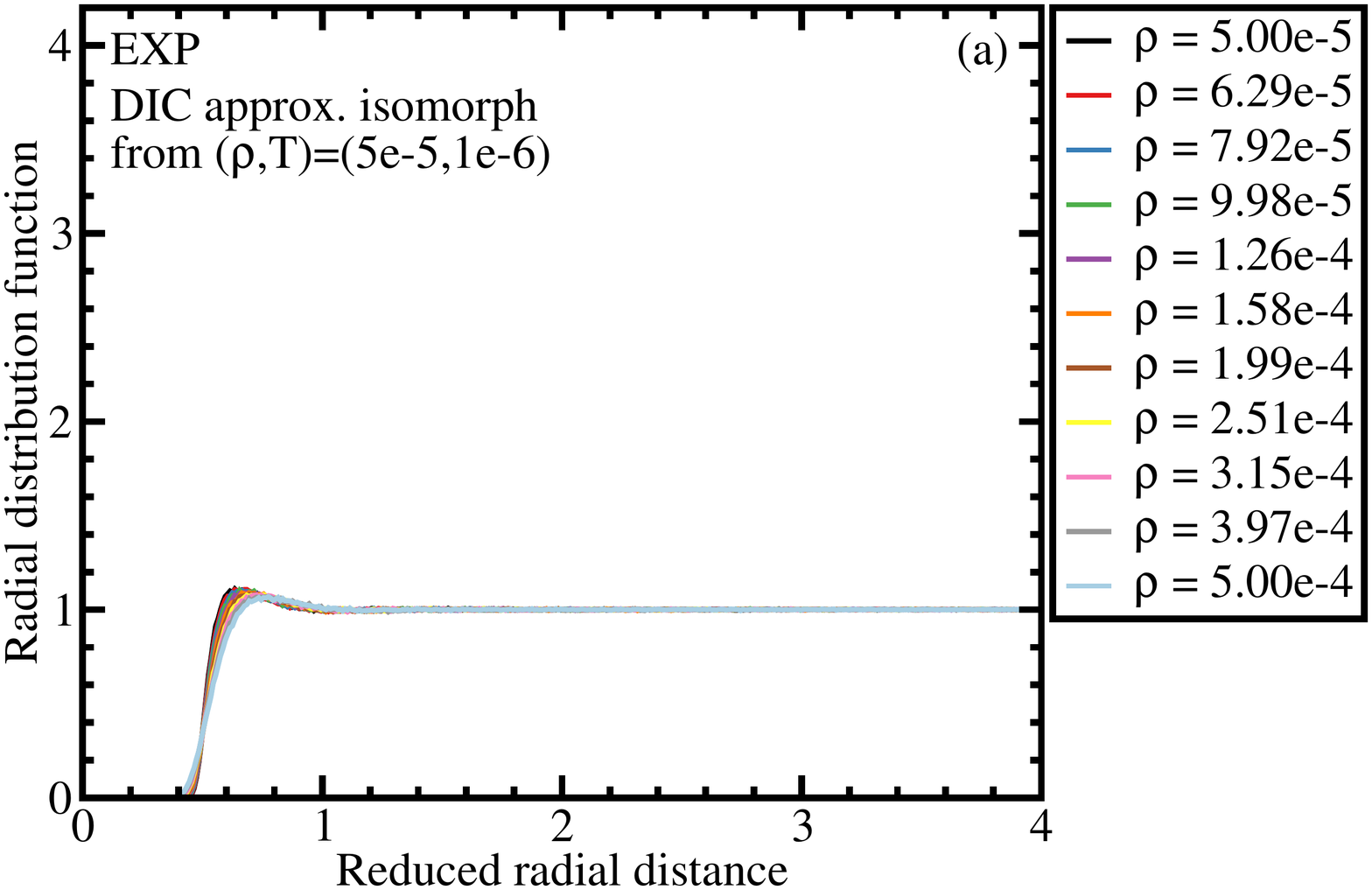}
	\includegraphics[width=0.4\textwidth]{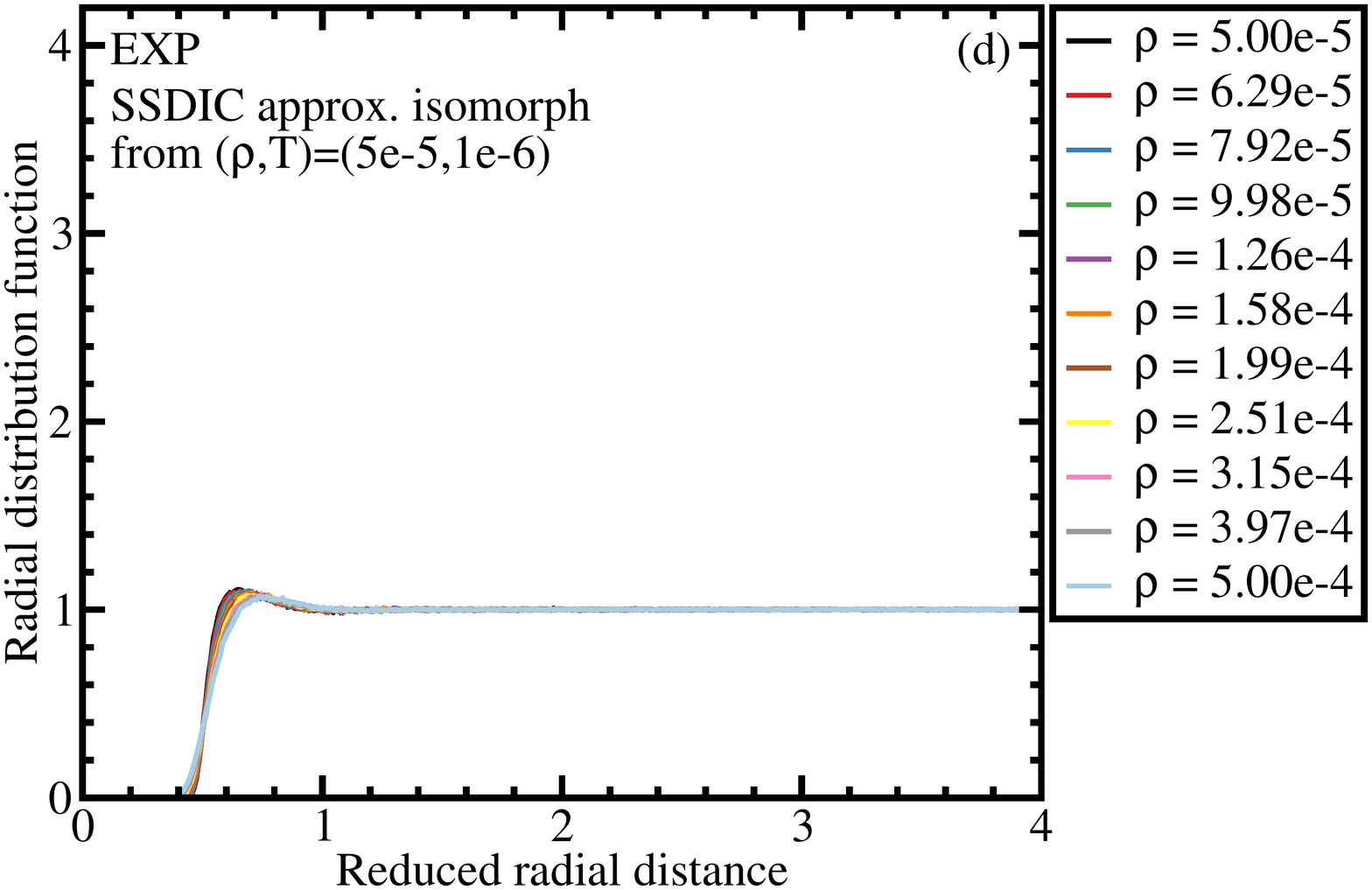}
	\includegraphics[width=0.4\textwidth]{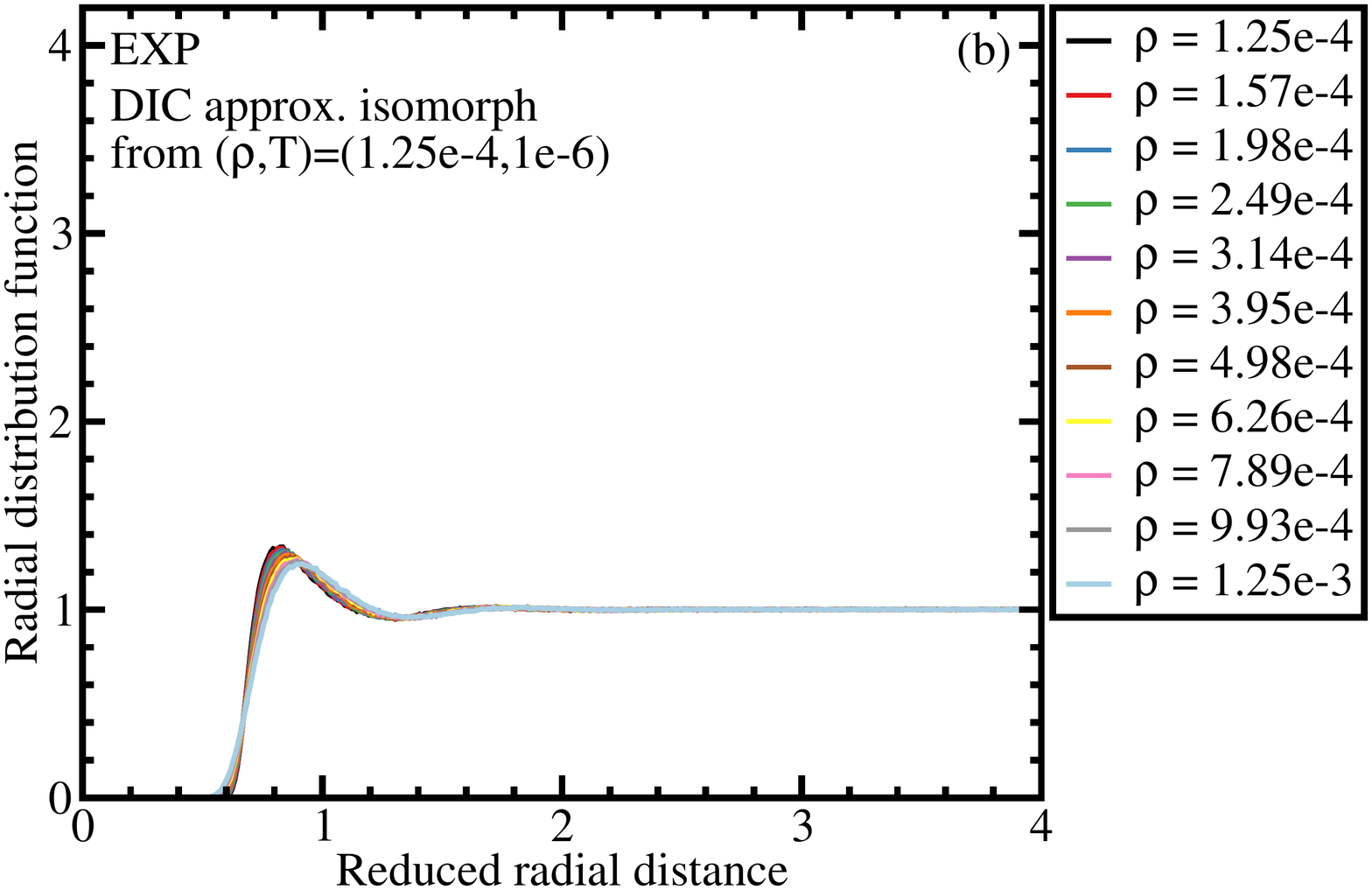}
	\includegraphics[width=0.4\textwidth]{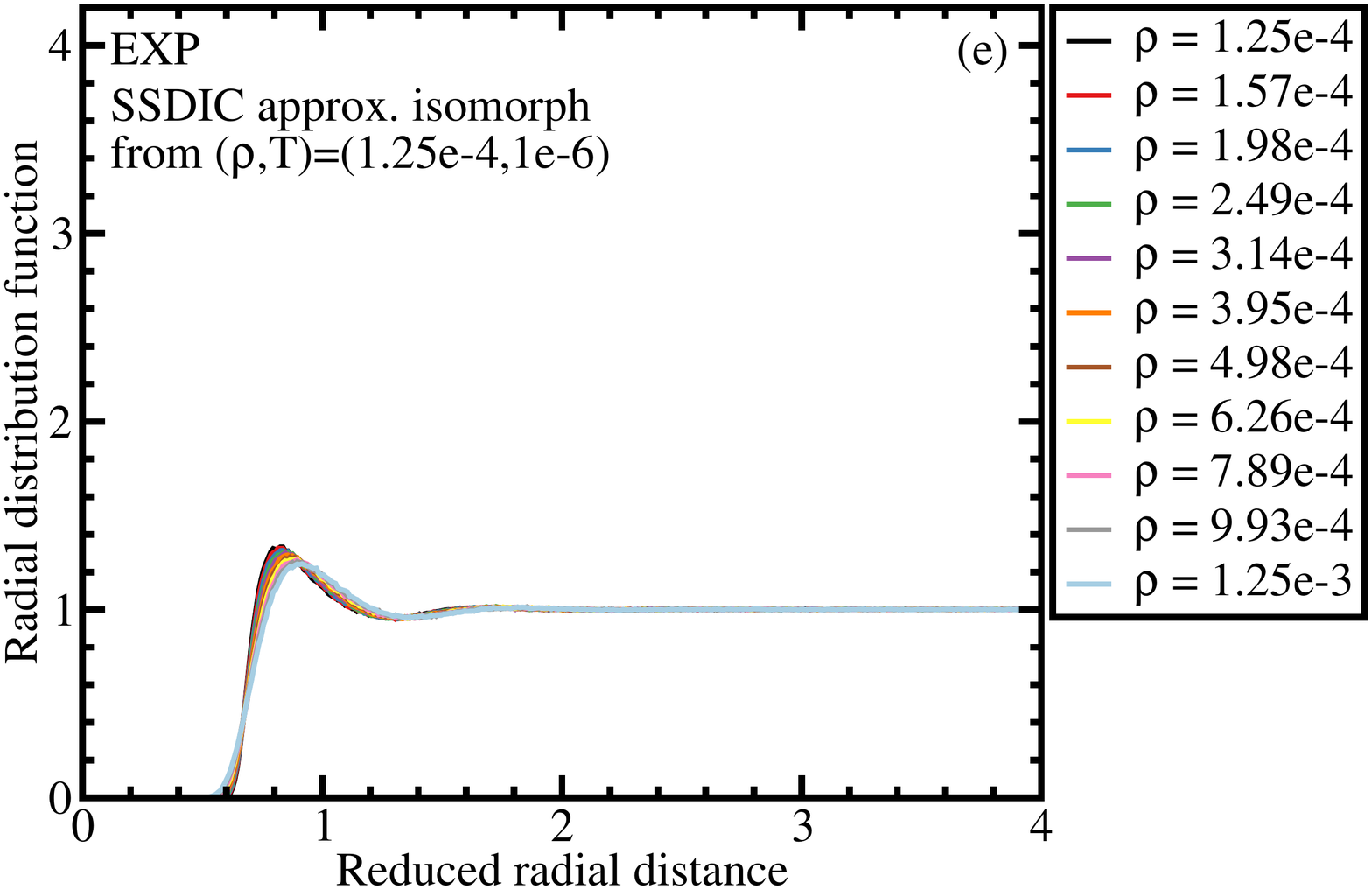}
	\includegraphics[width=0.4\textwidth]{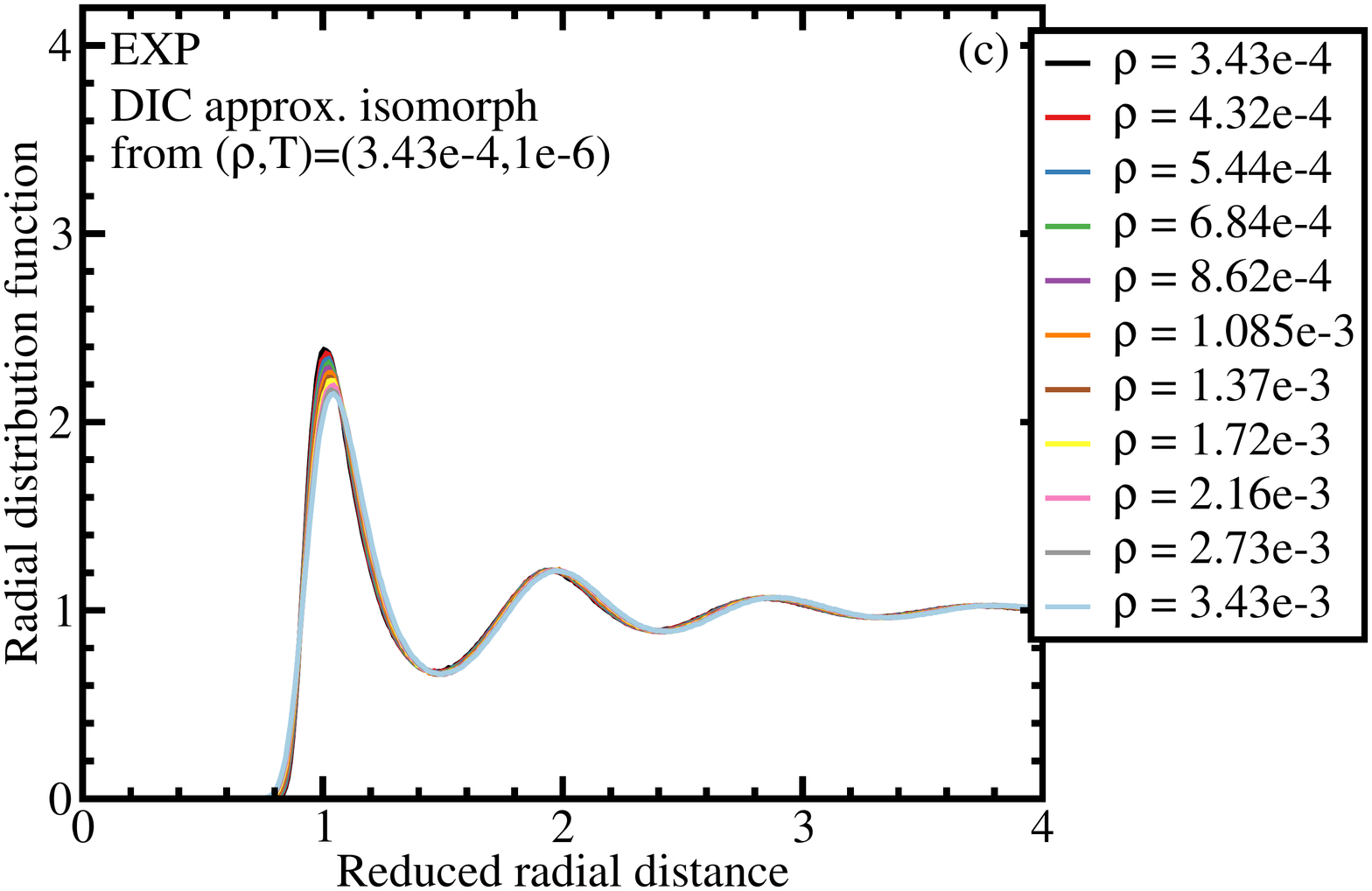}
	\includegraphics[width=0.4\textwidth]{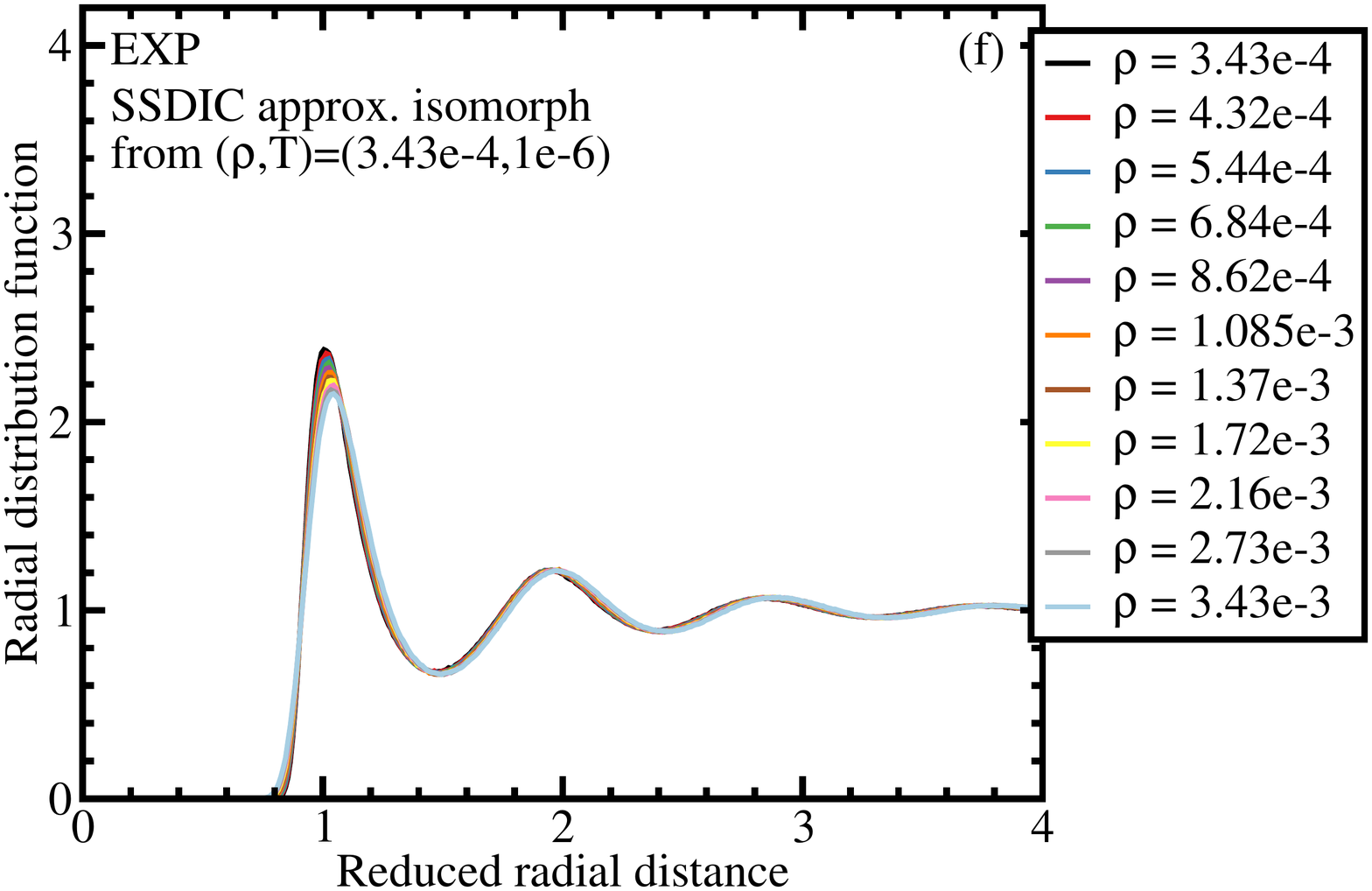}
	\caption{\label{f:DIC_structure}  
	\emph{Left:} RDFs for selected state points along each of three approximate isomorphs generated by ten consecutive applications of the DIC method. 
	(a) Gas-liquid isomorph, (b) dilute-liquid isomorph, (c) dense-liquid isomorph. 
	\emph{Right:} Structure for state points along each of three approximate isomorphs generated by the SSDIC method jumping in each case from the same starting state point. (d) Gas-liquid isomorph, (e) dilute-liquid isomorph, (f) dense-liquid isomorph. The two methods give similar results and compare well to the exact isomorph results of \fig{fi:isom-rdf} based on 230 small-step jumps.}
\end{figure}

\begin{figure}[!htbp]
	\centering
	\includegraphics[width=0.4\textwidth]{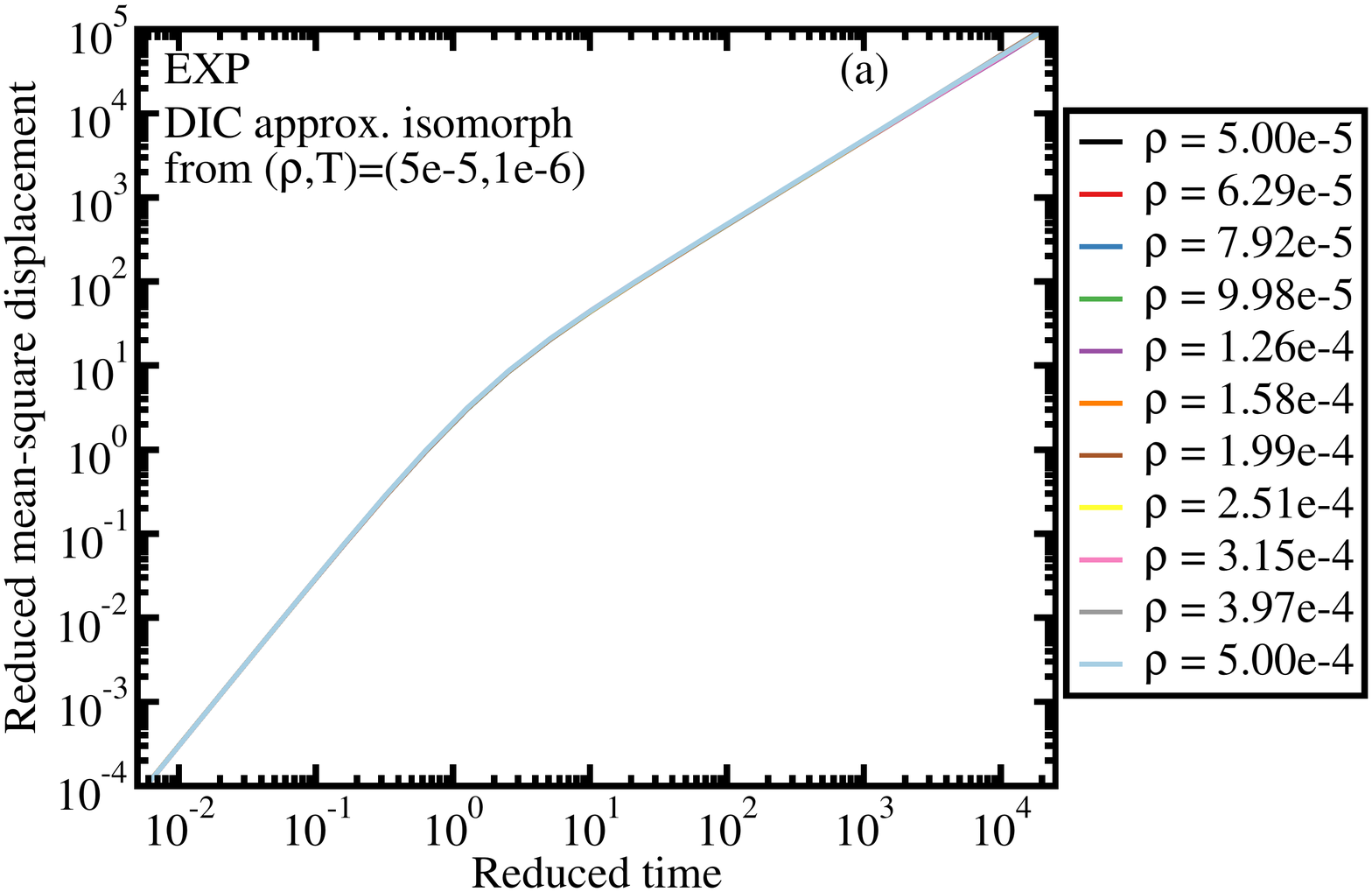}
	\includegraphics[width=0.4\textwidth]{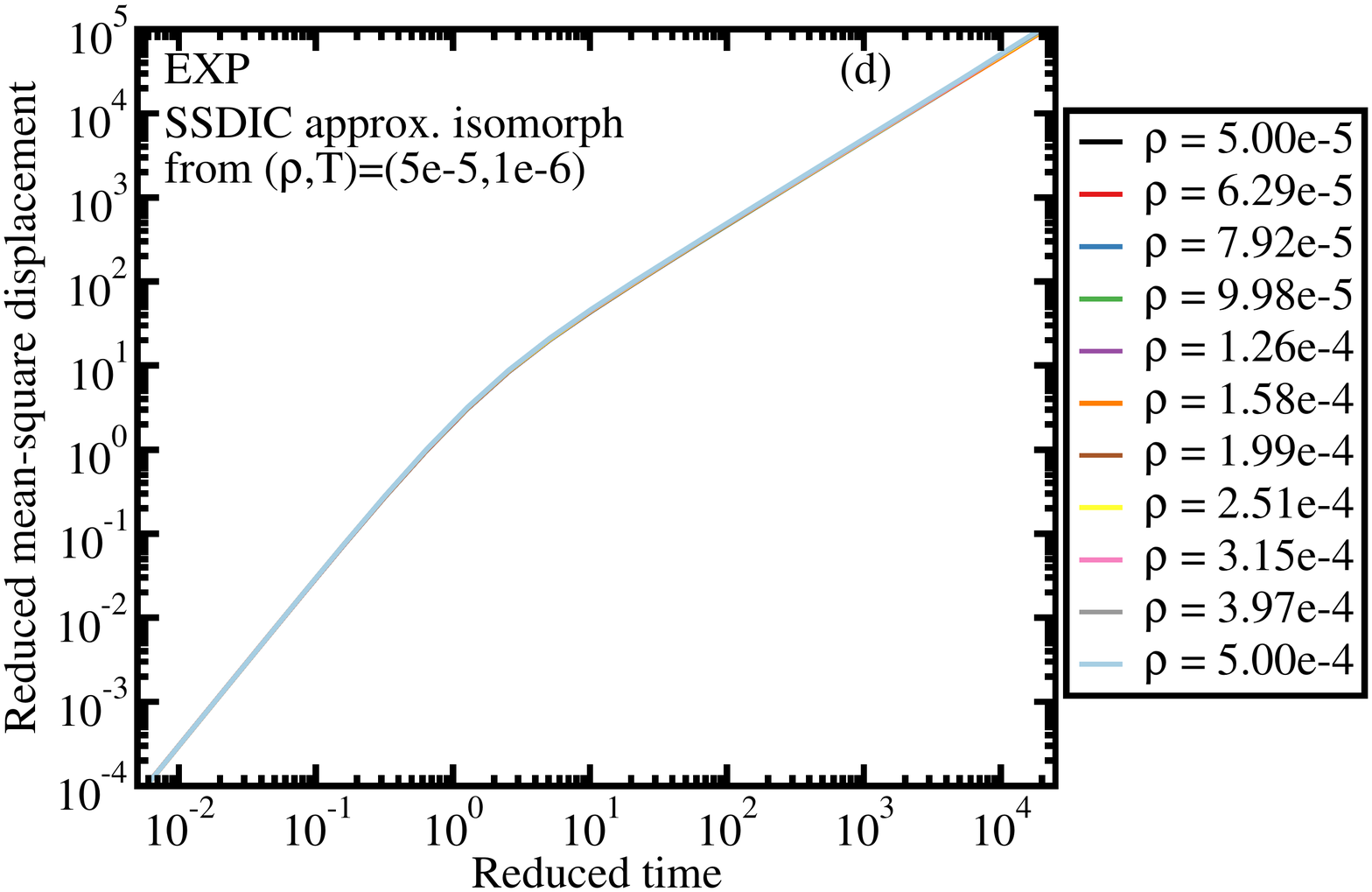}
	\includegraphics[width=0.4\textwidth]{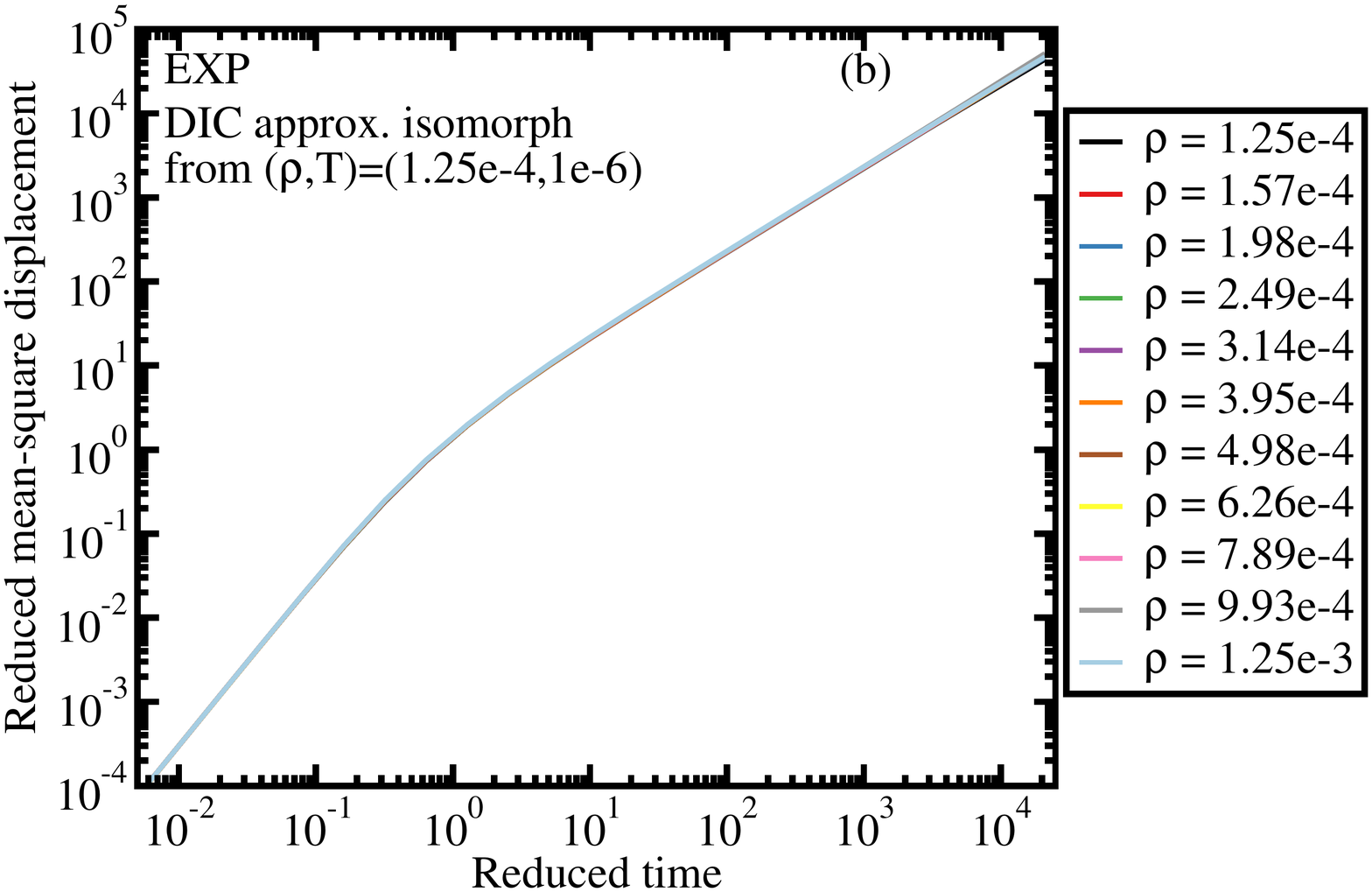}
	\includegraphics[width=0.4\textwidth]{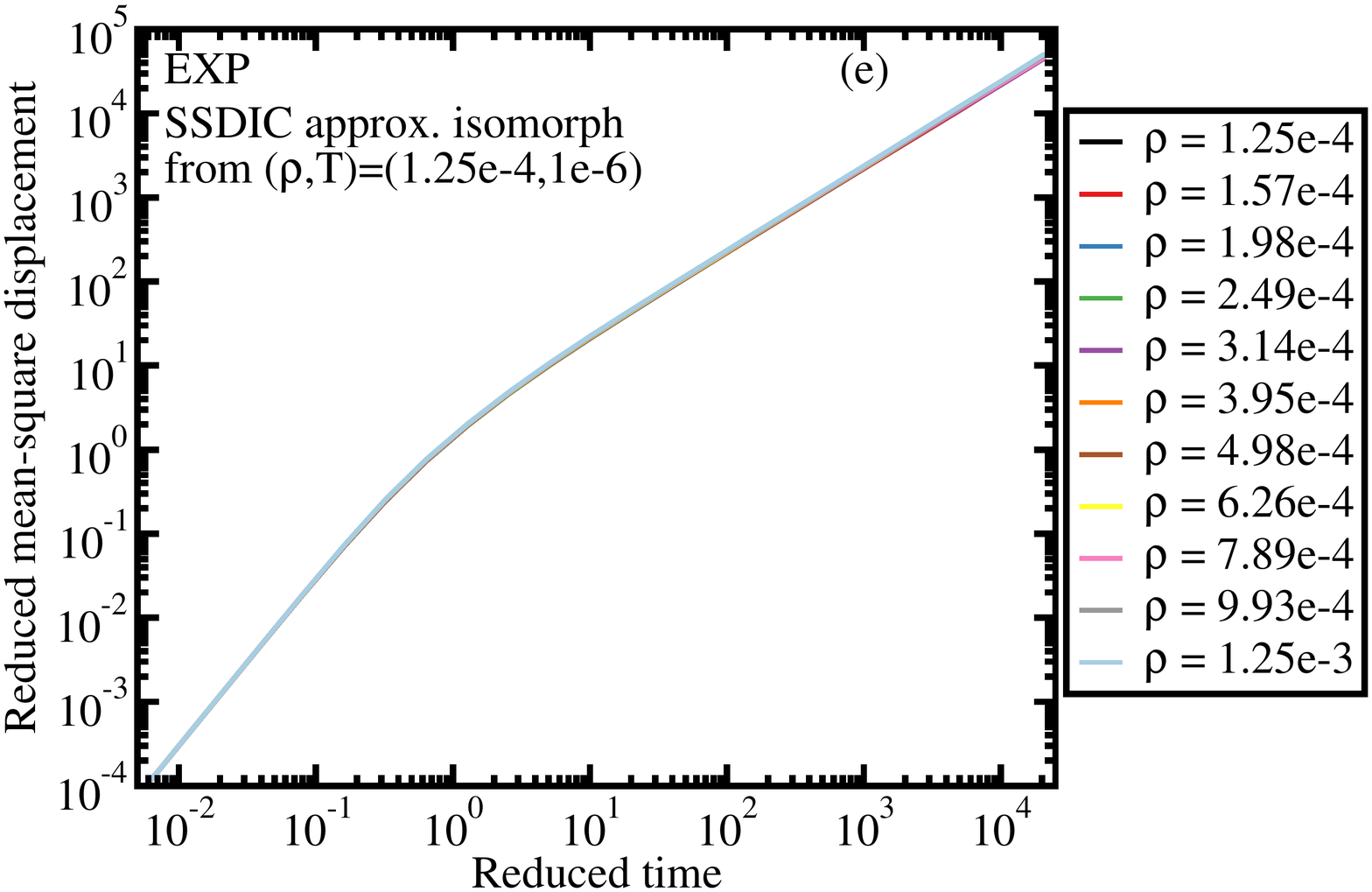}
	\includegraphics[width=0.4\textwidth]{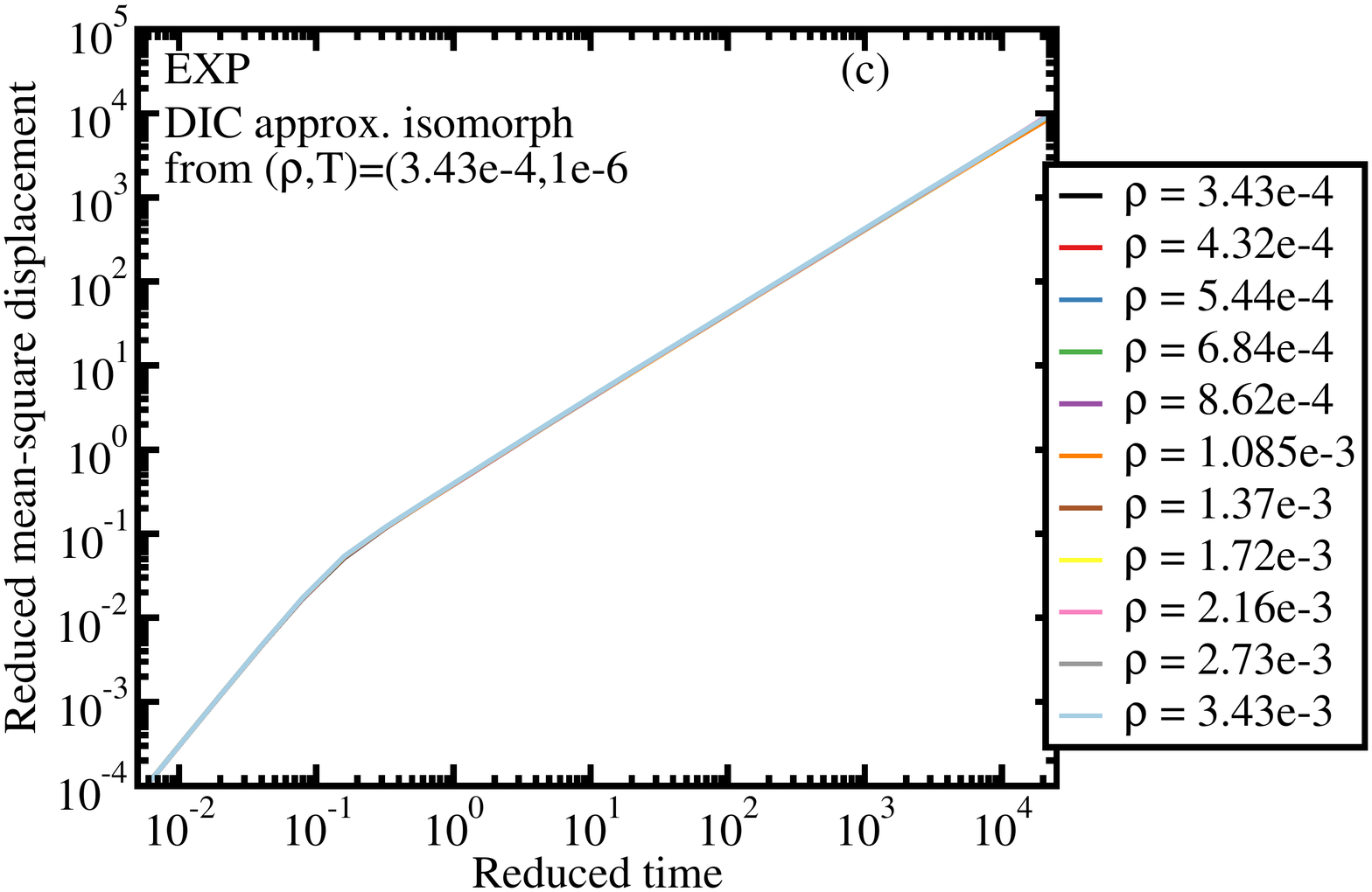}
	\includegraphics[width=0.4\textwidth]{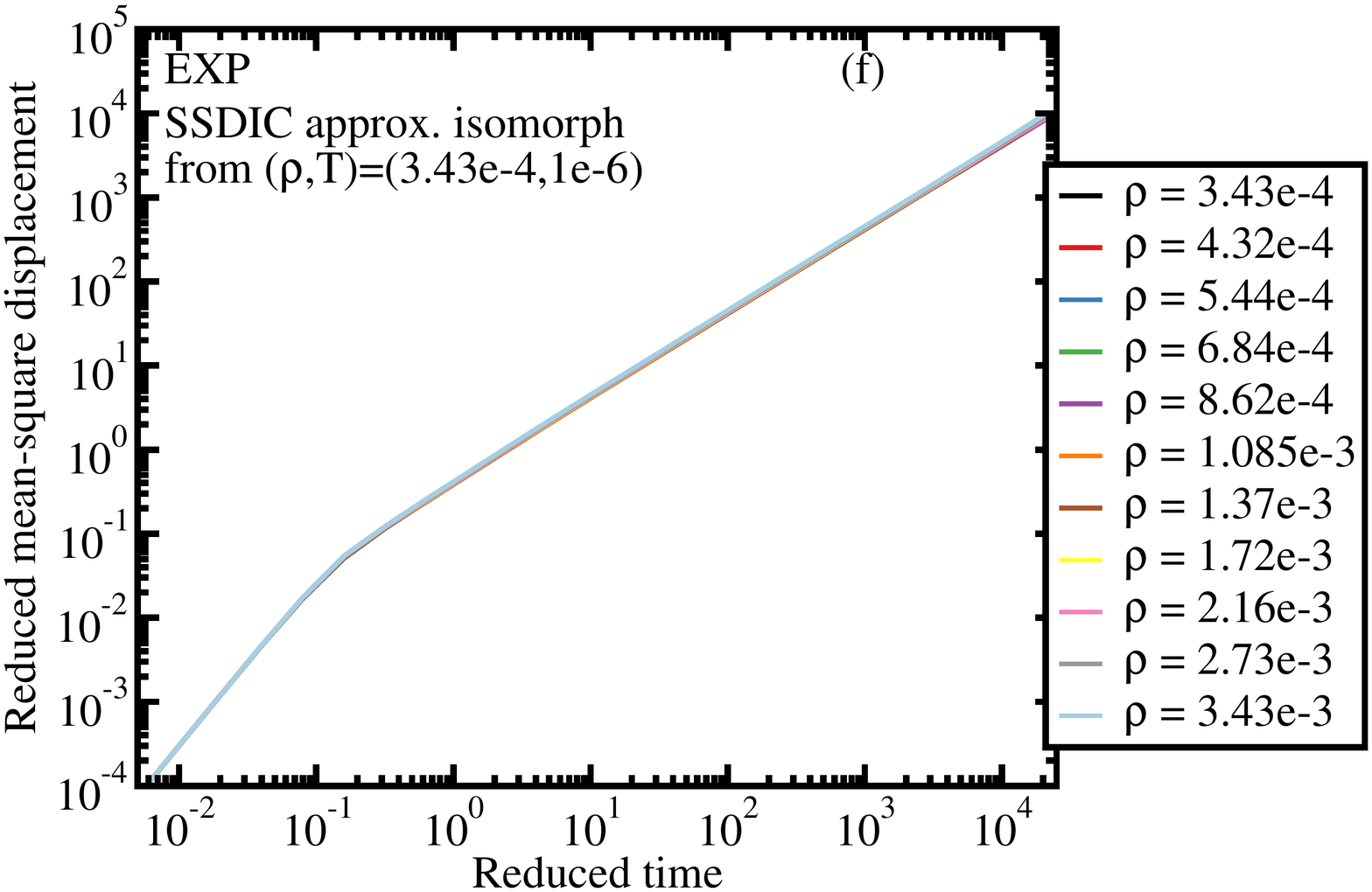}
	\caption{\label{f:DIC_MSD} 	
	\emph{Left:} Reduced MSD for selected state points along each of three approximate isomorphs generated by ten consecutive applications of the DIC method (same state points as in \fig{f:DIC_structure}, left). 
	(a) Gas-liquid isomorph, (b) dilute-liquid isomorph, (c) dense-liquid isomorph. 
	\emph{Right:} Reduced MSD for state points along each of three approximate isomorphs generated by the SSDIC method jumping in each case from the same starting state point. 
	(d) Gas-liquid isomorph, (e) dilute-liquid isomorph, (f) dense-liquid isomorph. The two methods give similar results and compare well to the exact isomorph results of \fig{fi:isom-msd} based on 230 small-step jumps.	
	}
\end{figure}

\begin{figure}[!htbp]
	\centering
	\includegraphics[width=0.4\textwidth]{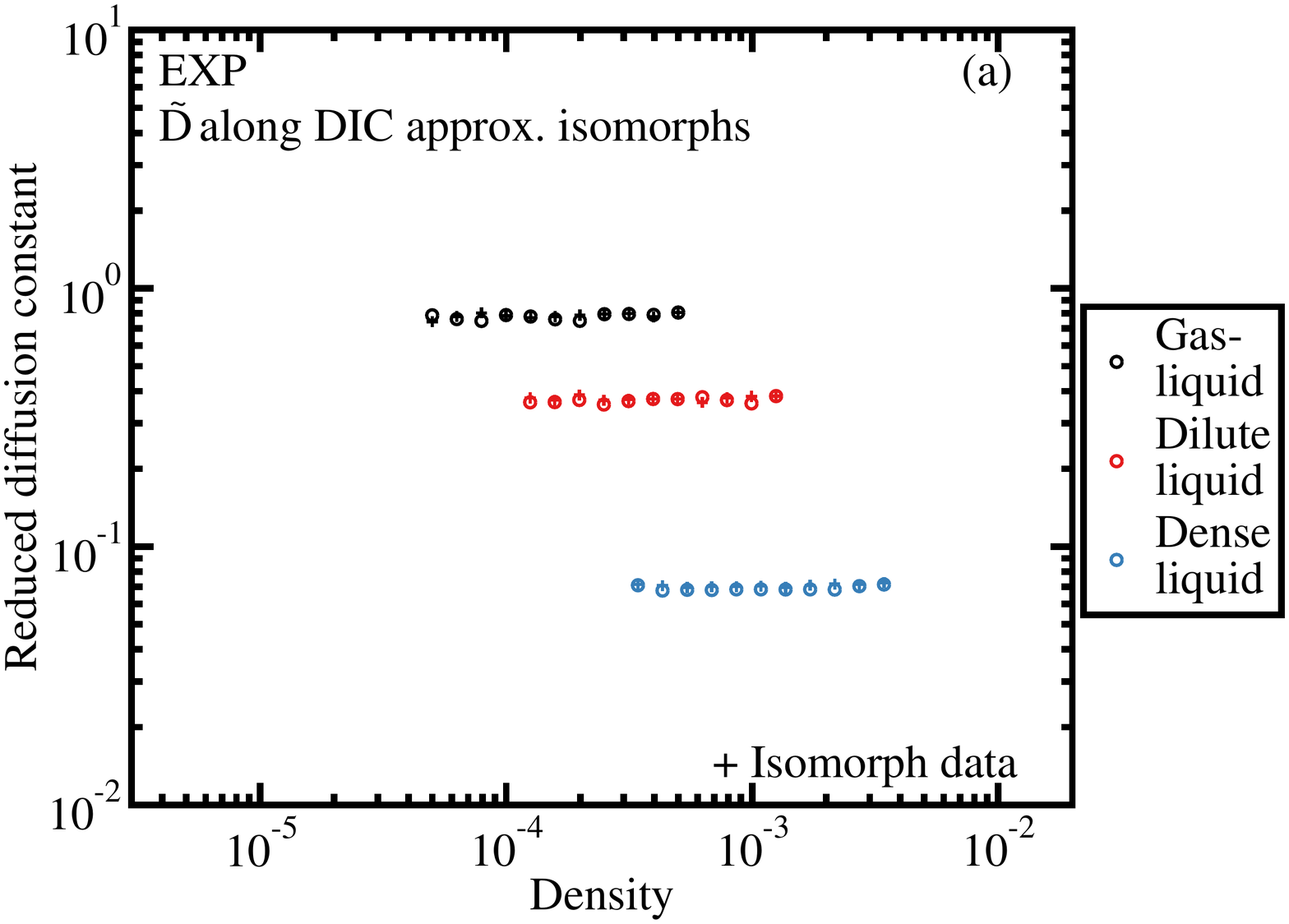}
	\includegraphics[width=0.4\textwidth]{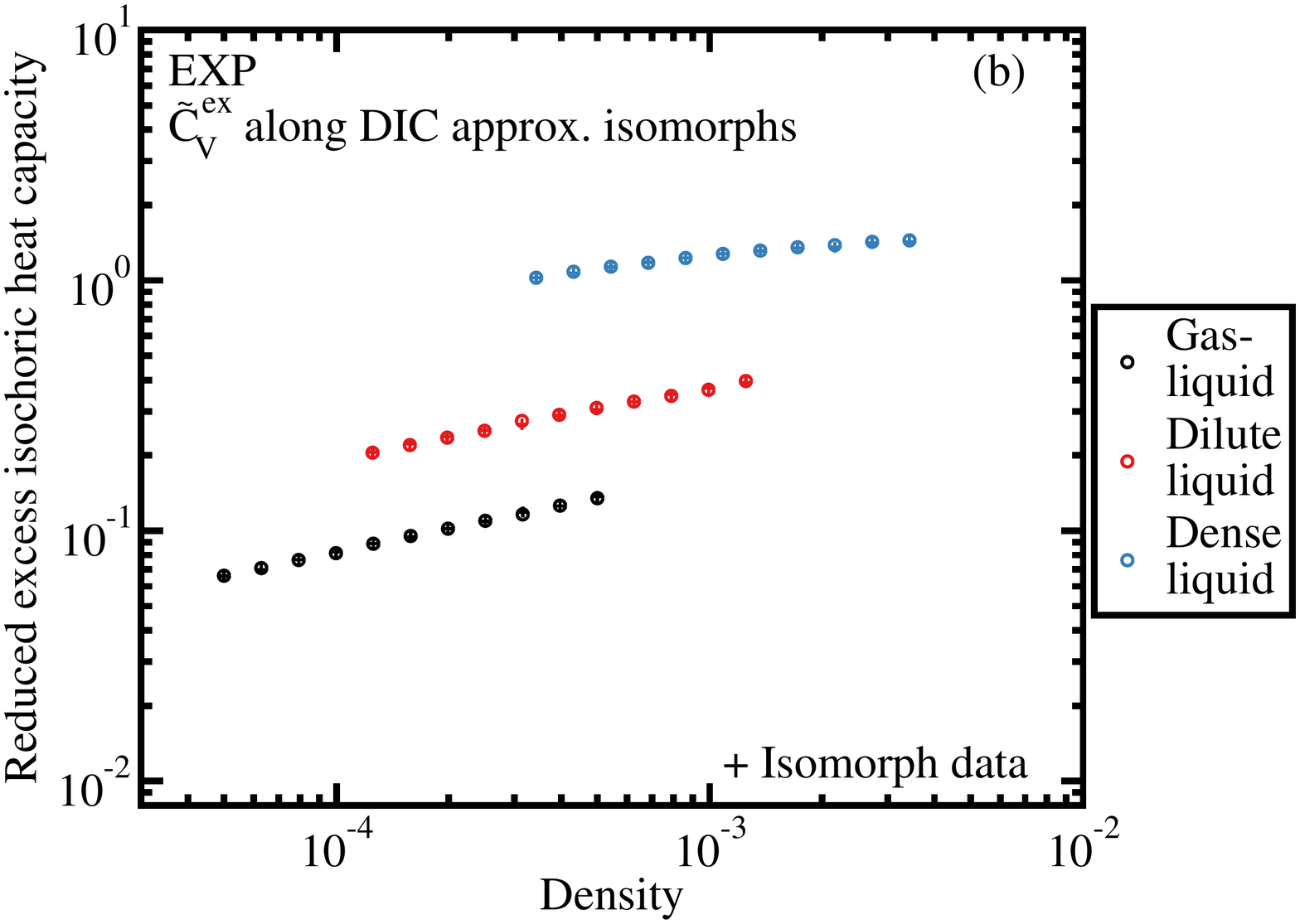}
	\includegraphics[width=0.4\textwidth]{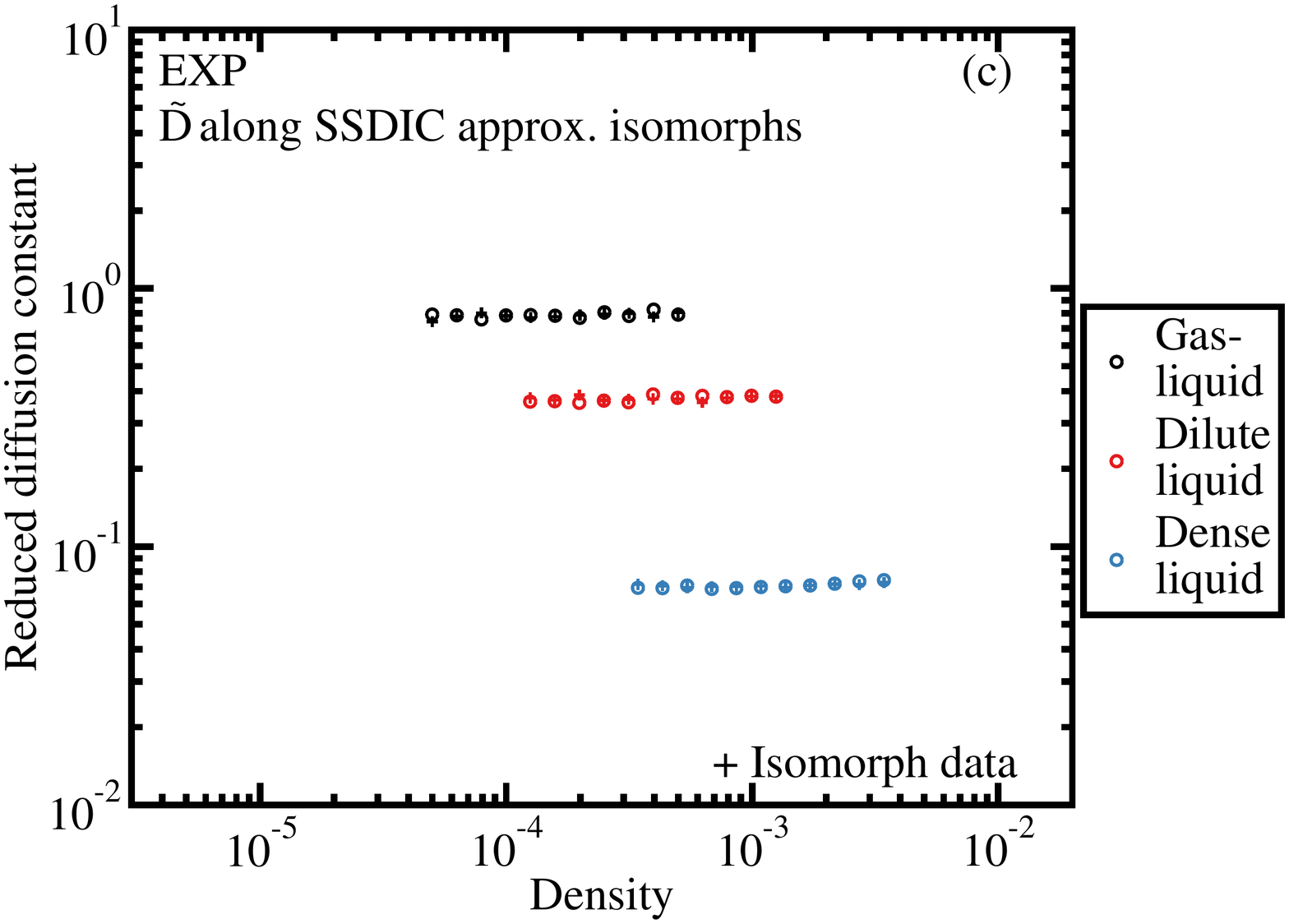}
	\includegraphics[width=0.4\textwidth]{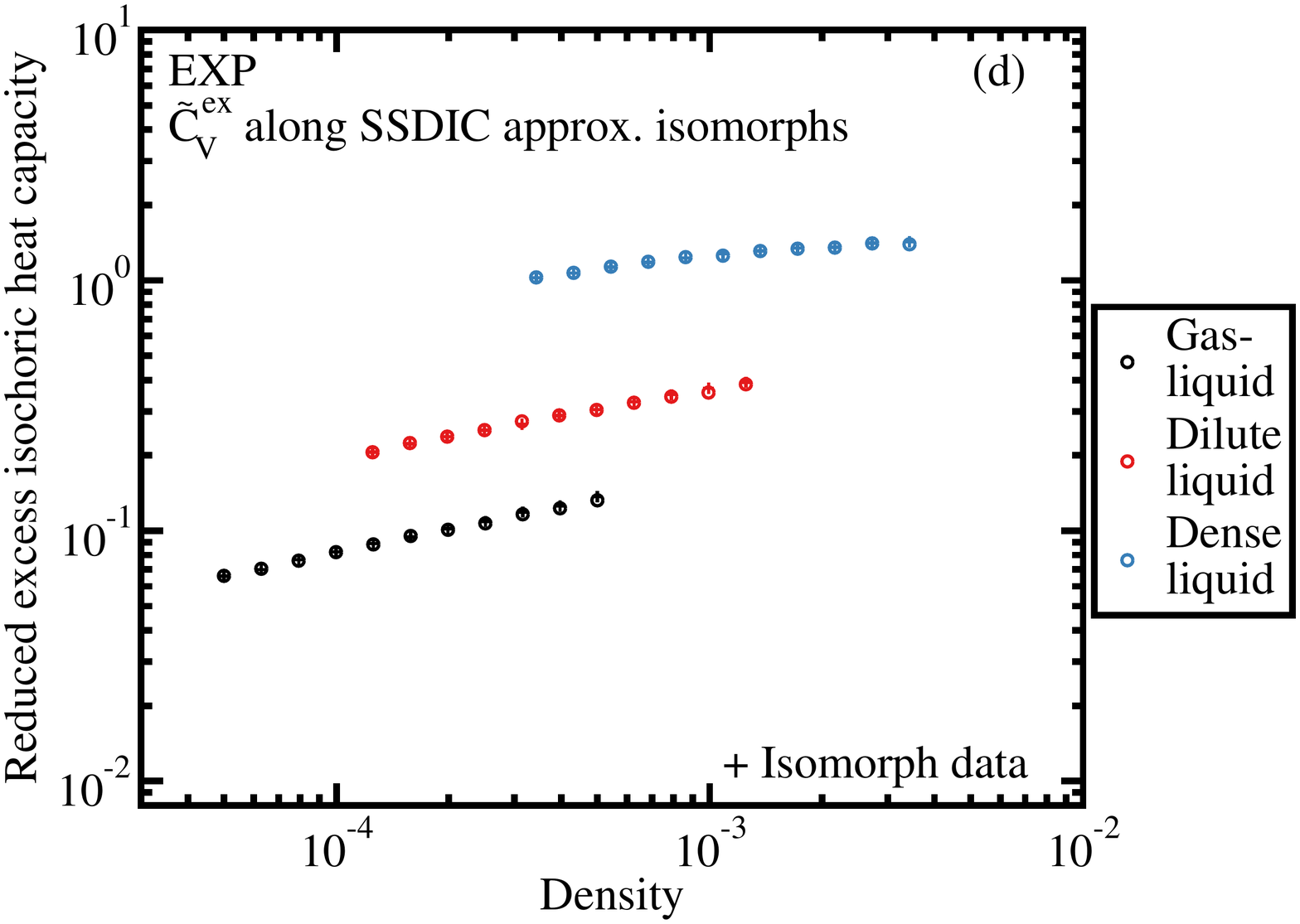}
	\caption{\label{DICvsSSDIC} DIC and SSDIC predicted data (circles) compared to the exact isomorph data (crosses).	
	(a) and (c) show DIC and SSDIC results for the reduced diffusion constant, which are hard to distinguish from each other and from the exact results of \fig{fi:isom-D}(a). (b) and (d) show DIC and SSDIC results for the reduced specific heat, which also compare well to the exact results of \fig{fi:isom-Cv}.
	}
\end{figure}

\Fig{fi:Delta_T} shows the relative temperature differences between approximate and ``exact'' isomorphs as a function of density with (a) giving the DIC method results and (b) the SSDIC method results. As the step size increases, the SSDIC method systematically overshoots the temperature. In this light it may seem surprising that good collapse is still maintained for the physics (\fig{f:DIC_structure} and \fig{f:DIC_MSD}). This is because the deviations in temperature from the small-step method temperatures obtained in Sec. \ref{sec:isom} are below $7\%$, even for temperature changes spanning two decades. 

\Fig{DICvsSSDIC} shows the reduced diffusion constant and the reduced specific heats along the DIC- and SSDIC-generated approximate isomorphs, respectively, in both cases showing results close to the those of the exact isomorphs.

\section{Discussion}\label{sec:sum}

As shown in Paper I, the EXP system has strong virial potential-energy correlations in the low-temperature part of its phase diagram. The present paper has demonstrated the existence of isomorphs in this part of the phase diagram. Three isomorphs were studied, one on the gas-liquid border, one in the dilute-liquid phase, and one in the condensed-liquid phase. Each isomorph covers one decade of density; they were generated by 230 simulations using \eq{gamma_eq} in a step-by-step numerical integration of \eq{dsexp}. Two approximate methods were studied for generating isomorphs numerically, the direct-isomorph-check (DIC) method and its single-state-point version (SSDIC). Both methods work well.

The virial potential-energy correlation coefficient $R$ is very close to unity at the lowest temperatures studied $T=10^{-6}$, in fact here above 99.8\% at all densities simulated (Paper I). From this one may be tempted to conclude that $R\rightarrow 1$ for $T\rightarrow 0$ at fixed density. We do not think this is the case, though. At any fixed density, the system eventually crystallizes upon cooling. Even though the crystal obeys $R\cong 1$ at low temperatures, there is no reason to expect $R\rightarrow 1$ for $T\rightarrow 0$ at a fixed density. The point is that the finite range of the EXP pair potential means that in a harmonic approximation there are not just nearest-neighbor springs, but also next-nearest neighbor springs, etc, and the different spring constants are not proportional to one another  when density is changed to investigate whether or not \eq{hs} applies rigorously. 

We conjecture, however, that $R\rightarrow 1$ for $T\rightarrow 0$ \textit{along any isomorph}. Here the physics is invariant, and as the temperature is lowered towards zero, the density-scaling exponent $\gamma$ increases towards infinity because the density decreases. The EXP system becomes more and more hard-sphere (HS) like, and \fig{fi:phase-diagram:cont}(b) shows that as $\gamma$ diverges, $R\rightarrow 1$ because $R$ was empirically found always to be larger than its gas limit (for which one rigorously has $R\rightarrow 1$ as $\rho\rightarrow 0$, the dashed curve of \fig{fi:phase-diagram:cont}(b)). 

For future work it would be interesting to undertake a systematic investigation of the crystal phase. We have observed that the stable crystal structure is bcc at relatively high temperatures, but that it transforms into the fcc structure below certain temperatures (roughly when $\gamma$ goes above two).

\section*{Appendix}

We give below the coordinates of selected state points along the three isomorphs determined by the ``small-step'' method via density changes of 1\%. For each isomorph every 23rd state point is listed.

\begin{table}[h!]
\begin{center}
\begin{tabular}{lcl}
\multicolumn{1}{c}{$\rho$} & \  & \multicolumn{1}{c}{$T$} \\ 
\hline 
5e-05 & \ & 1e-06 \\
6.2946e-05 & \ & 2.6930e-06 \\
7.9245e-05 & \ & 6.7383e-06 \\
9.9763e-05 & \ & 1.5752e-05 \\
0.00012559 & \ & 3.4567e-05 \\
0.00015811 & \ & 7.1554e-05 \\
0.00019905 & \ & 0.00014032 \\
0.00025059 & \ & 0.00026169 \\
0.00031548 & \ & 0.00046586 \\
0.00039716 & \ & 0.00079425 \\
5e-04 & \ & 0.0013009 \\
\end{tabular} 
\caption{Selected state points along the gas-liquid isomorph.}
\label{tb:gas-liquid}
\end{center}
\end{table}

\begin{table}[h!]
\begin{center}
\begin{tabular}{lcl}
\multicolumn{1}{c}{$\rho$} & \ & \multicolumn{1}{c}{$T$} \\ 
\hline 
0.00012500 & \ & 1e-06 \\
0.00015737 & \ & 2.6521e-06 \\
0.00019811 & \ & 6.5325e-06 \\
0.00024941 & \ & 1.5024e-05 \\
0.00031399 & \ & 3.2420e-05 \\
0.00039528 & \ & 6.5954e-05 \\
0.00049763 & \ & 0.00012696 \\
0.00062648 & \ & 0.00023222 \\
0.00078870 & \ & 0.00040490 \\
0.00099291 & \ & 0.00067539 \\
0.0012500 & \ & 0.0010808 \\
\end{tabular} 
\caption{Selected state points along the dilute-liquid isomorph. }
\label{tb:dilute-liquid}
\end{center}
\end{table}

\begin{table}[h!]
\begin{center}
\begin{tabular}{lcl}
\multicolumn{1}{c}{$\rho$} & \ & \multicolumn{1}{c}{$T$} \\ 
\hline 
0.00034300 & \ & 1e-06 \\
0.00043181 & \ & 2.4745e-06 \\
0.00054362 & \ & 5.6825e-06 \\
0.00068437 & \ & 1.2174e-05 \\
0.00086158 & \ & 2.4439e-05 \\
0.0010847 & \ &  4.6170e-05 \\
0.0013655 & \ &  8.2413e-05 \\
0.0017191 & \ &  0.00013950 \\
0.0021642 & \ &  0.00022469 \\
0.0027245 & \ &  0.00034553 \\
0.0034300 &\ &   0.00050885 \\
\end{tabular} 
\caption{Selected state points along the dense-liquid isomorph. }
\label{tb:dense-liquid}
\end{center}
\end{table}

\acknowledgments{We thank Lorenzo Costigliola for helpful discussions.
This work was supported by the VILLUM Foundation's \textit{Matter} grant (16515).}


\begin{thebibliography}{37}%
	\makeatletter
	\providecommand \@ifxundefined [1]{%
		\@ifx{#1\undefined}
	}%
	\providecommand \@ifnum [1]{%
		\ifnum #1\expandafter \@firstoftwo
		\else \expandafter \@secondoftwo
		\fi
	}%
	\providecommand \@ifx [1]{%
		\ifx #1\expandafter \@firstoftwo
		\else \expandafter \@secondoftwo
		\fi
	}%
	\providecommand \natexlab [1]{#1}%
	\providecommand \enquote  [1]{``#1''}%
	\providecommand \bibnamefont  [1]{#1}%
	\providecommand \bibfnamefont [1]{#1}%
	\providecommand \citenamefont [1]{#1}%
	\providecommand \href@noop [0]{\@secondoftwo}%
	\providecommand \href [0]{\begingroup \@sanitize@url \@href}%
	\providecommand \@href[1]{\@@startlink{#1}\@@href}%
	\providecommand \@@href[1]{\endgroup#1\@@endlink}%
	\providecommand \@sanitize@url [0]{\catcode `\\12\catcode `\$12\catcode
		`\&12\catcode `\#12\catcode `\^12\catcode `\_12\catcode `\%12\relax}%
	\providecommand \@@startlink[1]{}%
	\providecommand \@@endlink[0]{}%
	\providecommand \url  [0]{\begingroup\@sanitize@url \@url }%
	\providecommand \@url [1]{\endgroup\@href {#1}{\urlprefix }}%
	\providecommand \urlprefix  [0]{URL }%
	\providecommand \Eprint [0]{\href }%
	\providecommand \doibase [0]{http://dx.doi.org/}%
	\providecommand \selectlanguage [0]{\@gobble}%
	\providecommand \bibinfo  [0]{\@secondoftwo}%
	\providecommand \bibfield  [0]{\@secondoftwo}%
	\providecommand \translation [1]{[#1]}%
	\providecommand \BibitemOpen [0]{}%
	\providecommand \bibitemStop [0]{}%
	\providecommand \bibitemNoStop [0]{.\EOS\space}%
	\providecommand \EOS [0]{\spacefactor3000\relax}%
	\providecommand \BibitemShut  [1]{\csname bibitem#1\endcsname}%
	\let\auto@bib@innerbib\@empty
	%</preamble>
	\bibitem [{\citenamefont {Bacher}\ \emph {et~al.}(2018)\citenamefont {Bacher},
		\citenamefont {Schr{\o}der},\ and\ \citenamefont {Dyre}}]{EXP_I_arXiv}%
	\BibitemOpen
	\bibfield  {author} {\bibinfo {author} {\bibfnamefont {A.~K.}\ \bibnamefont
			{Bacher}}, \bibinfo {author} {\bibfnamefont {T.~B.}\ \bibnamefont
			{Schr{\o}der}}, \ and\ \bibinfo {author} {\bibfnamefont {J.~C.}\ \bibnamefont
			{Dyre}},\ }\bibfield  {title} {\enquote {\bibinfo {title} {The {EXP}
				pair-potential system. {I.} {Fluid} phase isotherms, isochores, and
				quasiuniversality},}\ }\href@noop {} {\bibfield  {journal} {\bibinfo
			{journal} {Companion Paper}\ } (\bibinfo {year} {2018})}\BibitemShut
	{NoStop}%
	\bibitem [{\citenamefont {Hoover}\ \emph {et~al.}(1972)\citenamefont {Hoover},
		\citenamefont {Young},\ and\ \citenamefont {Grover}}]{hoo72}%
	\BibitemOpen
	\bibfield  {author} {\bibinfo {author} {\bibfnamefont {W.~G.}\ \bibnamefont
			{Hoover}}, \bibinfo {author} {\bibfnamefont {D.~A.}\ \bibnamefont {Young}}, \
		and\ \bibinfo {author} {\bibfnamefont {R.}~\bibnamefont {Grover}},\
	}\bibfield  {title} {\enquote {\bibinfo {title} {Statistical mechanics of
				phase diagrams. i. inverse power potentials and the close-packed to
				body-packed cubic transition},}\ }\href@noop {} {\bibfield  {journal}
		{\bibinfo  {journal} {J. Chem. Phys.}\ }\textbf {\bibinfo {volume} {56}},\
		\bibinfo {pages} {2207--2210} (\bibinfo {year} {1972})}\BibitemShut {NoStop}%
	\bibitem [{\citenamefont {Hiwatari}\ \emph {et~al.}(1974)\citenamefont
		{Hiwatari}, \citenamefont {Matsuda}, \citenamefont {Ogawa}, \citenamefont
		{Ogita},\ and\ \citenamefont {Ueda}}]{hiw74}%
	\BibitemOpen
	\bibfield  {author} {\bibinfo {author} {\bibfnamefont {Y.}~\bibnamefont
			{Hiwatari}}, \bibinfo {author} {\bibfnamefont {H.}~\bibnamefont {Matsuda}},
		\bibinfo {author} {\bibfnamefont {T.}~\bibnamefont {Ogawa}}, \bibinfo
		{author} {\bibfnamefont {N.}~\bibnamefont {Ogita}}, \ and\ \bibinfo {author}
		{\bibfnamefont {A.}~\bibnamefont {Ueda}},\ }\bibfield  {title} {\enquote
		{\bibinfo {title} {Molecular dynamics studies on the soft-core model},}\
	}\href@noop {} {\bibfield  {journal} {\bibinfo  {journal} {Prog. Theor.
				Phys.}\ }\textbf {\bibinfo {volume} {52}},\ \bibinfo {pages} {1105--1123}
		(\bibinfo {year} {1974})}\BibitemShut {NoStop}%
	\bibitem [{\citenamefont {Heyes}\ and\ \citenamefont {Branka}(2007)}]{hey07}%
	\BibitemOpen
	\bibfield  {author} {\bibinfo {author} {\bibfnamefont {D.~M.}\ \bibnamefont
			{Heyes}}\ and\ \bibinfo {author} {\bibfnamefont {A.~C.}\ \bibnamefont
			{Branka}},\ }\bibfield  {title} {\enquote {\bibinfo {title} {Physical
				properties of soft repulsive particle fluids},}\ }\href@noop {} {\bibfield
		{journal} {\bibinfo  {journal} {Phys. Chem. Chem. Phys.}\ }\textbf {\bibinfo
			{volume} {9}},\ \bibinfo {pages} {5570--5575} (\bibinfo {year}
		{2007})}\BibitemShut {NoStop}%
	\bibitem [{\citenamefont {Heyes}\ and\ \citenamefont {Branka}(2008)}]{hey08}%
	\BibitemOpen
	\bibfield  {author} {\bibinfo {author} {\bibfnamefont {D.~M.}\ \bibnamefont
			{Heyes}}\ and\ \bibinfo {author} {\bibfnamefont {A.~C.}\ \bibnamefont
			{Branka}},\ }\bibfield  {title} {\enquote {\bibinfo {title} {Self-diffusion
				coefficients and shear viscosity of inverse power fluids: from hard- to
				soft-spheres},}\ }\href@noop {} {\bibfield  {journal} {\bibinfo  {journal}
			{Phys. Chem. Chem. Phys.}\ }\textbf {\bibinfo {volume} {10}},\ \bibinfo
		{pages} {4036--4044} (\bibinfo {year} {2008})}\BibitemShut {NoStop}%
	\bibitem [{\citenamefont {Lange}\ \emph {et~al.}(2009)\citenamefont {Lange},
		\citenamefont {Caballero}, \citenamefont {Puertas},\ and\ \citenamefont
		{Fuchs}}]{lan09}%
	\BibitemOpen
	\bibfield  {author} {\bibinfo {author} {\bibfnamefont {E.}~\bibnamefont
			{Lange}}, \bibinfo {author} {\bibfnamefont {J.~B.}\ \bibnamefont
			{Caballero}}, \bibinfo {author} {\bibfnamefont {A.~M.}\ \bibnamefont
			{Puertas}}, \ and\ \bibinfo {author} {\bibfnamefont {M.}~\bibnamefont
			{Fuchs}},\ }\bibfield  {title} {\enquote {\bibinfo {title} {Comparison of
				structure and transport properties of concentrated hard and soft sphere
				fluids},}\ }\href@noop {} {\bibfield  {journal} {\bibinfo  {journal} {J.
				Chem. Phys.}\ }\textbf {\bibinfo {volume} {130}},\ \bibinfo {pages} {174903}
		(\bibinfo {year} {2009})}\BibitemShut {NoStop}%
	\bibitem [{\citenamefont {Branka}\ and\ \citenamefont {Heyes}(2011)}]{bra11}%
	\BibitemOpen
	\bibfield  {author} {\bibinfo {author} {\bibfnamefont {A.~C.}\ \bibnamefont
			{Branka}}\ and\ \bibinfo {author} {\bibfnamefont {D.~M.}\ \bibnamefont
			{Heyes}},\ }\bibfield  {title} {\enquote {\bibinfo {title} {Pair correlation
				function of soft-sphere fluids},}\ }\href@noop {} {\bibfield  {journal}
		{\bibinfo  {journal} {J. Chem. Phys.}\ }\textbf {\bibinfo {volume} {134}},\
		\bibinfo {pages} {064115} (\bibinfo {year} {2011})}\BibitemShut {NoStop}%
	\bibitem [{\citenamefont {Pieprzyk}\ \emph {et~al.}(2014)\citenamefont
		{Pieprzyk}, \citenamefont {Heyes},\ and\ \citenamefont {Branka}}]{pie14}%
	\BibitemOpen
	\bibfield  {author} {\bibinfo {author} {\bibfnamefont {S.}~\bibnamefont
			{Pieprzyk}}, \bibinfo {author} {\bibfnamefont {D.~M.}\ \bibnamefont {Heyes}},
		\ and\ \bibinfo {author} {\bibfnamefont {A.~C.}\ \bibnamefont {Branka}},\
	}\bibfield  {title} {\enquote {\bibinfo {title} {Thermodynamic properties and
				entropy scaling law for diffusivity in soft spheres},}\ }\href {\doibase
		10.1103/PhysRevE.90.012106} {\bibfield  {journal} {\bibinfo  {journal} {Phys.
				Rev. E}\ }\textbf {\bibinfo {volume} {90}},\ \bibinfo {pages} {012106}
		(\bibinfo {year} {2014})}\BibitemShut {NoStop}%
	\bibitem [{\citenamefont {Ding}\ and\ \citenamefont {Mittal}(2015)}]{din15}%
	\BibitemOpen
	\bibfield  {author} {\bibinfo {author} {\bibfnamefont {Y.}~\bibnamefont
			{Ding}}\ and\ \bibinfo {author} {\bibfnamefont {J.}~\bibnamefont {Mittal}},\
	}\bibfield  {title} {\enquote {\bibinfo {title} {Equilibrium and
				nonequilibrium dynamics of soft sphere fluids},}\ }\href {\doibase
		10.1039/C5SM00637F} {\bibfield  {journal} {\bibinfo  {journal} {Soft Matter}\
		}\textbf {\bibinfo {volume} {11}},\ \bibinfo {pages} {5274--5281} (\bibinfo
		{year} {2015})}\BibitemShut {NoStop}%
	\bibitem [{\citenamefont {Born}\ and\ \citenamefont {Meyer}(1932)}]{bor32}%
	\BibitemOpen
	\bibfield  {author} {\bibinfo {author} {\bibfnamefont {M}~\bibnamefont
			{Born}}\ and\ \bibinfo {author} {\bibfnamefont {J.~E.}\ \bibnamefont
			{Meyer}},\ }\bibfield  {title} {\enquote {\bibinfo {title} {{Zur
					Gittertheorie der Ionenkristalle}},}\ }\href@noop {} {\bibfield  {journal}
		{\bibinfo  {journal} {Z. Phys.}\ }\textbf {\bibinfo {volume} {75}},\ \bibinfo
		{pages} {1--18} (\bibinfo {year} {1932})}\BibitemShut {NoStop}%
	\bibitem [{\citenamefont {Foiles}\ \emph {et~al.}(1986)\citenamefont {Foiles},
		\citenamefont {Baskes},\ and\ \citenamefont {Daw}}]{foi86}%
	\BibitemOpen
	\bibfield  {author} {\bibinfo {author} {\bibfnamefont {S.~M.}\ \bibnamefont
			{Foiles}}, \bibinfo {author} {\bibfnamefont {M.~I.}\ \bibnamefont {Baskes}},
		\ and\ \bibinfo {author} {\bibfnamefont {M.~S.}\ \bibnamefont {Daw}},\
	}\bibfield  {title} {\enquote {\bibinfo {title} {Embedded-atom-method
				functions for the fcc metals cu, ag, au, ni, pd, pt, and their alloys},}\
	}\href {\doibase 10.1103/PhysRevB.33.7983} {\bibfield  {journal} {\bibinfo
			{journal} {Phys. Rev. B}\ }\textbf {\bibinfo {volume} {33}},\ \bibinfo
		{pages} {7983--7991} (\bibinfo {year} {1986})}\BibitemShut {NoStop}%
	\bibitem [{\citenamefont {Lad}\ \emph {et~al.}(2017)\citenamefont {Lad},
		\citenamefont {Jakse},\ and\ \citenamefont {Pasturel}}]{lad17}%
	\BibitemOpen
	\bibfield  {author} {\bibinfo {author} {\bibfnamefont {K.~N.}\ \bibnamefont
			{Lad}}, \bibinfo {author} {\bibfnamefont {N.}~\bibnamefont {Jakse}}, \ and\
		\bibinfo {author} {\bibfnamefont {A.}~\bibnamefont {Pasturel}},\ }\bibfield
	{title} {\enquote {\bibinfo {title} {How closely do many-body potentials
				describe the structure and dynamics of {Cu-Zr} glass-forming alloy?}}\ }\href
	{\doibase 10.1063/1.4979125} {\bibfield  {journal} {\bibinfo  {journal} {J.
				Chem. Phys.}\ }\textbf {\bibinfo {volume} {146}},\ \bibinfo {pages} {124502}
		(\bibinfo {year} {2017})}\BibitemShut {NoStop}%
	\bibitem [{\citenamefont {Yukawa}(1935)}]{yuk35}%
	\BibitemOpen
	\bibfield  {author} {\bibinfo {author} {\bibfnamefont {H.}~\bibnamefont
			{Yukawa}},\ }\bibfield  {title} {\enquote {\bibinfo {title} {On the
				interaction of elementary particles},}\ }\href@noop {} {\bibfield  {journal}
		{\bibinfo  {journal} {Proc. Phys.-Math. Soc. Jpn.}\ }\textbf {\bibinfo
			{volume} {17}},\ \bibinfo {pages} {48--57} (\bibinfo {year}
		{1935})}\BibitemShut {NoStop}%
	\bibitem [{\citenamefont {Rowlinson}(1989)}]{row89}%
	\BibitemOpen
	\bibfield  {author} {\bibinfo {author} {\bibfnamefont {J.~S.}\ \bibnamefont
			{Rowlinson}},\ }\bibfield  {title} {\enquote {\bibinfo {title} {The {Yukawa}
				potential},}\ }\href@noop {} {\bibfield  {journal} {\bibinfo  {journal}
			{Physica A}\ }\textbf {\bibinfo {volume} {156}},\ \bibinfo {pages} {15--34}
		(\bibinfo {year} {1989})}\BibitemShut {NoStop}%
	\bibitem [{\citenamefont {Kac}\ \emph {et~al.}(1963)\citenamefont {Kac},
		\citenamefont {Uhlenbeck},\ and\ \citenamefont {Hemmer}}]{kac63}%
	\BibitemOpen
	\bibfield  {author} {\bibinfo {author} {\bibfnamefont {M.}~\bibnamefont
			{Kac}}, \bibinfo {author} {\bibfnamefont {G.~E.}\ \bibnamefont {Uhlenbeck}},
		\ and\ \bibinfo {author} {\bibfnamefont {P.~C.}\ \bibnamefont {Hemmer}},\
	}\bibfield  {title} {\enquote {\bibinfo {title} {On the van der {Waals}
				theory of the vapor-liquid equilibrium. i. discussion of a one-dimensional
				model},}\ }\href {\doibase http://dx.doi.org/10.1063/1.1703946} {\bibfield
		{journal} {\bibinfo  {journal} {J. Math. Phys.}\ }\textbf {\bibinfo {volume}
			{4}},\ \bibinfo {pages} {216--228} (\bibinfo {year} {1963})}\BibitemShut
	{NoStop}%
	\bibitem [{\citenamefont {Schr{\o}der}\ and\ \citenamefont
		{Dyre}(2014)}]{sch14}%
	\BibitemOpen
	\bibfield  {author} {\bibinfo {author} {\bibfnamefont {T.~B.}\ \bibnamefont
			{Schr{\o}der}}\ and\ \bibinfo {author} {\bibfnamefont {J.~C.}\ \bibnamefont
			{Dyre}},\ }\bibfield  {title} {\enquote {\bibinfo {title} {Simplicity of
				condensed matter at its core: Generic definition of a {Roskilde}-simple
				system},}\ }\href {\doibase http://dx.doi.org/10.1063/1.4901215} {\bibfield
		{journal} {\bibinfo  {journal} {J. Chem. Phys.}\ }\textbf {\bibinfo {volume}
			{141}},\ \bibinfo {eid} {204502} (\bibinfo {year} {2014})}\BibitemShut
	{NoStop}%
	\bibitem [{\citenamefont {Schr{\o}der}\ \emph
		{et~al.}(2009{\natexlab{a}})\citenamefont {Schr{\o}der}, \citenamefont
		{Pedersen}, \citenamefont {Bailey}, \citenamefont {Toxvaerd},\ and\
		\citenamefont {Dyre}}]{sch09}%
	\BibitemOpen
	\bibfield  {author} {\bibinfo {author} {\bibfnamefont {T.~B.}\ \bibnamefont
			{Schr{\o}der}}, \bibinfo {author} {\bibfnamefont {U.~R.}\ \bibnamefont
			{Pedersen}}, \bibinfo {author} {\bibfnamefont {N.~P.}\ \bibnamefont
			{Bailey}}, \bibinfo {author} {\bibfnamefont {S.}~\bibnamefont {Toxvaerd}}, \
		and\ \bibinfo {author} {\bibfnamefont {J.~C.}\ \bibnamefont {Dyre}},\
	}\bibfield  {title} {\enquote {\bibinfo {title} {{Hidden Scale Invariance in
					Molecular van der Waals Liquids: A Simulation Study}},}\ }\href@noop {}
	{\bibfield  {journal} {\bibinfo  {journal} {Phys. Rev. E}\ }\textbf {\bibinfo
			{volume} {80}},\ \bibinfo {pages} {041502} (\bibinfo {year}
		{2009}{\natexlab{a}})}\BibitemShut {NoStop}%
	\bibitem [{\citenamefont {Dyre}(2014)}]{dyr14}%
	\BibitemOpen
	\bibfield  {author} {\bibinfo {author} {\bibfnamefont {J.~C.}\ \bibnamefont
			{Dyre}},\ }\bibfield  {title} {\enquote {\bibinfo {title} {{Hidden Scale
					Invariance in Condensed Matter}},}\ }\href@noop {} {\bibfield  {journal}
		{\bibinfo  {journal} {J. Phys. Chem. B}\ }\textbf {\bibinfo {volume} {118}},\
		\bibinfo {pages} {10007--10024} (\bibinfo {year} {2014})}\BibitemShut
	{NoStop}%
	\bibitem [{\citenamefont {Pedersen}\ \emph {et~al.}(2008)\citenamefont
		{Pedersen}, \citenamefont {Bailey}, \citenamefont {Schr{\o}der},\ and\
		\citenamefont {Dyre}}]{ped08}%
	\BibitemOpen
	\bibfield  {author} {\bibinfo {author} {\bibfnamefont {U.~R.}\ \bibnamefont
			{Pedersen}}, \bibinfo {author} {\bibfnamefont {N.~P.}\ \bibnamefont
			{Bailey}}, \bibinfo {author} {\bibfnamefont {T.~B.}\ \bibnamefont
			{Schr{\o}der}}, \ and\ \bibinfo {author} {\bibfnamefont {J.~C.}\ \bibnamefont
			{Dyre}},\ }\bibfield  {title} {\enquote {\bibinfo {title} {{Strong
					Pressure-Energy Correlations in van der Waals Liquids}},}\ }\href@noop {}
	{\bibfield  {journal} {\bibinfo  {journal} {Phys. Rev. Lett.}\ }\textbf
		{\bibinfo {volume} {100}},\ \bibinfo {pages} {015701} (\bibinfo {year}
		{2008})}\BibitemShut {NoStop}%
	\bibitem [{\citenamefont {Bailey}\ \emph
		{et~al.}(2008{\natexlab{a}})\citenamefont {Bailey}, \citenamefont {Pedersen},
		\citenamefont {Gnan}, \citenamefont {Schr{\o}der},\ and\ \citenamefont
		{Dyre}}]{I}%
	\BibitemOpen
	\bibfield  {author} {\bibinfo {author} {\bibfnamefont {N.~P.}\ \bibnamefont
			{Bailey}}, \bibinfo {author} {\bibfnamefont {U.~R.}\ \bibnamefont
			{Pedersen}}, \bibinfo {author} {\bibfnamefont {N.}~\bibnamefont {Gnan}},
		\bibinfo {author} {\bibfnamefont {T.~B.}\ \bibnamefont {Schr{\o}der}}, \ and\
		\bibinfo {author} {\bibfnamefont {J.~C.}\ \bibnamefont {Dyre}},\ }\bibfield
	{title} {\enquote {\bibinfo {title} {{Pressure-Energy Correlations in
					Liquids. I. Results from Computer Simulations}},}\ }\href@noop {} {\bibfield
		{journal} {\bibinfo  {journal} {J. Chem. Phys.}\ }\textbf {\bibinfo {volume}
			{129}},\ \bibinfo {pages} {184507} (\bibinfo {year}
		{2008}{\natexlab{a}})}\BibitemShut {NoStop}%
	\bibitem [{\citenamefont {Bacher}\ \emph {et~al.}(2014)\citenamefont {Bacher},
		\citenamefont {Schr{\o}der},\ and\ \citenamefont {Dyre}}]{bac14a}%
	\BibitemOpen
	\bibfield  {author} {\bibinfo {author} {\bibfnamefont {A.~K.}\ \bibnamefont
			{Bacher}}, \bibinfo {author} {\bibfnamefont {T.~B.}\ \bibnamefont
			{Schr{\o}der}}, \ and\ \bibinfo {author} {\bibfnamefont {J.~C.}\ \bibnamefont
			{Dyre}},\ }\bibfield  {title} {\enquote {\bibinfo {title} {Explaining why
				simple liquids are quasi-universal},}\ }\href@noop {} {\bibfield  {journal}
		{\bibinfo  {journal} {Nat. Commun.}\ }\textbf {\bibinfo {volume} {5}},\
		\bibinfo {pages} {5424} (\bibinfo {year} {2014})}\BibitemShut {NoStop}%
	\bibitem [{\citenamefont {Gnan}\ \emph {et~al.}(2009)\citenamefont {Gnan},
		\citenamefont {Schr{\o}der}, \citenamefont {Pedersen}, \citenamefont
		{Bailey},\ and\ \citenamefont {Dyre}}]{IV}%
	\BibitemOpen
	\bibfield  {author} {\bibinfo {author} {\bibfnamefont {N.}~\bibnamefont
			{Gnan}}, \bibinfo {author} {\bibfnamefont {T.~B.}\ \bibnamefont
			{Schr{\o}der}}, \bibinfo {author} {\bibfnamefont {U.~R.}\ \bibnamefont
			{Pedersen}}, \bibinfo {author} {\bibfnamefont {N.~P.}\ \bibnamefont
			{Bailey}}, \ and\ \bibinfo {author} {\bibfnamefont {J.~C.}\ \bibnamefont
			{Dyre}},\ }\bibfield  {title} {\enquote {\bibinfo {title} {{Pressure-Energy
					Correlations in Liquids. IV. ``Isomorphs'' in Liquid Phase Diagrams}},}\
	}\href@noop {} {\bibfield  {journal} {\bibinfo  {journal} {J. Chem. Phys.}\
		}\textbf {\bibinfo {volume} {131}},\ \bibinfo {pages} {234504} (\bibinfo
		{year} {2009})}\BibitemShut {NoStop}%
	\bibitem [{\citenamefont {Rosenfeld}(1977)}]{ros77}%
	\BibitemOpen
	\bibfield  {author} {\bibinfo {author} {\bibfnamefont {Y.}~\bibnamefont
			{Rosenfeld}},\ }\bibfield  {title} {\enquote {\bibinfo {title} {{Relation
					between the Transport Coefficients and the Internal Entropy of Simple
					Systems}},}\ }\href@noop {} {\bibfield  {journal} {\bibinfo  {journal} {Phys.
				Rev. A}\ }\textbf {\bibinfo {volume} {15}},\ \bibinfo {pages} {2545--2549}
		(\bibinfo {year} {1977})}\BibitemShut {NoStop}%
	\bibitem [{\citenamefont {Ingebrigtsen}\ \emph
		{et~al.}(2012{\natexlab{a}})\citenamefont {Ingebrigtsen}, \citenamefont
		{Schr\o{}der},\ and\ \citenamefont {Dyre}}]{ing12}%
	\BibitemOpen
	\bibfield  {author} {\bibinfo {author} {\bibfnamefont {T.~S.}\ \bibnamefont
			{Ingebrigtsen}}, \bibinfo {author} {\bibfnamefont {T.~B.}\ \bibnamefont
			{Schr\o{}der}}, \ and\ \bibinfo {author} {\bibfnamefont {J.~C.}\ \bibnamefont
			{Dyre}},\ }\bibfield  {title} {\enquote {\bibinfo {title} {What is a simple
				liquid?}}\ }\href@noop {} {\bibfield  {journal} {\bibinfo  {journal} {Phys.
				Rev. X}\ }\textbf {\bibinfo {volume} {2}},\ \bibinfo {pages} {011011}
		(\bibinfo {year} {2012}{\natexlab{a}})}\BibitemShut {NoStop}%
	\bibitem [{\citenamefont {Dyre}(2016)}]{dyr16}%
	\BibitemOpen
	\bibfield  {author} {\bibinfo {author} {\bibfnamefont {J.~C.}\ \bibnamefont
			{Dyre}},\ }\bibfield  {title} {\enquote {\bibinfo {title} {Simple liquids'
				quasiuniversality and the hard-sphere paradigm},}\ }\href@noop {} {\bibfield
		{journal} {\bibinfo  {journal} {J. Phys. Condens. Matter}\ }\textbf {\bibinfo
			{volume} {28}},\ \bibinfo {pages} {323001} (\bibinfo {year}
		{2016})}\BibitemShut {NoStop}%
	\bibitem [{\citenamefont {Ingebrigtsen}\ \emph {et~al.}(2011)\citenamefont
		{Ingebrigtsen}, \citenamefont {Toxvaerd}, \citenamefont {Heilmann},
		\citenamefont {Schr{\o}der},\ and\ \citenamefont {Dyre}}]{NVU_I}%
	\BibitemOpen
	\bibfield  {author} {\bibinfo {author} {\bibfnamefont {T.~S.}\ \bibnamefont
			{Ingebrigtsen}}, \bibinfo {author} {\bibfnamefont {S.}~\bibnamefont
			{Toxvaerd}}, \bibinfo {author} {\bibfnamefont {O.~J.}\ \bibnamefont
			{Heilmann}}, \bibinfo {author} {\bibfnamefont {T.~B.}\ \bibnamefont
			{Schr{\o}der}}, \ and\ \bibinfo {author} {\bibfnamefont {J.~C.}\ \bibnamefont
			{Dyre}},\ }\bibfield  {title} {\enquote {\bibinfo {title} {{{\it NVU}
					dynamics. I. Geodesic Motion on the Constant-Potential-Energy
					Hypersurface}},}\ }\href@noop {} {\bibfield  {journal} {\bibinfo  {journal}
			{J. Chem. Phys.}\ }\textbf {\bibinfo {volume} {135}},\ \bibinfo {pages}
		{104101} (\bibinfo {year} {2011})}\BibitemShut {NoStop}%
	\bibitem [{\citenamefont {Andrade}(1931)}]{and31}%
	\BibitemOpen
	\bibfield  {author} {\bibinfo {author} {\bibfnamefont {E.~N.~C.}\
			\bibnamefont {Andrade}},\ }\bibfield  {title} {\enquote {\bibinfo {title}
			{{Viscosity of Liquids}},}\ }\href@noop {} {\bibfield  {journal} {\bibinfo
			{journal} {Nature}\ }\textbf {\bibinfo {volume} {128}},\ \bibinfo {pages}
		{835} (\bibinfo {year} {1931})}\BibitemShut {NoStop}%
	\bibitem [{\citenamefont {Hansen}\ and\ \citenamefont
		{McDonald}(2013)}]{han13}%
	\BibitemOpen
	\bibfield  {author} {\bibinfo {author} {\bibfnamefont {J.-P.}\ \bibnamefont
			{Hansen}}\ and\ \bibinfo {author} {\bibfnamefont {I.~R.}\ \bibnamefont
			{McDonald}},\ }\href@noop {} {\emph {\bibinfo {title} {{Theory of Simple
					Liquids: With Applications to Soft Matter}}}},\ \bibinfo {edition} {4th}\
	ed.\ (\bibinfo  {publisher} {Academic, New York},\ \bibinfo {year}
	{2013})\BibitemShut {NoStop}%
	\bibitem [{\citenamefont {Pedersen}\ \emph {et~al.}(2016)\citenamefont
		{Pedersen}, \citenamefont {Costigliola}, \citenamefont {Bailey},
		\citenamefont {Schr{\o}der},\ and\ \citenamefont {Dyre}}]{ped16}%
	\BibitemOpen
	\bibfield  {author} {\bibinfo {author} {\bibfnamefont {U.~R.}\ \bibnamefont
			{Pedersen}}, \bibinfo {author} {\bibfnamefont {L.}~\bibnamefont
			{Costigliola}}, \bibinfo {author} {\bibfnamefont {N.~P.}\ \bibnamefont
			{Bailey}}, \bibinfo {author} {\bibfnamefont {T.~B}\ \bibnamefont
			{Schr{\o}der}}, \ and\ \bibinfo {author} {\bibfnamefont {J.~C.}\ \bibnamefont
			{Dyre}},\ }\bibfield  {title} {\enquote {\bibinfo {title} {Thermodynamics of
				freezing and melting},}\ }\href@noop {} {\bibfield  {journal} {\bibinfo
			{journal} {Nat. Commun.}\ }\textbf {\bibinfo {volume} {7}},\ \bibinfo {pages}
		{12386} (\bibinfo {year} {2016})}\BibitemShut {NoStop}%
	\bibitem [{\citenamefont {Ingebrigtsen}\ \emph
		{et~al.}(2012{\natexlab{b}})\citenamefont {Ingebrigtsen}, \citenamefont
		{Schr{\o}der},\ and\ \citenamefont {Dyre}}]{ing12b}%
	\BibitemOpen
	\bibfield  {author} {\bibinfo {author} {\bibfnamefont {T.~S.}\ \bibnamefont
			{Ingebrigtsen}}, \bibinfo {author} {\bibfnamefont {T.~B.}\ \bibnamefont
			{Schr{\o}der}}, \ and\ \bibinfo {author} {\bibfnamefont {J.~C.}\ \bibnamefont
			{Dyre}},\ }\bibfield  {title} {\enquote {\bibinfo {title} {{Isomorphs in
					Model Molecular Liquids}},}\ }\href@noop {} {\bibfield  {journal} {\bibinfo
			{journal} {J. Phys. Chem. B}\ }\textbf {\bibinfo {volume} {116}},\ \bibinfo
		{pages} {1018--1034} (\bibinfo {year} {2012}{\natexlab{b}})}\BibitemShut
	{NoStop}%
	\bibitem [{\citenamefont {Veldhorst}\ \emph {et~al.}(2014)\citenamefont
		{Veldhorst}, \citenamefont {Dyre},\ and\ \citenamefont
		{Schr{\o}der}}]{vel14}%
	\BibitemOpen
	\bibfield  {author} {\bibinfo {author} {\bibfnamefont {A.~A.}\ \bibnamefont
			{Veldhorst}}, \bibinfo {author} {\bibfnamefont {J.~C.}\ \bibnamefont {Dyre}},
		\ and\ \bibinfo {author} {\bibfnamefont {T.~B.}\ \bibnamefont
			{Schr{\o}der}},\ }\bibfield  {title} {\enquote {\bibinfo {title} {{Scaling of
					the Dynamics of Flexible Lennard-Jones Chains}},}\ }\href@noop {} {\bibfield
		{journal} {\bibinfo  {journal} {J. Chem. Phys.}\ }\textbf {\bibinfo {volume}
			{141}},\ \bibinfo {pages} {054904} (\bibinfo {year} {2014})}\BibitemShut
	{NoStop}%
	\bibitem [{\citenamefont {Schr{\o}der}\ \emph
		{et~al.}(2009{\natexlab{b}})\citenamefont {Schr{\o}der}, \citenamefont
		{Bailey}, \citenamefont {Pedersen}, \citenamefont {Gnan},\ and\ \citenamefont
		{Dyre}}]{III}%
	\BibitemOpen
	\bibfield  {author} {\bibinfo {author} {\bibfnamefont {T.~B.}\ \bibnamefont
			{Schr{\o}der}}, \bibinfo {author} {\bibfnamefont {N.~P.}\ \bibnamefont
			{Bailey}}, \bibinfo {author} {\bibfnamefont {U.~R.}\ \bibnamefont
			{Pedersen}}, \bibinfo {author} {\bibfnamefont {N.}~\bibnamefont {Gnan}}, \
		and\ \bibinfo {author} {\bibfnamefont {J.~C.}\ \bibnamefont {Dyre}},\
	}\bibfield  {title} {\enquote {\bibinfo {title} {{Pressure-Energy
					Correlations in Liquids. III. Statistical Mechanics and Thermodynamics of
					Liquids with Hidden Scale Invariance}},}\ }\href@noop {} {\bibfield
		{journal} {\bibinfo  {journal} {J. Chem. Phys.}\ }\textbf {\bibinfo {volume}
			{131}},\ \bibinfo {pages} {234503} (\bibinfo {year}
		{2009}{\natexlab{b}})}\BibitemShut {NoStop}%
	\bibitem [{\citenamefont {Roland}\ \emph {et~al.}(2005)\citenamefont {Roland},
		\citenamefont {Hensel-Bielowka}, \citenamefont {Paluch},\ and\ \citenamefont
		{Casalini}}]{rol05}%
	\BibitemOpen
	\bibfield  {author} {\bibinfo {author} {\bibfnamefont {C.~M.}\ \bibnamefont
			{Roland}}, \bibinfo {author} {\bibfnamefont {S.}~\bibnamefont
			{Hensel-Bielowka}}, \bibinfo {author} {\bibfnamefont {M.}~\bibnamefont
			{Paluch}}, \ and\ \bibinfo {author} {\bibfnamefont {R.}~\bibnamefont
			{Casalini}},\ }\bibfield  {title} {\enquote {\bibinfo {title} {{Supercooled
					Dynamics of Glass-Forming Liquids and Polymers under Hydrostatic
					Pressure}},}\ }\href@noop {} {\bibfield  {journal} {\bibinfo  {journal} {Rep.
				Prog. Phys.}\ }\textbf {\bibinfo {volume} {68}},\ \bibinfo {pages}
		{1405--1478} (\bibinfo {year} {2005})}\BibitemShut {NoStop}%
	\bibitem [{\citenamefont {Bailey}\ \emph
		{et~al.}(2008{\natexlab{b}})\citenamefont {Bailey}, \citenamefont {Pedersen},
		\citenamefont {Gnan}, \citenamefont {Schr{\o}der},\ and\ \citenamefont
		{Dyre}}]{II}%
	\BibitemOpen
	\bibfield  {author} {\bibinfo {author} {\bibfnamefont {N.~P.}\ \bibnamefont
			{Bailey}}, \bibinfo {author} {\bibfnamefont {U.~R.}\ \bibnamefont
			{Pedersen}}, \bibinfo {author} {\bibfnamefont {N.}~\bibnamefont {Gnan}},
		\bibinfo {author} {\bibfnamefont {T.~B.}\ \bibnamefont {Schr{\o}der}}, \ and\
		\bibinfo {author} {\bibfnamefont {J.~C.}\ \bibnamefont {Dyre}},\ }\bibfield
	{title} {\enquote {\bibinfo {title} {{Pressure-Energy Correlations in
					Liquids. II. Analysis and Consequences}},}\ }\href@noop {} {\bibfield
		{journal} {\bibinfo  {journal} {J. Chem. Phys.}\ }\textbf {\bibinfo {volume}
			{129}},\ \bibinfo {pages} {184508} (\bibinfo {year}
		{2008}{\natexlab{b}})}\BibitemShut {NoStop}%
	\bibitem [{\citenamefont {Veldhorst}\ \emph {et~al.}(2015)\citenamefont
		{Veldhorst}, \citenamefont {Schr{\o}der},\ and\ \citenamefont
		{Dyre}}]{vel15}%
	\BibitemOpen
	\bibfield  {author} {\bibinfo {author} {\bibfnamefont {A.~A.}\ \bibnamefont
			{Veldhorst}}, \bibinfo {author} {\bibfnamefont {T.~B}\ \bibnamefont
			{Schr{\o}der}}, \ and\ \bibinfo {author} {\bibfnamefont {J.~C.}\ \bibnamefont
			{Dyre}},\ }\bibfield  {title} {\enquote {\bibinfo {title} {Invariants in the
				{Yukawa} system's thermodynamic phase diagram},}\ }\href@noop {} {\bibfield
		{journal} {\bibinfo  {journal} {Phys. Plasmas}\ }\textbf {\bibinfo {volume}
			{22}},\ \bibinfo {pages} {073705} (\bibinfo {year} {2015})}\BibitemShut
	{NoStop}%
	\bibitem [{\citenamefont {Costigliola}\ \emph {et~al.}(2016)\citenamefont
		{Costigliola}, \citenamefont {Schr\o{}der},\ and\ \citenamefont
		{Dyre}}]{cos16a}%
	\BibitemOpen
	\bibfield  {author} {\bibinfo {author} {\bibfnamefont {L.}~\bibnamefont
			{Costigliola}}, \bibinfo {author} {\bibfnamefont {T.~B.}\ \bibnamefont
			{Schr\o{}der}}, \ and\ \bibinfo {author} {\bibfnamefont {J.~C.}\ \bibnamefont
			{Dyre}},\ }\bibfield  {title} {\enquote {\bibinfo {title} {Communication:
				Studies of the lennard-jones fluid in 2, 3, and 4 dimensions highlight the
				need for a liquid-state 1/d expansion},}\ }\href@noop {} {\bibfield
		{journal} {\bibinfo  {journal} {J. Chem. Phys.}\ }\textbf {\bibinfo {volume}
			{144}},\ \bibinfo {pages} {231101} (\bibinfo {year} {2016})}\BibitemShut
	{NoStop}%
	\bibitem [{\citenamefont {Ingebrigtsen}\ \emph
		{et~al.}(2012{\natexlab{c}})\citenamefont {Ingebrigtsen}, \citenamefont
		{B{\o}hling}, \citenamefont {Schr{\o}der},\ and\ \citenamefont
		{Dyre}}]{ing12a}%
	\BibitemOpen
	\bibfield  {author} {\bibinfo {author} {\bibfnamefont {T.~S.}\ \bibnamefont
			{Ingebrigtsen}}, \bibinfo {author} {\bibfnamefont {L.}~\bibnamefont
			{B{\o}hling}}, \bibinfo {author} {\bibfnamefont {T.~B.}\ \bibnamefont
			{Schr{\o}der}}, \ and\ \bibinfo {author} {\bibfnamefont {J.~C.}\ \bibnamefont
			{Dyre}},\ }\bibfield  {title} {\enquote {\bibinfo {title} {{Thermodynamics of
					Condensed Matter with Strong Pressure-Energy Correlations}},}\ }\href@noop {}
	{\bibfield  {journal} {\bibinfo  {journal} {J. Chem. Phys.}\ }\textbf
		{\bibinfo {volume} {136}},\ \bibinfo {pages} {061102} (\bibinfo {year}
		{2012}{\natexlab{c}})}\BibitemShut {NoStop}%
\end{thebibliography}
\end{document}